\PassOptionsToPackage{table}{xcolor}
\documentclass[acmsmall,screen]{acmart}

\setcopyright{none}
\renewcommand\footnotetextcopyrightpermission[1]{}
\fancypagestyle{arxivstyle}{%
  \fancyhf{}%
  \fancyfoot[C]{\thepage}%
}

\makeatletter
\def\input@path{{./}{latex/}}
\makeatother

\title{A Survey of Code Review Benchmarks and Evaluation Practices in Pre-LLM and LLM Era}

\author{Taufiqul Islam Khan}
\affiliation{
  \institution{University of Manitoba, Department of Computer Science}
  \country{Canada}
}
\email{khanti@myumanitoba.ca}

\author{Shaowei Wang}
\affiliation{
  \institution{University of Manitoba, Department of Computer Science}
  \country{Canada}
}
\email{shaowei.wang@umanitoba.ca}

\author{Haoxiang Zhang}
\affiliation{
  \institution{Huawei Canada, Centre for Software Excellence}
  \country{Canada}
}
\email{haoxiang.zhang@acm.org}

\author{Tse-Hsun Chen}
\affiliation{%
  \institution{Concordia University}
  \city{Montreal}
  \country{Canada}
}
\email{peterc@encs.concordia.ca}


\usepackage{amsmath,amssymb,mathtools}
\usepackage{graphicx}
\usepackage{wrapfig}
\usepackage{longtable}
\usepackage{ltablex}
\keepXColumns
\usepackage{multirow}
\usepackage{booktabs}
\usepackage[table]{xcolor}
\usepackage{colortbl}
\usepackage{array}
\usepackage{tabulary}     
\usepackage[flushleft]{threeparttable}
\usepackage{tablefootnote}
\usepackage{balance}      
\usepackage{adjustbox} 
\usepackage{graphicx}
\usepackage{subcaption}
\DeclareUnicodeCharacter{2265}{$\geq$}

\usepackage{tikz}
\usepackage{tcolorbox}
\usepackage{circledsteps}

\usepackage[ruled, lined, linesnumbered, commentsnumbered, longend]{algorithm2e}

\SetCommentSty{mycommfont}

\graphicspath{{./}{latex/}{figures/}{latex/figures/}}

\begin{document}

\begin{abstract}
Code review is a critical practice in modern software engineering, helping developers detect defects early, improve code quality, and facilitate knowledge sharing. With the rapid advancement of large language models (LLMs), a growing body of work has explored automated support for code review. However, progress in this area is hindered by the lack of a systematic understanding of existing benchmarks and evaluation practices. Current code review datasets are scattered, vary widely in design, and provide limited insight into what review capabilities are actually being assessed. In this paper, we present a comprehensive survey of code review benchmarks spanning both the Pre-LLM and LLM eras (2015–2025). We analyze 104 research papers (60 Pre-LLM era and 44 LLM era) and extract key metadata, including datasets, evaluation metrics, data sources, and target tasks. Based on this analysis, we propose a multi-level taxonomy that organizes code review research into five domains and 18 fine-grained tasks. Our study reveals an increased emphasis on generative peer-review results, increased multilingual coverage, and a decline in standalone change-understanding tasks. We further identify critical gaps in current benchmarks, such as missing macro-level and structural review tasks, reliance on misaligned or uncalibrated evaluation metrics, and the limited use of execution-grounded assessment. We derive four taxonomy-driven design principles to address them: broader task coverage, execution-grounded evaluation, calibrated multi-reference assessment, and cross-language generalization testing. This survey provides a structured foundation for developing more realistic and comprehensive benchmarks for LLM-based code review.

\end{abstract}

\maketitle
\pagestyle{arxivstyle}
\thispagestyle{arxivstyle}

\noindent\textit{This article has been accepted for publication in the Empirical Software Engineering (EMSE) journal. This is the authors' accepted version of the manuscript.}

\section{Introduction}\label{sec:intro}

Code review is an important practice in software engineering where developers inspect code changes before they are merged into the main branch. It has become an essential activity in modern software development as reviewing code changes, especially early on, helps developers find code defects or design issues before such failures grow into bigger problems. Code review makes software easier to maintain and safer to release. In addition, code review facilitates collaborative learning and helps train new team members~\citep{bacchelli2013expectations}. When developers read and comment on a code change, they may suggest clearer or more efficient ways to write the code. These discussions, explanations, and suggestions help teammates understand the whole project better and learn good coding practices. Research has shown that early code review can greatly reduce defects before testing and deployment~\citep{mcintosh2016empirical,bosu2016process}. 

However, manual code review is time-consuming and often requires significant effort from development teams \citep{bosu2016process}. The recent emergence of Large Language Models (LLMs) has led to the development of various automation approaches, which show promising results in streamlining the code review process \citep{sun2025bitsai,wang2024unity}. Since LLMs are typically trained on vast code repositories, including commit messages and discussions, they are well-equipped to review pull requests, suggest improvements, and identify potential issues within code changes.

To properly measure the proposed code review approaches, we need reliable benchmarks to analyze and understand. Benchmarks demonstrate how well approaches perform on real code review scenarios. They provide a standard dataset and evaluation metrics so that approaches can be compared in a fair and consistent way. 
A few code review benchmarks have been created recently, such as CodeReviewer~\citep{LP1}, CodeReviewQA~\citep{LP9}, and CodeFuse-CR-Bench~\citep{LP10}, but these resources are scattered and are not well organized or widely known. In other areas, such as code generation~\citep{related2}, there are survey papers that summarize available benchmarks and evaluation strategies, which help researchers understand which datasets to use and how to measure performance in a particular problem statement. For code review, we still have limited knowledge about how existing benchmarks are built, what review skills they test, and how closely they reflect real code review practice. This gap presents the need for a more systematic study of code review benchmarks for LLMs.

To address this gap, we conducted a comprehensive survey that examines code review benchmarks across both the Pre-LLM and LLM eras from 2015 to present. Our goal is to gather datasets that have been used in prior code review papers, thereby identifying the code review tasks that have been measured and analyzed, and understanding how these datasets are used to measure the code review tasks. We also compare traditional benchmarks in Pre-LLM era with newer LLM-focused ones to understand how the datasets and evaluation strategy of code review have changed over time. To conduct a systematic survey of code review datasets and evaluation metrics, we have collected 60 Pre-LLM code review papers and 44 LLM era papers published between Jan. 2015 and Dec. 2025. These papers cover a wide range of areas within code review. We analyzed key metadata from each study, which includes the datasets, evaluation metrics, and the specific code review tasks involved in that research. 

Our survey provides a comprehensive analysis of the code review landscape, offering the following key contributions:

\begin{itemize}
    \item \textbf{A Multi-Level Taxonomy of Code Review Tasks:} We categorize existing studies into five high-level domains—Review Prioritization/Selection, Change Understanding and Analysis, Peer Review, Review Assessment and Analysis, and Code Refinement—further subdivided into 18 specific sub-tasks. This framework enables researchers to identify the most suitable benchmarks for specific research objectives.

    \item \textbf{Systematic Classification of Benchmarks and Metrics:} We classify prior research by evaluation metrics, data sources, and granularity levels. This systematic mapping assists researchers in selecting appropriate metrics and facilitates more robust cross-study comparisons.

    \item \textbf{Shift from Pre-LLM to LLM Era.} We observe a clear shift in research focus between the two eras. 
    In the Pre-LLM era, Change Understanding and Analysis was a central research direction, supported by 16 of the 60 Pre-LLM datasets (27\%); in the LLM era, it appears as a standalone task in only one of the 31 LLM-era datasets (3\%). This does not indicate that change understanding has disappeared as a capability—rather, tasks such as change decomposition and impact analysis are now less frequently framed as independent benchmark objectives. Instead, these capabilities are often evaluated implicitly within broader LLM era tasks, such as issue localization and security detection.

    At the same time, peer review tasks have become the dominant focus in the LLM era, where 22 of the 31 LLM-era datasets (71\%) support at least one Peer Review sub-task, compared with 14 of the 60 Pre-LLM datasets (23\%). By contrast, Pre-LLM datasets were more evenly distributed across categories. This shift is consistent with a stronger emphasis on generative peer-review outputs, such as review comment generation, and with the consolidation of multiple review activities into broader review settings in some studies. However, this trend is not uniform across all benchmarks: CodeReviewQA, for instance, decomposes review reasoning into finer-grained capabilities, including change-type recognition, change localization, and solution identification. Our finding is therefore not that all LLM era benchmarks are monolithic or end-to-end, but rather that the LLM era shows a stronger emphasis on generative review outputs, while only a subset of benchmarks preserve or reintroduce fine-grained task decomposition.

    Finally, we observe a shift from language-specific studies toward cross-language generalization. While 58\% of Pre-LLM datasets focus on a single programming language, 16\% of LLM era datasets cover nine or more programming languages, indicating growing attention to multilingual benchmark design.

    \item \textbf{Future Directions for Improving LLM Benchmarks:} We identify limitations of current benchmarks for LLM and propose future directions, in terms of improving task coverage (e.g., developing LLM benchmark for macro-level review responsibilities such as impact analysis and commit decomposition), transitioning from Static to Dynamic Evaluation (e.g., developing benchmark include runtime evaluation such as build success rate, test case verification), and granular benchmarking via Task Taxonomy (e.g., measure an LLM’s issue resolving capabilities for sub-tasks).
\end{itemize}

Our contribution is not to claim the first systematic survey of code review research in general; rather, we provide a benchmark-centered and LLM era-aware analysis of code review datasets, evaluation metrics, task taxonomy, model settings, and cross-era evaluation shifts.
\section{Background \& Related work}\label{sec:relatedWorkSafeguard}

\subsection{Code Review process}


\begin{figure*}  
    \centering
    \includegraphics[width=\linewidth]{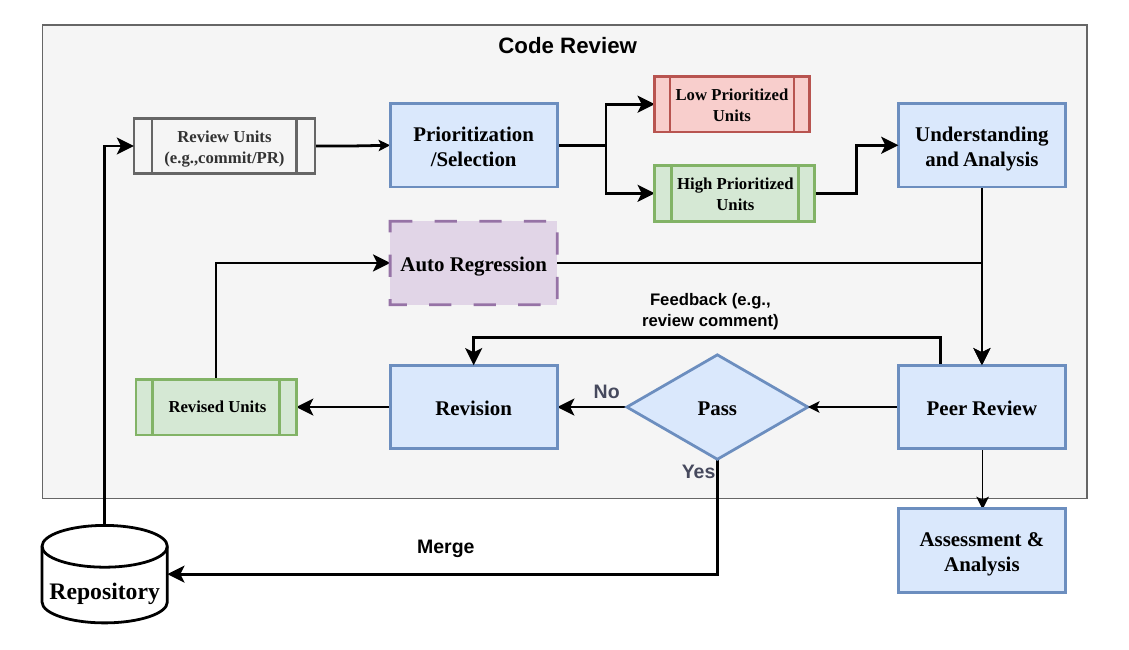}
    \vspace{-0.1in}
    \caption{Code Review Process.}
    \label{fig:code_review_process}
\end{figure*}


Figure~\ref{fig:code_review_process} presents an overview of the code review process. The code review process starts when a developer submits a review unit, which could be of different granularity, such as chunk-level, file-level, commit-level, or pull request (PR)-level. The code review process includes the following steps: 

\textbf{Prioritization/Selection.} While doing the analysis, some commits appear not to contain much valuable information or work that would be prioritized in a low rank. Those commits demonstrate that every change is not equally important. Therefore, before formal review, the first task is to decide which commits are worth reviewing and are given a high priority. This priority can come from information such as the issue description, the part of the code that was changed, or how urgent the fix is. Low-impact changes are often deferred, while commits involving major bug fixes, new features, or critical code paths are often prioritized.

\textbf{Understanding and Analysis.} Next, the high-priority code changes are moved forward to the next step. During this step, useful background information is collected to help reviewers understand the change. This includes details about the affected files, related issues, and any historical information that may help explain the motivation behind the update. A short summary of the code modifications is also prepared so the reviewer can quickly see what was changed, why it was changed, and how large the update is. 

\textbf{Peer Review.} With this context, the reviewer starts the actual evaluation of the commit. They check whether a code change is logically correct, follows project standards, and does not introduce new risks. During this stage, reviewers leave comments and suggestions pointing out unclear logic, potential bugs, style issues, or areas that need improvement.

\textbf{Assessment \& Analysis.} After peer review feedback is produced, the resulting review artifacts and interactions can be further examined to understand the quality and effectiveness of the review process. This stage includes analyzing whether review comments are constructive, whether they correctly refer to the relevant code, whether the feedback is useful to developers, and whether the review discussion reflects productive collaboration. In this way, assessment and analysis complements peer review by evaluating the feedback and interactions generated during the review process.

\textbf{Revision.} Once the review is complete, a decision is made about whether the commit is acceptable. If the reviewer finds the change correct and complete, it proceeds to merging. If issues are found, the commit is returned to the developer for revisions. The developer then revisits the units in the code repository, updates the implementation based on the feedback, and makes any necessary adjustments to related tests or documentation. Typically, automated regression tests may run to ensure the revised code does not break existing functionality. The updated code changes then enter the cycle again, going through review until it satisfies all the requirements and is finally merged into the main branch of the codebase. This cycle continues until the change becomes stable, correct, and ready for integration.

\subsection{Survey on coding tasks benchmark}

\citet{related1} critically review evaluation practices for LLM-based code generation, arguing that benchmarks and metrics remain immature despite rapid model progress. They have surveyed major models and common benchmarks (e.g., HumanEval, APPS, MBPP) and analyzed how tasks are sourced, cleaned, decontaminated, and structured across different granularities and languages. They compare similarity metrics (BLEU, ROUGE, CodeBLEU) with execution-based ones (pass rates, pass@k), noting their weak correlation with true correctness and developer usefulness. They call for richer, scenario-driven benchmarks with meaningful metadata. However, unlike code generation, no systematic benchmark study yet exists for code review tasks, leaving evaluation inconsistent and incomplete.

\citet{related2} surveys how benchmarks for code LLMs and agents align with the SDLC phases, covering tasks such as NL-to-code generation, completion, translation, bug fixing, test generation, understanding, and security analysis. They catalog representative datasets, inputs/outputs, languages, and evaluation metrics, and identify gaps in multi-round interactions, cross-phase workflows, and tasks requiring coordinated capabilities. While the benchmark ecosystem is broad, it remains unbalanced. A major missing area is code review: despite its central role in quality and maintenance, there is no systematic SDLC-aware benchmark landscape for review tasks, limiting fair comparison and realistic evaluation of LLM-based review assistance.

\citet{related4} reviewed 112 code-review studies from 2011 to 2019, examining their methodologies and evaluation methods. They find evaluation-focused work dominates, while solution proposals remain limited. Most studies rely on Gerrit/GitHub data, yet only about half release replicable datasets, hindering comparability. The authors identify 457 metrics across sixteen categories, showing fragmented definitions of review quality and behavior. Dataset choices are also inconsistent, with most papers building their own. The study concludes that the field lacks a standardized benchmark for code-review tasks.  


Several recent works have surveyed modern code review more broadly. \citep{related6} conducted a systematic literature review and taxonomy of modern code review. Their work provides an important overview of modern code review research and organizes prior studies at the level of research contribution and topic. More recently, \cite{related3} surveyed modern code review progress, challenges, and opportunities. Their survey analyzes 231 papers and organizes the field around Improvement Techniques and Understanding Studies, while also summarizing research trends, methodologies, tools, empirical insights, challenges, and future opportunities. Their primary focus is not the benchmark and evaluation landscape itself. In contrast, our survey specifically examines how code review tasks have been benchmarked and evaluated. We extract and compare benchmark-related metadata, including target task, dataset, data source, granularity, programming language coverage, evaluation metrics, and LLM era evaluation settings.
Recent work has also begun to propose new benchmarks for LLM-based code review. \cite{LP19} introduce SWR-Bench, a PR-centric benchmark consisting of 1,000 manually verified pull requests from GitHub. Their benchmark provides full project context and evaluates generated reviews using an objective LLM-based method that checks whether issues from a structured ground truth are covered in the generated review.Zeng et al. focus on constructing and studying one specific PR-centric benchmark, whereas our work surveys the broader benchmark ecosystem across multiple code review tasks and eras. 

Our work is not positioned as the first systematic survey of code review in general. Instead, it is positioned as a benchmark-centered and LLM era aware synthesis of code review evaluation practices. Existing surveys provide broad mappings of modern code review research, but they do not primarily examine how code review benchmarks are constructed across Pre-LLM and LLM era studies, what fine-grained review capabilities they evaluate, how evaluation protocols differ across eras, or how LLM era issues such as prompting, fine-tuning, LLM-as-Judge evaluation, dataset reuse, and benchmark contamination affect evaluation validity. Therefore, our survey is complementary to prior work: we focus specifically on code review benchmarks and evaluation practices, compare how they differ between the Pre-LLM and LLM eras, and organize them into a fine-grained taxonomy of five high-level domains and 18 sub-tasks.


\begin{table}[ht]
\centering
\scriptsize
\setlength{\tabcolsep}{1pt}
\renewcommand{\arraystretch}{1}
\setlength{\arrayrulewidth}{0.8pt}
\setlength{\aboverulesep}{0pt}
\setlength{\belowrulesep}{0pt}
\rowcolors{2}{gray!10}{white}
\resizebox{\columnwidth}{!}{%
\begin{tabular}{>{\scriptsize\arraybackslash}p{1.4cm}|>{\scriptsize\arraybackslash}p{2.8cm}|>{\scriptsize\arraybackslash}p{1.4cm}|>{\scriptsize\arraybackslash}p{2.8cm}|>{\scriptsize\arraybackslash}p{2.5cm}|>{\scriptsize\arraybackslash}p{2.0cm}}
\toprule
{\textbf{Literature}} &
{\textbf{Research Focus}} &
{\textbf{Search Period}} &
{\textbf{Classification Scheme}} &
{\textbf{Evaluation Dimensions}} &
{\textbf{Primary Studies}} \\
\midrule

{\citet{related4}} &
{Benchmarking potential of code review research: methodology, datasets, and metrics} &
{2011--2019} &
{7 contribution and methodology types; replication categories (private, partial, public); 16 metric categories across 9 code review research topics} &
{Benchmarking feasibility (concluded not yet mature); dataset replicability (${\sim}50\%$ publicly available); metric fragmentation; dataset inconsistency} &
{112 papers} \\

{\citep{related6}} &
{Broad systematic review of modern code review: foundational studies, proposals, and evaluations} &
{Up to 2019} &
{Foundational Studies (7 subcategories); Proposals (3 phases, 9 subtypes); Evaluations (4 types)} &
{Offline evaluations dominate (64.6\%); reviewer recommendation most studied; few user studies; precision and recall as primary metrics} &
{139 papers} \\

{\citet{related1}} &
{LLM code generation benchmarks and evaluation metrics} &
{N/A} &
{Task types (document-to-code, code-to-code); generation granularity (statement, function, class, program); benchmark dimensions: source, processing, structure, metadata} &
{Correctness (pass@k, AvgTPR); similarity (BLEU, CodeBLEU); misalignment between metrics and practice} &
{N/A} \\

{\cite{related3}} &
{MCR automation techniques and empirical understanding studies} &
{2013--2024} &
{Improvement Techniques (code change analysis, reviewer recommendation, review comment analysis); Understanding Studies (influence factors, metrics, root cause analysis, user study)} &
{MCR automation effectiveness; empirical insights on technique performance; research trends and open challenges} &
{231 papers} \\

{\citet{related2}} &
{LLM and agent benchmarks across all SDLC phases} &
{2022--2025} &
{5 SDLC phases; two-tier framework: general dimensions (anti-contamination, authenticity, interaction mode, evaluation paradigm) and phase-specific dimensions} &
{Phase coverage imbalance (${\sim}61\%$ on implementation); benchmark alignment with real-world SE tasks; code review as an under-studied phase} &
{461 papers} \\

{ \cite{LP19}} &
{Pull-request-level benchmark for LLM-based code review (SWR-Bench)} &
{N/A} &
{Single task (line-level issue localization); structured ground truth with LLM-based objective evaluation} &
{Coverage of ground-truth issues in generated reviews; Precision, Recall, F-measure} &
{N/A} \\

{Khan et al. (Ours)} &
{Code review benchmarks and evaluation practices across Pre-LLM and LLM eras} &
{2015--2025} &
{5 high-level domains; 18 fine-grained sub-tasks (Review Prioritization and Selection, Change Understanding and Analysis, Peer Review, Review Assessment and Analysis, Code Refinement)} &
{Era-level benchmark evolution; macro-level task coverage gaps; metric adequacy (lexical vs.\ execution-based); cross-language generalization} &
{104 papers} \\
\bottomrule
\end{tabular}%
}
\caption{{Comparison of related works and our survey}}
\label{tab:survey_comparison}
\end{table}
\section{Methodology}\label{sec:method}

 We conduct our investigation around three research questions (RQ1–RQ3). Together, these questions examine how code review approaches have been benchmarked and evaluated in both the Pre-LLM and LLM eras, helping us reveal established practices, new trends, and gaps that still need to be addressed.

\begin{itemize}
    \item \textbf{RQ1: What tasks and evaluation strategies were used for code review studies in the pre-LLM era?}  
    
    This RQ explores what were the common tasks and how such tasks were evaluated in code review studies before the introduction of LLMs, when traditional machine-learning or heuristic-based methods dominated. Understanding the tasks, datasets, and evaluation metrics used in this earlier era helps us establish a baseline for how the software engineering community historically investigated code review quality, defect detection, reviewer support, and other core skills. By mapping these foundational tasks and their evaluation strategies, we can better identify what remains relevant in the LLM era, what the limitations are that motivated the shift toward LLM-based techniques, and where current benchmarks may still lack coverage.

    \item \textbf{RQ2: What task and evaluation strategies are used for code review studies in the LLM era?} 
    
    This question focuses on understanding how the tasks and their evaluation approaches for code review have evolved in the LLM era. By examining the tasks and datasets designed specifically for LLM-based systems, we identify what tasks the SE research community is focusing on and expect these models to demonstrate their abilities. Studying the tasks and their evaluation setups used in these works helps reveal how LLM performance is currently measured, and where gaps or inconsistencies may still exist in benchmarking code review.

    \item \textbf{RQ3: How have tasks and evaluation strategies in code review evolved from the Pre-LLM to the LLM era?}  
    
    The emergence of LLMs has fundamentally changed the landscape of code review research, necessitating a critical examination of how task definitions, dataset characteristics, and evaluation strategies have evolved. By systematically comparing the Pre-LLM era with the LLM era, we can understand the trend when integrating LLMs in code review. Crucially, understanding this transition allows us to identify the inherent limitations of current LLM era datasets, such as the ``vanishing'' of code review activities. Identifying the evolution of code review research is crucial for understanding whether the current direction of code review research is unintentionally overlooking important ways of measuring how well a model performs code review tasks.
\end{itemize}

\subsection{Paper Collection}
\label{subsec:Paper Collection}
\begin{figure}[!ht]
    \centering
    \includegraphics[width=\textwidth]{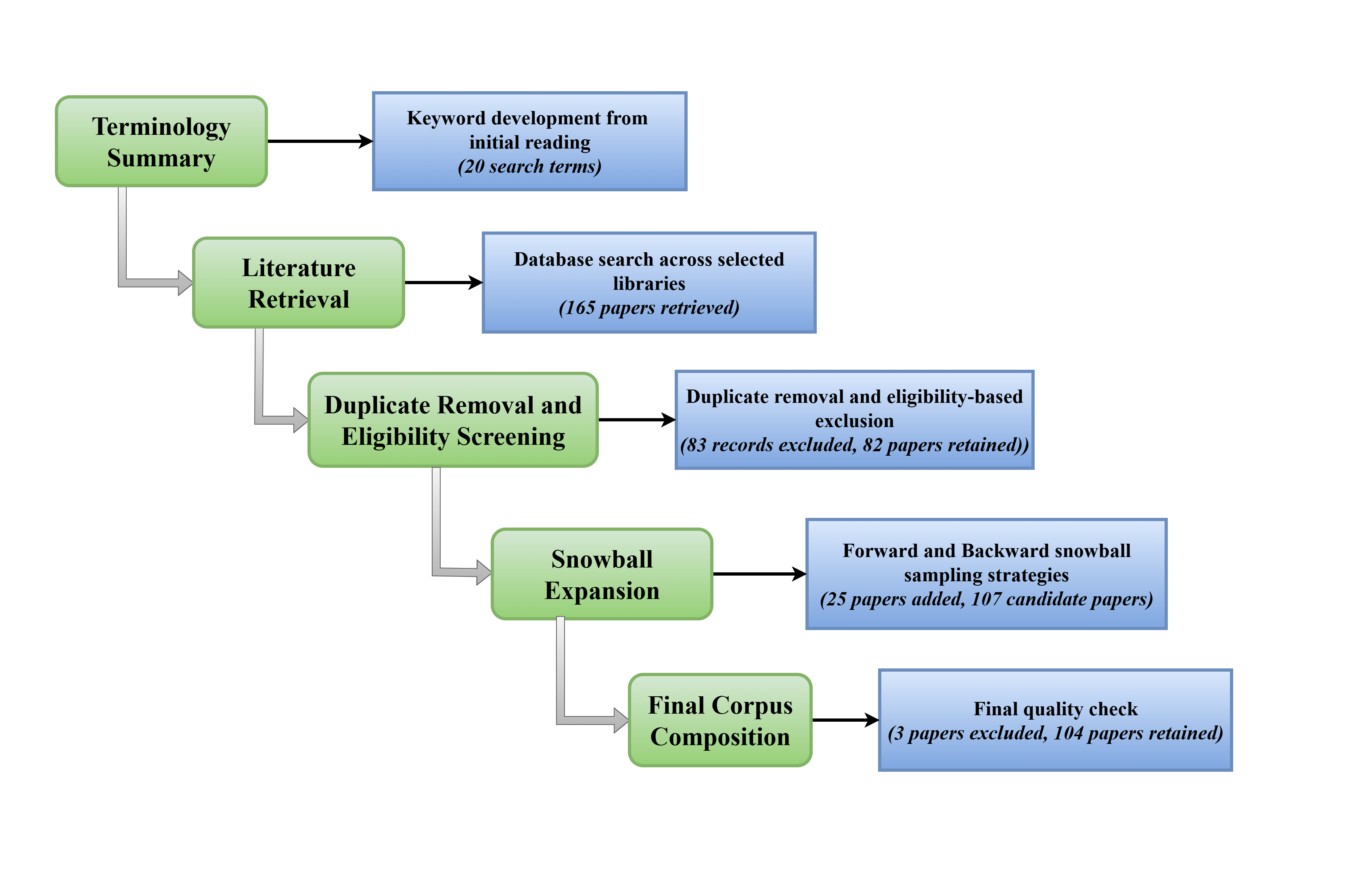}
    \caption{Overview of the literature collection process.}
    \label{fig:paper_collection}
\end{figure}

To ensure a comprehensive and systematic collection of research on code review, we adopted a rigorous multi-stage literature collection process. Our workflow consisted of multiple steps, which are illustrated and briefly described in Figure~\ref{fig:paper_collection}.

\textbf{Terminology Summary. }We began by conducting preliminary readings and compiling a set of keyword combinations related to code review approaches. To avoid constructing the keyword list in an ad hoc manner, we developed the search terms through a three-step process. Firstly, the first two authors independently conducted preliminary readings of foundational papers in code review automation to identify the core concepts and terminology used in the field. This produced an initial pool of candidate terms covering major task types, such as review comment generation, defect prediction, and code revision; methodological paradigms, such as machine learning for code review and LLM-based review; and artifact types, such as pull requests, patches, diffs, and commits.

Secondly, the candidate terms were consolidated and refined through discussion between the first two authors. During this step, redundant expressions were removed, and additional terms were added to cover underrepresented task areas identified during preliminary reading, such as review prioritization and reviewer assignment. This produced a working set of search terms that was used to conduct an initial pilot search across the selected databases.

Thirdly, the first two authors examined the papers retrieved in the pilot search and used this reading to further refine the keyword list. The pilot search was used as a keyword-validation step rather than as a separate corpus-counting stage. Specifically, the retrieved papers were inspected to determine whether the working search terms captured relevant code review tasks, datasets, benchmarks, and evaluation studies. The pilot search results introduced additional terminology and task framings that were not fully captured by the initial candidate terms. For example, terms such as reviewer recommendation, review effort prediction, and change understanding emerged from reading the pilot search results and were added to improve coverage of the code review research landscape. This iterative refinement continued until additional readings no longer suggested new terms that meaningfully extended coverage.

The final keyword list contained 20 search terms. Some representative examples include “code-review benchmarks,” “code review automation,” “machine learning for code review,” “LLM for code review,” “AI-assisted review,” “pull-request review datasets,” “review comment generation,” “defect detection in code review,” “code refinement benchmarks,” “review recommendation,” “change-understanding tasks,” “PR-level analysis,” “review prioritization,” “LLM agents for code review,” “review comment quality evaluation,” and “automated reviewer assignment.

\textbf{Literature Retrieval.}
Using these predefined keywords, we conducted automated searches across several major academic databases, such as IEEE Xplore, ACM Digital Library, Elsevier ScienceDirect, ACL Anthology, arXiv, Google Scholar, and SpringerLink. These sources were selected because they provide broad coverage of core venues in software engineering research as well as emerging work on AI-assisted benchmarking and code review automation. To reflect the evolution of modern code review and related benchmark research, we limited our search to publications from 2015 onward. In total, our search retrieved 165 publications related to code review from 1st Jan. 2015 to 31st Dec. 2025.

\textbf{Duplicate and Borderline Cases.}
If the same work appeared in multiple databases, or appeared both in arXiv and in a formally published venue, we retained only one version. When available, we prioritized the most current and officially published version, such as the conference or journal paper over the arXiv preprint. During this process, we removed 13 duplicate or overlapping papers.

\textbf{Exclusion Criteria.}
Papers were excluded if they met any of the following conditions:

\begin{enumerate}
\item \textbf{Not directly related to code review:} Papers focused solely on general code generation, bug localization, program repair, vulnerability detection, or bug fixing without a code review or review-feedback context were excluded. This criterion removed 30 papers.

\item \textbf{No clear task definition:} Papers lacking a clear definition of the code review task or research objective, or lacking identifiable inputs and outputs, were excluded. This criterion removed 10 papers.

\item \textbf{No identifiable code-review-specific context:} Papers were excluded when they were broader empirical studies or user studies on AI-assisted software engineering without an identifiable code-review-specific study context. This criterion removed 12 papers.

\item \textbf{No clear evaluation details:} Papers that did not report dataset information, benchmark descriptions, dataset size, or evaluation metrics were excluded, as these details are central to our survey's goals. We also excluded papers whose referenced dataset links were no longer accessible or functional. This criterion removed 18 papers.
\end{enumerate}

\textbf{More specifically, to concrete the exclusion criteria}, during screening the first two authors checked each candidate paper using a predefined eligibility checklist. We assessed whether the paper: (1) clearly defined a code review or review-related task; (2) was sufficiently connected to code review artifacts or context; (3) reported identifiable code-review-specific data points, such as pull requests, patches, review comments, review decisions, or review-based code revisions; (4) described core benchmark metadata, including data source, project or language coverage, granularity, and dataset size; and (5) provided sufficient evaluation details, such as evaluation setup, metrics, and procedures, to support reproducibility. Papers that did not satisfy these relevance or reporting requirements were excluded. If multiple papers used the same dataset, we recorded this as dataset reuse rather than treating it as an exclusion criterion. 

\textbf{Influence on Inclusion Decisions.} Quality assessment directly influenced inclusion decisions. Specifically, papers that did not report usable dataset information or evaluation details were excluded from the pool regardless of their publication venue or citation count. However, we did not apply a formal numerical scoring rubric.

\textbf{Inclusion Criteria.} Building on the structured filtering process described above, we applied the following inclusion criteria to identify studies relevant to the scope and objectives of this survey. After screening titles, abstracts and where necessary full texts, 82 papers satisfied these criteria:
\begin{itemize}
    \item The study must address code review, review automation, or code review-related tasks (e.g., comment generation, review recommendation, defect detection during review).
    \item Priority is given to papers accepted by leading SE and AI venues (e.g., TOSEM, TSE, EMSE, ICSE, FSE, ASE, ISSTA, ACL, etc.).
  \item {For arXiv preprints, we applied the same eligibility criteria used for other papers.}
    \item Papers must disclose their dataset details or evaluation approaches. The papers without such details were filtered from the pool.
\end{itemize}

 \textbf{Snowball Expansion.}
Recognizing that keyword searches alone may miss important contributions, we conducted both backward and forward snowballing from our selected papers. This step allowed us to incorporate an additional 25 studies, capturing influential work that may not explicitly match our initial keyword set. This process ensured breadth, completeness, and coverage of relevant benchmark- and review-related tasks.

\textbf{Final Corpus Composition.} Table~\ref{tab:screening-summary} summarizes the screening process used to construct the final corpus. Our initial database search retrieved 165 records. After duplicate removal and eligibility screening based on the exclusion criteria described above, 82 papers were retained. We then added 25 additional studies through snowballing, resulting in 107 candidate papers. During a final quality check, we re-applied the same eligibility criteria to ensure that all retained studies were directly relevant to our survey. This final check excluded 3 additional papers. Therefore, the final revised corpus contains 104 papers, including 60 Pre-LLM era papers and 44 LLM era papers.


\setlength{\tabcolsep}{1pt}
\renewcommand\arraystretch{0.8}
\rowcolors{2}{gray!10}{white}

\begin{table}[ht]
\centering
\scriptsize
\renewcommand{\arraystretch}{1} 
\resizebox{\columnwidth}{!}{%

\begin{tabular}{|>{\scriptsize\arraybackslash}p{6.24cm}|>{\centering\scriptsize\arraybackslash}p{3cm}|}
\hline
{\textbf{Screening step}} & {\textbf{Number of papers}}\\
\hline

{Records retrieved from database search} & {165}\\
\hline 
{Papers retained after duplicate removal and eligibility screening} & {82}\\
\hline
{Additional papers included through snowballing} & {25}\\
\hline
{Candidate papers before final quality check} & {107}\\
\hline
{Papers excluded during final quality check} & {3}\\
\hline
{\textbf{Final  corpus}} & {\textbf{104}}\\
\hline

\end{tabular}
}
\caption{{Summary of paper screening and final corpus construction.}}
\label{tab:screening-summary}
\end{table}







\textbf{Counting Policy for Papers, Tasks, and Datasets.} The overall corpus size is counted at the paper level: each study is counted once in the final corpus, even if it addresses multiple code review tasks. However, the task-level and sub-task-level analyses follow a multi-label coding policy. If a paper explicitly addresses more than one sub-task, it is assigned to each relevant sub-task. For example, a study that evaluates both code revision and refactoring-oriented changes may be assigned to both Code Revision and Code Refactoring (Rework). Therefore, the sum of sub-task-level counts may exceed the number of unique papers in the corpus. Dataset-level counts are treated separately from paper-level counts because the same dataset may be reused across multiple studies or may support multiple sub-tasks. As a result, the number of unique datasets within a broader task category may be smaller than the sum of datasets reported across its individual sub-tasks. This distinction allows us to report the unique literature corpus while also capturing the full range of tasks and datasets covered by the surveyed studies.

In the LLM era corpus, we also observe dataset reuse among studies on review comment generation. For example, \cite{LP6} relies on a dataset that is reused by several subsequent studies \cite{LP33,LP34,LP35,LP36,LP37,LP38,LP39,LP40,LP41}. Similarly, \cite{LP3} represents another group of works \cite{LP29,LP30,LP31,LP32} that employ closely related or overlapping datasets originating from the same CodeReviewer-style corpus. These papers are counted as distinct studies in the corpus because they present separate methods, experiments, or evaluation settings, but their datasets are not treated as independent unique datasets when interpreting dataset-level trends. Consequently, the 60 Pre-LLM papers correspond to 60 unique datasets, whereas the 44 LLM-era papers correspond to 31 unique datasets, because 13 LLM-era studies reuse a dataset that is already counted.

\textbf{Classification of Pre-LLM and LLM era Studies.} The distinction between Pre-LLM and LLM era studies in our survey is based on the actual use of large language models in the proposed approach, benchmark, or evaluation pipeline, rather than publication year alone.  Specifically, we classify a study as LLM era if it proposes, evaluates, or benchmarks a code review method that directly uses a large-scale generative language model, such as GPT-family models, LLaMA-family models, or similar instruction-tuned code or natural-language models, as a core component of the system or evaluation setup. We also include studies whose primary benchmark target is the evaluation of such LLM-based code review capabilities. For example if a paper compares LLMs against traditional machine learning, information retrieval, or encoder-based transformer baselines it is also considered as LLM era paper. The classification is therefore made at the study level, not at the level of every individual model used in the experiments. Accordingly, models such as BERT, RoBERTa, CodeBERT, GraphCodeBERT, BM25 etc models may appear in LLM era studies as baselines or comparison models. In contrast, studies that rely on traditional machine learning, heuristic-based methods, information retrieval techniques, rule-based systems, or non-generative models are classified as Pre-LLM era, even if they were published during the recent LLM period. This criterion allows us to distinguish studies by methodological design rather than by publication date alone.

\subsection{Statistics of Papers}

Table \ref{tab:venue-summary} summarizes the distribution of all papers included in our survey across different publication venues and lists the corresponding literature for each venue. The table highlights how research on code review automation and related topics is spread across major conferences and journals, with arXiv, ICSE, SANER, FSE, MSR, and other flagship venues contributing the largest number of publications. arXiv represents a larger portion of the recent LLM era literature, reflecting the rapid dissemination of emerging work in this area. This distribution helps illustrate where code review benchmarking studies have been published, which venues are most active in publishing work on code review related technologies, and how diverse the underlying literature base is. By organizing the papers in this way, the table provides a clear overview of the breadth of prior work and the venues that have shaped current developments in this area.


\setlength{\tabcolsep}{1pt}
\renewcommand\arraystretch{1.3}
{\setlength{\arrayrulewidth}{0.8pt}
\rowcolors{2}{gray!10}{white}
\setlength{\aboverulesep}{0pt}
\setlength{\belowrulesep}{0pt}
\begin{table}[ht]
\centering
\scriptsize
\renewcommand{\arraystretch}{1} 
\resizebox{\columnwidth}{!}{%
\begin{tabular}{|>{\centering\scriptsize\arraybackslash}p{3cm}|>{\centering\scriptsize\arraybackslash}p{1.4cm}|>{\centering\scriptsize\arraybackslash}p{1.2cm}|>{\scriptsize\arraybackslash}p{8cm}|}
\hline
\textbf{Domain} & \textbf{Venue} & \textbf{Papers No.} & \textbf{Literature} \\
\hline

Software Engineering (SE) & ICSE & 12 & \citet{LP16}; \citet{LP17}; \citet{LP25}; \citet{LP32}; \citet{LP34}; \citet{NP22}; \citet{NP23}; \citet{NP26}; \citet{NP38}; \citet{NP39}; \citet{NP50}; \citet{NP8} \\ \cmidrule{2-4}
& FSE & 4 & \citet{LP13}; \citet{LP1}; \citet{NP35}; \citet{NP43}\\ \cmidrule{2-4}
& ASE & 2 & \citet{NP30}; \citet{NP36} \\ \cmidrule{2-4}
& ICSME & 4 & \citet{NP19}; \citet{NP20}; \citet{NP31}; \citet{NP45} \\ \cmidrule{2-4}
&  SANER & 6 & \citet{LP2}; \citet{NP10}; \citet{NP24}; \citet{NP25}; \citet{NP40}; \citet{NP41}  \\ \cmidrule{2-4}
& MSR & 7 & \citet{LP14}; \citet{LP23}; \citet{LP40}; \citet{NP18}; \citet{NP27}; \citet{NP32}; \citet{NP4}\\ \cmidrule{2-4}
& ICPC & 2 & \citet{NP56}; \citet{NP59} \\ \cmidrule{2-4}
& ESEM & 3 & \citet{NP12}; \citet{NP37}; \citet{NP44}  \\ \cmidrule{2-4}
& APSEC & 1 & \citet{NP57}\\ \cmidrule{2-4}
& ISEC & 1 & \citet{NP16} \\ \cmidrule{2-4}
& ICSSP & 1 & \citet{NP52} \\ \cmidrule{2-4}
& SWQD & 1 & \citet{NP34} \\ \cmidrule{2-4}
& SCAM & 1 & \citet{LP7} \\ \cmidrule{2-4}
& VISSOFT & 1 & \citet{NP58} \\ \cmidrule{2-4}
& IWSC & 1 & \citet{NP21} \\ \cmidrule{2-4}
& FORGE & 1 & \citet{LP18} \\ \cmidrule{2-4}
& ISSRE & 2 & \citet{LP3}; \citet{NP51}\\ \cmidrule{2-4}
& FASE & 1 & \citet{LP31}\\ \cmidrule{2-4}

 & EMSE & 4 & \citet{NP11}; \citet{NP5}; \citet{NP7}; \citet{NP9} \\ \cmidrule{2-4}
& TOSEM & 3 & \citet{LP38}; \citet{NP53}; \citet{NP61} \\ \cmidrule{2-4}
& IST & 3 & \citet{LP36}; \citet{NP17}; \citet{NP6} \\ \cmidrule{2-4}
& JSS & 1 & \citet{NP60} \\ \cmidrule{2-4}
& SPE & 1 & \citet{NP15}\\ \hline
AI/ML & ICML & 1 & \citet{LP22} \\ \cmidrule{2-4}
& AAAI & 1 & \citet{NP1} \\ \cmidrule{2-4}

  &  ACL & 2 & \citet{LP5}; \citet{LP9} \\ \cmidrule{2-4}
& EMNLP & 1 & \citet{LP4} \\ \hline

Data Mining \& Data Science & KDD & 1 & \citet{NP42} \\ \hline

HCI & VL/HCC & 1 & \citet{NP55} \\ \hline

Others & ARXIV & 22 & \citet{LP10}; \citet{LP11}; \citet{LP12}; \citet{LP19}; \citet{LP20}; \citet{LP21}; \citet{LP24}; \citet{LP26}; \citet{LP27}; \citet{LP29}; \citet{LP30}; \citet{LP33}; \citet{LP35}; \citet{LP37}; \citet{LP39}; \citet{LP41}; \citet{LP6}; \citet{LP42}; \citet{LP43}; \citet{LP44}; \citet{LP45}; \citet{LP46}
 \\ \cmidrule{2-4}
& JCST & 2 & \citet{NP29}; \citet{NP33}\\ \cmidrule{2-4}   
& COMPSAC & 2 & \citet{NP2}; \citet{NP3} \\ \cmidrule{2-4}
& CCSE & 1 & \citet{LP28} \\ \cmidrule{2-4}
& EIT & 1 & \citet{NP49}   \\ \cmidrule{2-4}   
& FIE & 1 & \citet{LP8} \\ \cmidrule{2-4}     
& EAAI & 1 & \citet{NP48} \\ \cmidrule{2-4} 
& IEEE ACCESS & 1 & \citet{LP15}\\ \cmidrule{2-4}
& FEDCSIS & 1 & \citet{NP14}\\ \cmidrule{2-4}
& SOFSEM & 1 & \citet{NP28} \\ \cmidrule{2-4}
& SENSORS & 1 & \citet{NP47}\\ 
\hline

\end{tabular}
}
\caption{Distribution of papers per venue, grouped by domain and corresponding literature.}
\label{tab:venue-summary}
\end{table}}

\begin{figure}[!ht]
    \centering
    \includegraphics[width=\textwidth]{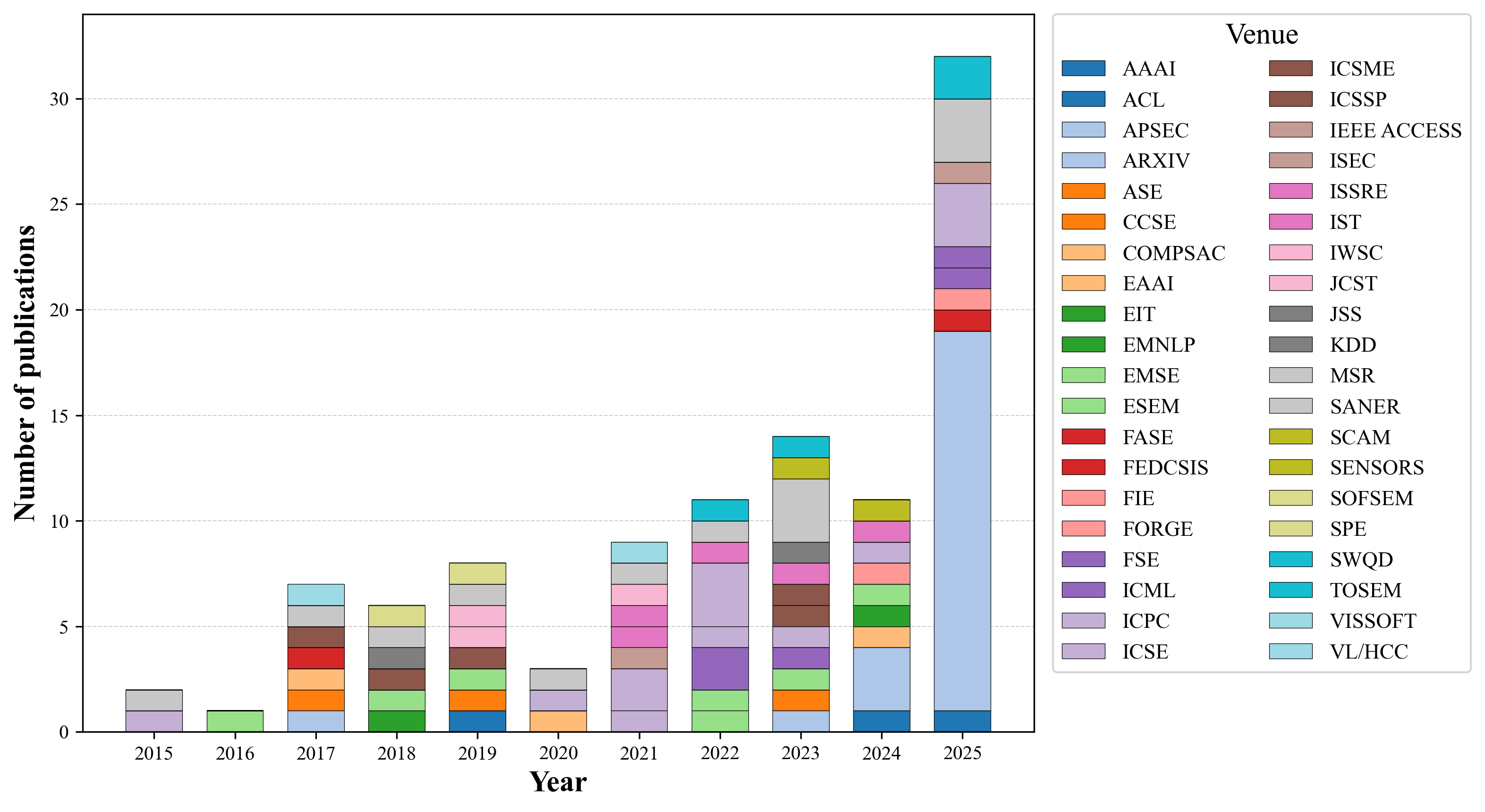}
    \caption{Venue–year distribution of the analyzed studies.}
    \label{fig:venue_year_distribution}
\end{figure}

Figure \ref{fig:venue_year_distribution} presents a year-by-year distribution of code-review benchmark publications across a wide range of venues, illustrating both the growth of the field and the diversification of publication venues over time. Each bar corresponds to a specific year, and the colored segments inside the bars represent the number of publications at each venue in that year. The height of a bar indicates how many benchmark-related papers were published overall during that year, while the variation in colors shows how many different conferences, journals, and workshops contributed to this research.

From left to right, the figure shows that early studies (2015--2018) were limited, with only a small number of papers each year and contributions mainly from traditional software engineering venues such as MSR, ICPC, ICSME, and ICSSP. As time progresses, particularly starting around 2020, the number of papers increases and venues become more diverse, indicating that more papers are being published and papers increasingly appear in additional venues. This pattern suggests increasing venue-level diversity in code review benchmarking research.

The period from 2022 onward shows a clear acceleration. These years display a broader mix of venues, including software engineering conferences (ICSE, FSE, and SANER), journals (EMSE and JSS), AI/NLP venues (EMNLP and ACL), and applied or interdisciplinary outlets (COMPSAC and FIE). This suggests that code review benchmarking has become increasingly visible across both traditional software engineering venues and AI/ML/NLP venues.

The most dramatic change appears in 2025, where the bar rises sharply compared to all previous years. A significant portion of this increase comes from arXiv publications, indicating a surge in preprint activity that is consistent with rapid dissemination of LLM-based work. At the same time, established venues still contribute smaller but steady numbers of papers. This combination indicates a change in publication patterns, with benchmark datasets and evaluation studies increasingly appearing through both established peer-reviewed venues and preprint channels such as arXiv.

\subsection{Constructing the Taxonomy of Code Review-related Tasks}

To construct the taxonomy of code review-related tasks, the first two authors perform open coding. The process follows a three-stage process:

\begin{itemize}
    \item \textbf{Phase 1: Open/Initial Coding} The goal of this phase was to identify the primary code review-related task addressed by each paper. To establish a consistent coding scheme, the first two authors randomly sampled 50 papers from the corpus and independently performed open coding. As an initial reference, we considered the task taxonomy from the previous survey paper \citet{related3} to guide label naming and ensure alignment with prior code review research. However, we did not directly adopt their taxonomy as a fixed coding scheme. Instead, we used it as a reference point to compare our emerging labels with existing categories and to identify task areas that were either missing or too broadly represented.

    Based on this comparison, our final taxonomy reuses and extends the previous  taxonomy. First, some labels were retained because they directly matched the tasks observed in our corpus, such as “Change Decomposition” and “Defective Change Prediction.” Second, we added standalone categories for tasks that were absent from, or differently scoped in, the previous study , such as post-review Code Refinement. Third, we added sub-tasks that were not covered by the previous survey paper, such as code compliance and review summarization. Finally, we split broad MCR categories, such as Change Understanding, into finer sub-tasks based on recurring patterns in our corpus.
    
    Using this taxonomy-development strategy, each author assigned a concise low-level task label that captured the central objective of each paper, based on its abstract and introduction primarily. When the task description was unclear, the methodology and evaluation sections were additionally inspected to determine the intended task and expected outputs. Each author also examined the evaluation and experimental setting sections to extract relevant evaluation details, including the dataset and evaluation metrics.
    
    After independently coding the initial 50 papers, the two authors met to compare labels, discuss discrepancies, and consolidate semantically similar labels. Through this process, an initial coding guideline and set of low-level task labels were established. The guideline included the task name, a brief description of the task objective, representative examples, and rules for distinguishing related tasks.
    
    Using this derived coding scheme, the two authors then independently labeled the remaining papers. For each paper, the authors inspected the abstract and introduction to identify the target task, and consulted the methodology and evaluation sections when additional context was needed. Along with the task label, the authors recorded the task input/output, target artifacts, and evaluation setup (e.g., datasets and evaluation metrics) to facilitate subsequent analysis.
    
    For papers addressing multiple code review-related tasks, we adopted a multi-label strategy. A paper could receive multiple low-level labels if it explicitly proposed and evaluated more than one distinct task. In subsequent analyses, each task label was treated as an independent observation.
    
    The coding process remained iterative throughout the study. As additional papers were analyzed, previously assigned labels and coding rules were periodically revisited to ensure consistency. When new task types emerged that were not adequately captured by the existing coding scheme, the two authors discussed whether to introduce a new label or merge the task into an existing category. Any updates to the coding guideline were retrospectively applied to previously coded papers to maintain consistency across the dataset.
    
    After independently labeling all papers, the two authors compared their assigned labels and resolved disagreements through discussion. A third senior researcher acted as an arbitrator, reviewing the paper and the disputed labels to make the final decision. Most disagreements were related to task granularity or papers that addressed more than one code review-related task. For example, papers that generated revised code from review feedback were discussed as either Code Revision or Code Refactoring (Rework), depending on whether the main objective was functional correction or behavior-preserving restructuring. Similarly, papers evaluating the usefulness of review comments were distinguished from Code Compliance by checking whether the evaluation focused on the usefulness of the feedback or on the consistency between the code and a natural-language artifact (e.g., a review comment) that describes or constrains it. To resolve such cases, the authors re-examined the paper’s stated objective, input/output artifacts, and evaluation setup. If a paper explicitly proposed and evaluated more than one distinct task, the multi-label coding policy was applied; otherwise, the label corresponding to the primary evaluated task was retained. The inter-rater agreement before reconciliation achieved a Cohen’s Kappa of 0.81, indicating substantial agreement. This arbitration process helped improve the consistency, transparency, and coverage of the final taxonomy.

    \item \textbf{Phase 2: Selective Coding} After the low-level labels were finalized, the first two authors collaboratively examined the semantic relationships among them. Using a bottom-up approach, they grouped conceptually related labels into broader categories representing higher-level code review tasks, based on similarities in task purpose and their roles within the overall code review process.

    For example, low-level labels such as Relevant Change Identification and Impact Analysis both focus on understanding and analyzing code changes. As a result, they were grouped under the higher-level task category of Change Understanding and Analysis. The resulting clusters were iteratively refined through discussion until all labels were assigned to a high-level category and the taxonomy structure stabilized. The final taxonomy therefore reflects both the fine-grained task distinctions observed in the literature and the broader conceptual relationships among code review research topics.
    \end{itemize} 

We construct the following high-level tasks:
\begin{itemize}
    \item \textbf{Review Prioritization / Selection} refers to the task of deciding which code changes should be reviewed and in what order so that limited review effort is used efficiently. It is commonly formulated as a prediction or ranking problem in both pre-LLM and LLM-based approaches. Given information such as the patch or diff content (e.g., changed files, size, and complexity), developer and reviewer history, review comments, and available test or CI signals, a model estimates the urgency or importance of each change. The output may be a binary or multi-class label (e.g., review needed, merged, or abandoned), a numerical estimate (such as the expected number of review rounds or delay), or a ranked list of patches indicating which changes should receive attention first. 
    \item \textbf{Change Understanding / Analysis} encompasses tasks that support reviewers in interpreting the intent, impact, and structure of code changes and review requests. In both pre-LLM and LLM eras, these tasks are typically formulated as prediction, classification, or ranking problems, where systems analyze information such as code diffs or patches, surrounding code context, commit messages, reviewer comments, and basic project or reviewer metadata. The outputs may be labels or scores (e.g., high-risk or high-effort), ranked or highlighted change regions, concise structured summaries, or, in reviewer-intent–focused settings, a change-type label indicating whether the requested modification involves adding, deleting, or modifying code. 
    \item \textbf{Peer Review} refers to a broad class of automated tasks designed to support or emulate human code review by analyzing code changes and related review information. Across both pre-LLM and LLM eras, these tasks are formulated as prediction, ranking, recommendation, classification, localization, or text-generation problems, using inputs such as code changes (patches or diffs at file, method, commit, or pull-request level), surrounding code context, review or issue text, review metadata, process information, and reviewer history. Depending on the formulation, the outputs include defect or rework labels, relevance scores or ranked recommendations (e.g., similar past examples or useful review comments), identified code locations requiring attention, or natural-language review feedback that reflects a reviewer’s judgment. 
    \item \textbf{Review Assessment / Analysis} refers to a set of tasks that examine the quality, tone, and effectiveness of review-related artifacts and interactions in code review, as well as the consistency between code and the natural-language artifacts or conventions that describe or constrain it. Across both pre-LLM and LLM eras, these tasks are typically formulated as classification or scoring problems that analyze inputs such as review comment text, code changes or diffs, commit messages, pull request descriptions, linked issues, coding policies or conventions and basic reviewer, author, or process metadata. The outputs are labels or scores that characterize aspects such as usefulness, clarity, relevance, consistency, toxicity, or overall review adequacy, and in some settings include indicators of whether feedback was later addressed by code updates.
    \item \textbf{Code Refinement} refers to tasks that support the code revision stage after review by helping developers improve an existing change based on review outcomes. In the surveyed benchmarks, this task is mainly formulated as a code generation or code transformation problem, where the goal is to generate a revised version of a method or code fragment that matches the reviewer-approved implementation, and in some cases as a classification problem that distinguishes refactoring changes from behavior-changing edits or identifies consistent naming. The typical inputs include the original code before revision, the revised code used as supervision, code diffs or diff markers, and sometimes review comments or basic pull-request metadata. The outputs are revised code snippets, refactored methods, or simple labels indicating the type of change or recommended names. 
\end{itemize}

Table~\ref{tab:tax} presents the resultant taxonomy of code review tasks, including high-level task, sub-task, and their definitions. We end up with five high-level tasks, which align with the code review process as shown in Figure~\ref{fig:code_review_process}.
\setlength{\aboverulesep}{0pt}
\setlength{\belowrulesep}{0pt}

\begin{table}[ht]
\centering
\scriptsize
\setlength{\tabcolsep}{1pt}
\renewcommand{\arraystretch}{1}
\setlength{\arrayrulewidth}{0.8pt}
\setlength{\aboverulesep}{0pt}
\setlength{\belowrulesep}{0pt}
\rowcolors{2}{gray!10}{white}
\resizebox{\columnwidth}{!}{%
\begin{tabular}{>{\scriptsize\arraybackslash}p{1.6cm}|>{\scriptsize\arraybackslash}p{2.1cm}|>{\scriptsize\arraybackslash}p{5.0cm}|>{\scriptsize\arraybackslash}p{2.0cm}|>{\scriptsize\arraybackslash}p{1.8cm}}
\toprule
\textbf{High-level task} & \textbf{Sub-task} & \textbf{Description} &
{\textbf{Input Type}} &
{\textbf{Output Type}} \\ \midrule

Review Prioritization / Selection & Review Prioritization & Determining the urgency and necessity of a review to optimize the engineering team's bandwidth.& {Diff/patch; reviewer-developer history} & {Ranking / Classification} \\ \cmidrule{2-5}
    & Change Quality Prediction & Predicting the quality of code changes, whether they require review or would be accepted or not.& {Diff/patch; change metadata} & {Classification} \\ \hline

Change Understanding and Analysis & Relevant Change Identification & Contextualizing the current changes by tracing their relationship to historical versions or related files in the repository. & {Diff; historical commits/files} & {Ranking} \\ \cmidrule{2-5}
 & Impact Analysis & Determining the ``blast radius'' of a change by identifying which downstream components or dependencies might be impacted. & {Diff; dependency/codebase structure} & {Localization / Classification} \\ \cmidrule{2-5}
 & Change Decomposition & Breaking down a monolithic Pull Request into smaller, logical ``atomic'' commits/slicers to reduce reviewer cognitive load. & {Monolithic PR/diff} & {Text Generation} \\ \cmidrule{2-5}
 & Visualization & Utilizing graphs or diff-enhancements to provide a mental map of how the code structure has evolved. & {Diff/commit history} & {Text Generation} \\ \hline

Peer Review & Issue Localization & Pinpointing the specific lines of code or architectural layers where a reported defect or logical error originates. & {Code change; defect report} & {Localization} \\ \cmidrule{2-5}
 & Review Comment Generation & Generating natural-language feedback that explains code issues to the author. & {Diff; surrounding context} & {Text Generation} \\ \cmidrule{2-5}
 & Issue Labeling / Classification & Categorize review comments into specific issue types (e.g., Bug, Readability, Design, etc.). & {Review comment text} & {Classification} \\ \cmidrule{2-5}
 & Refactoring Identification & Spotting ``code smells'' or technical debt that should be addressed. & {Code diff/snippet} & {Classification} \\ \cmidrule{2-5}
 & Defective Change Prediction & Flagging code changes that contain defects. & {Code diff; change metadata} & {Classification} \\ \cmidrule{2-5}
 & Security Detection & Scanning the proposed changes for vulnerabilities such as injection flaws or credential leaks. & {Source code/diff} & {Classification} \\ \hline

Review Assessment and Analysis & Review Quality Evaluation & Assessing the utility and constructiveness of a reviewer's feedback. & {Review comment text} & {Classification} \\ \cmidrule{2-5}
 & {Code Compliance} & {Examining whether code is consistent with a natural-language artifact (e.g., review comment, commit message, policy) or coding convention that describes or constrains it.} & {Review comment, commit message, policy, or coding convention; code snippet/diff} & {Classification} \\ \cmidrule{2-5}
 & Review Summarization & Condensing complex discussions into a concise status report. & {Review discussion thread} & {Text Generation} \\ \cmidrule{2-5}
 & Sentiment / Toxicity Analysis & Monitoring the emotional tone of review interactions. & {Review comment text} & {Classification} \\ \hline

Code Refinement & Code Revision & Implementing direct fixes to the logic based on the specific issues raised. & {Original code; review comments} & {Text Generation} \\ \cmidrule{2-5}
 & Code Refactoring (Rework) & Restructuring the implementation to improve long-term maintainability. & {Original code} & {Text Generation / Classification} \\ \hline
\end{tabular}
}
\caption{The definition of each code review-related task.}
\label{tab:tax}
\end{table}

\begin{figure}[!ht]
    \centering
    \includegraphics[width=\textwidth]{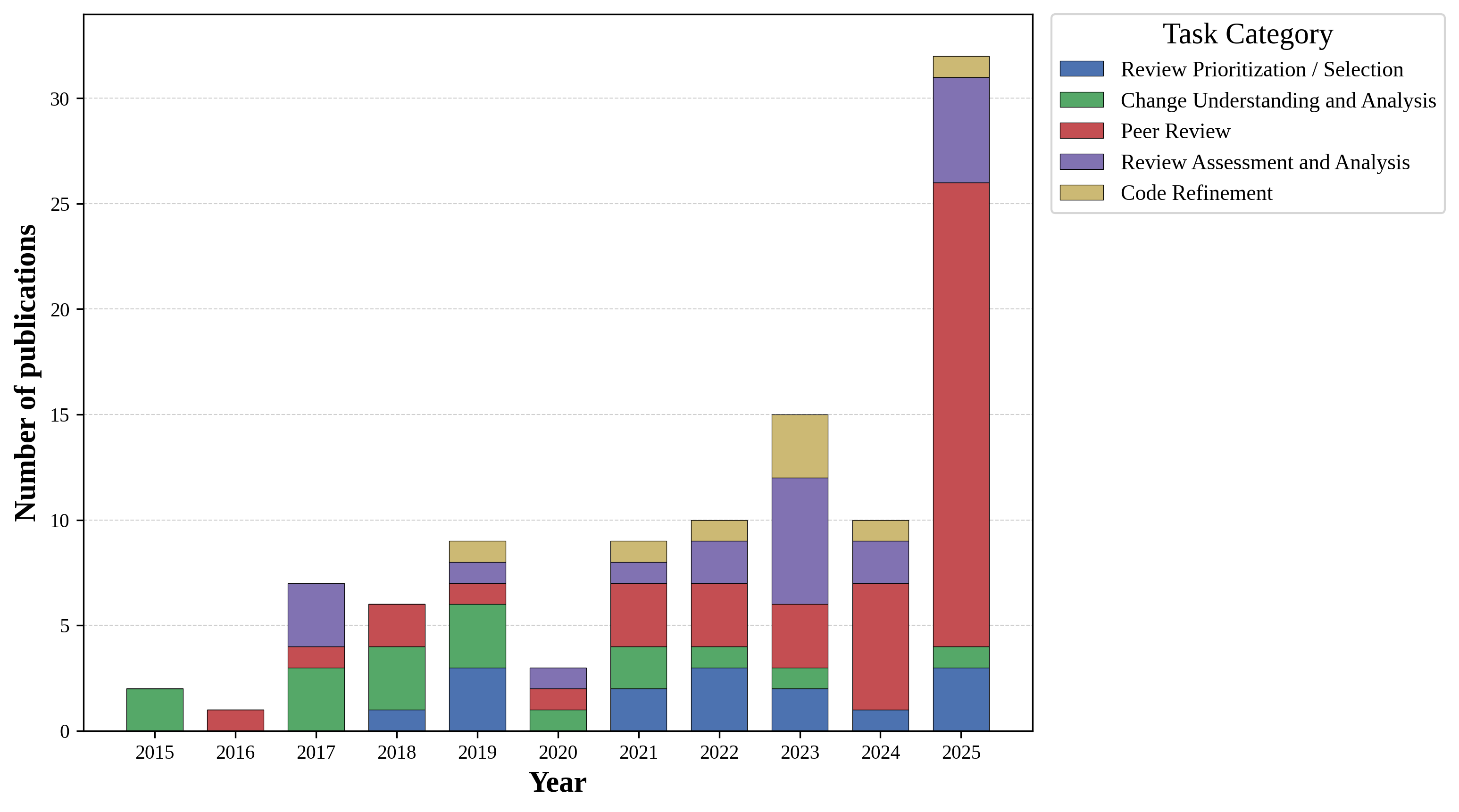}
    \caption{Year-by-Year Distribution of Surveyed Studies across Five High-level Task Categories}
    \label{fig:task_year_distribution.png}
\end{figure}

Figure~\ref{fig:task_year_distribution.png} provides a year-by-year view of how the surveyed studies are distributed across the five high-level task categories. The stacked bars show that Change Understanding and Analysis is more prominent in the earlier part of the corpus but becomes rare as a standalone category in recent years. In contrast, Peer Review grows sharply, particularly in 2025, when it accounts for the majority of publications. To avoid double-counting multi-task studies, each paper is counted once under a single high-level task category according to the categorization rule used for this analysis. Consequently, the annual bar heights sum to the revised corpus size of 104 unique papers across the full timeline. Secondary task labels associated with multi-task papers are analyzed separately and are not double-counted in this figure.

\section{RQ1: What tasks and evaluation strategies were used for code review studies in the Pre-LLM era }\label{sec:rq1}

\begin{figure}[!ht]
    \centering
    \includegraphics[width=0.7\textwidth]{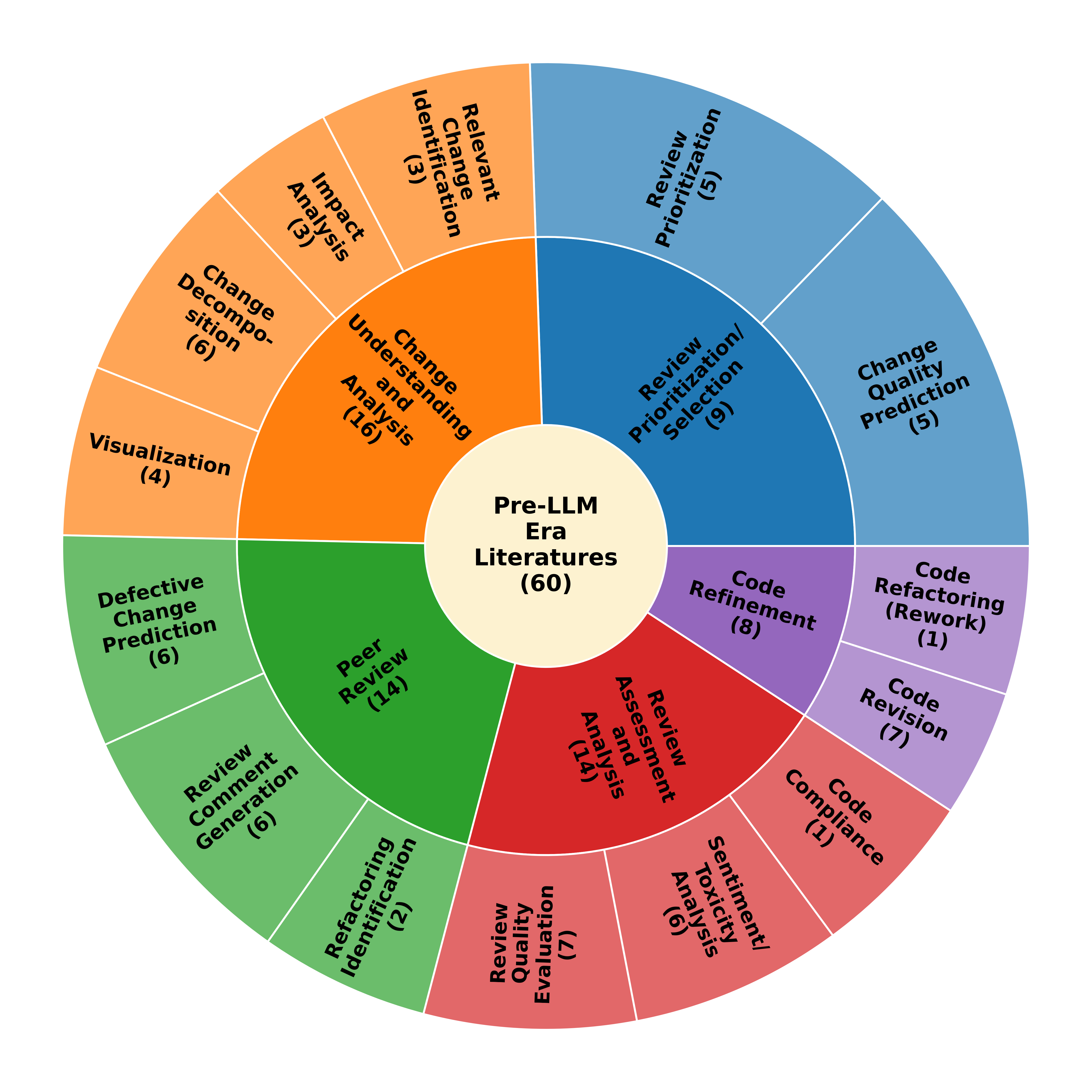}
    \caption{Distribution of literature across different code review-related tasks in Pre-LLM era.}
    \label{fig:non-llm-sunburst}
\end{figure}

In this research question, we examine which tasks and datasets have been used to evaluate code review systems in the Pre-LLM era. We also analyze how these systems have been measured, focusing on the evaluation metrics applied and the specific code review skills assessed.
Figure~\ref{fig:non-llm-sunburst} presents an overview of our taxonomy from the Pre-LLM era literature, illustrating how the 60 papers map out the landscape of code review research organized across five major high-level tasks and their sub-tasks. Each sub-task value in Figure~\ref{fig:non-llm-sunburst} is the number of studies that address that sub-task, and each high-level value counts a study once if it addresses at least one of its sub-tasks. Because a study may address more than one sub-task, a high-level value can be smaller than the sum of its sub-task values, and the values across tasks can exceed the 60 studies shown at the centre.

\subsection{Review Prioritization / Selection}
\label{sec:reviewprioritize}

\subsubsection{Change Quality Prediction:}

Change Quality Prediction helps decide the code change quality and predict whether a change would be accepted or not. Studies~\citep{NP1,NP5,NP6,NP7, NP11} constructed dataset to predict if a change would be accepted (merged) at the commit or pull request level, leveraging various information (e.g., code diffs, patch metadata, file history, developer experience, collaboration networks, and text), and evaluate performance using classification metrics such as AUC and F-measure. Table~\ref{tab:changeAcceptancePrediction} presents the details of each dataset and its evaluation metrics. 


\setlength{\tabcolsep}{1pt}
\renewcommand\arraystretch{0.8}
\rowcolors{2}{gray!10}{white}

\begin{table}[ht]
\centering
\scriptsize
\renewcommand{\arraystretch}{1} 
\resizebox{\columnwidth}{!}{%
\begin{tabular}{|>{\scriptsize\arraybackslash}p{1.4cm}|>{\scriptsize\arraybackslash}p{2cm}|>{\scriptsize\arraybackslash}p{1cm}|>{\scriptsize\arraybackslash}p{1.2cm}|>{\scriptsize\arraybackslash}p{2cm}|>{\scriptsize\arraybackslash}p{1.7cm}|>{\scriptsize\arraybackslash}p{1.3cm}|>{\scriptsize\arraybackslash}p{4.5cm}|}
\hline
\textbf{Literature} &
\textbf{Task} &
\textbf{Source} &
\textbf{Project No.} &
\textbf{Language} &
\textbf{Granularity} &
\textbf{Data Points} &
\textbf{Evaluation Metrics} \\
\hline

\citet{NP1} & Change acceptance prediction &
Open Source &
6 &
Java &
Chunk level &
35{,}640 changed hunks &
Classification Metrics: F-measure; AUC; t-test; Mann–Whitney U. \\
\hline

\citet{NP5} & Change acceptance prediction &
Open Source &
3 &
Ruby, Go, and 7 others &
Commit level &
166{,}215 Code Changes &
Classification Metrics: AUC; ER@K; Precision; Recall; F-measure\\ \hline

\citet{NP6} & Change acceptance prediction &
Open Source &
3 &
N/A &
Commit level &
146{,}612 Code Changes&
Classification Metrics: AUC; ER@K; Precision; Recall; F-measure
\\ \hline

\citet{NP7} & Change acceptance prediction/ranking &
Open Source &
74 &
Java &
PR level &
126{,}180 PRs &
Ranking Metrics: NDCG@k\\ \hline

\citet{NP11} & Change acceptance prediction &
Open Source &
3 &
N/A &
Commit level &
146{,}612 code review requests &
Classification Metrics: AUC\par
Ranking: Scott–Knott ESD \\ \hline

\end{tabular}
}
\caption{Datasets and Evaluation Metrics for Change Quality Prediction in Pre-LLM era.}
\label{tab:changeAcceptancePrediction}
\end{table}

\subsubsection{Review Prioritization: }

Review prioritization helps decide which code changes should be reviewed first and how much effort is needed. Prior studies look at code changes and past history to find risky or important patches, and help reviewers focus on urgent work. Table ~\ref{tab:reviewPrio} summarizes Pre-LLM research on review prioritization. 

For example, two studies~\citep{NP10,NP8} constructed datasets for review prioritization at two granularities, i.e., commit level and line level, respectively. \citet{NP9} constructed a dataset for review round prediction on 19{,}964 patches, using both classification metrics and regression metrics (e.g., MAE) to estimate review effort. Finally, two studies~\citep{NP11,NP4} constructed datasets for predicting whether a change would require high review effort at the chunk level, highlighting the importance of balancing merge likelihood and inspection effort under limited review capacity. Most of the datasets are sourced from open source projects, and only one is sourced from private project. In addition, all the datasets are at the commit or PR level.

\begin{table}[ht]
\centering
\scriptsize
\renewcommand{\arraystretch}{1} 
\resizebox{\columnwidth}{!}{%
\begin{tabular}{|>{\scriptsize\arraybackslash}p{1.4cm}|>{\scriptsize\arraybackslash}p{2cm}|>{\scriptsize\arraybackslash}p{1cm}|>{\scriptsize\arraybackslash}p{1.2cm}|>{\scriptsize\arraybackslash}p{2cm}|>{\scriptsize\arraybackslash}p{1.7cm}|>{\scriptsize\arraybackslash}p{1.3cm}|>{\scriptsize\arraybackslash}p{4.5cm}|}
\hline
\textbf{Literature} &
\textbf{Task} &
\textbf{Source} &
\textbf{Project No.} &
\textbf{Language} &
\textbf{Granularity} &
\textbf{Data Points} &
\textbf{Evaluation Metrics} \\
\hline

\citet{NP8} & Review prioritization &
Private Projects &
1 &
N/A &
Commit level &
N/A &
Classification Metrics: AUC, TPR/FPR \\ \hline

\citeauthor{NP9} & Review round count prediction &
Open Source &
3 &
Java and Python &
Commit level &
19{,}964 Patches &
Classification Metrics: F-measure\par Regression metrics: MAE, R²\\ \hline

\citet{NP10} & Review prioritization &
Open Source &
3 &
Java and Python &
Chunk level &
19{,}964 Patches &
Classification Metrics: Precision; Recall; Accuracy; F-measure\\ \hline

\citet{NP11} & Review effort prediction &
Open Source &
4 &
N/A &
Commit level &
146{,}612 code review requests &
Classification Metrics: ACC; Popt; Recall\par
Ranking: Scott–Knott ESD \\ \hline

\citet{NP4} & Review effort prediction &
Private Projects &
4 &
C\#, C, Java, Python &
Chunk level &
More than 100000 changes&
Classification Metrics: Precision; Recall; F-measure; AUC; \\
\hline

\end{tabular}
}
\caption{Datasets and Evaluation Metrics for Review Prioritization in Pre-LLM era.}
\label{tab:reviewPrio}
\end{table}

\subsection{Change Understanding and Analysis}

\subsubsection{Relevant Change Identification:} 
Change Identification aims to identify what changed in the code and understand those changes, even across files or versions. It checks for similar, missing, or inconsistent edits and highlights changes that may need more review. Table~\ref{tab:ChangeIdentificationPre-LLM} presents the dataset constructed to evaluate change identification in Pre-LLM era. 
\citet{NP2} studied inconsistent clone edit detection as a review support task, where the goal is to identify inconsistent updates among cloned methods; their approach was evaluated at the method level. 
\citet{NP3} focused on detecting incomplete edits in similar code, aiming to help reviewers find missed updates across related files. Finally, \citet{NP49} studied a clone evolution detection task that tracks clone changes across revisions to support review and maintenance. The dataset in this category usually focuses on low level, such as file, method, or chunk. All the datasets are in Java and sourced from open source projects.

\begin{table}[ht]
\centering
\scriptsize
\renewcommand{\arraystretch}{1} 
\resizebox{\columnwidth}{!}{%
\begin{tabular}{|>{\scriptsize\arraybackslash}p{1.4cm}|>{\scriptsize\arraybackslash}p{2cm}|>{\scriptsize\arraybackslash}p{1cm}|>{\scriptsize\arraybackslash}p{1.2cm}|>{\scriptsize\arraybackslash}p{2cm}|>{\scriptsize\arraybackslash}p{1.7cm}|>{\scriptsize\arraybackslash}p{1.3cm}|>{\scriptsize\arraybackslash}p{4.5cm}|}
\hline
\textbf{Literature} &
\textbf{Task} &
\textbf{Source} &
\textbf{Project No.} &
\textbf{Language} &
\textbf{Granularity} &
\textbf{Data Points} &
\textbf{Evaluation Metrics} \\
\hline

\citet{NP2} & Inconsistent clone edit detection &
Open Source &
6 &
Java &
Method level &
$14{,}842$ clone instances&
Classification Metrics: Precision; Recall; F-measure. \\
\hline

\citet{NP3} & Detect incomplete edits in similar code &
Open Source &
4 &
Java &
File level &
2.3 million Java files&
Classification Metrics: Precision; Recall; F-measure. \\
\hline

\citet{NP49} & Clone evolution detection &
Open Source &
3 &
Java &
Chunk level &
$657$ Revision pair &
Classification Metrics: Accuracy. \\
\hline

\end{tabular}
}
\caption{Datasets and Evaluation Metrics for Relevant Change Identification in Pre-LLM era.}
\label{tab:ChangeIdentificationPre-LLM}
\end{table} 

\subsubsection{Change Decomposition}
Change Decomposition breaks a large code change into smaller, meaningful, independent parts. It groups related edits so reviewers can understand complex changes and prioritize the most impactful changes first that may need deeper review. Table~\ref{tab:Benchmarks for Change Decomposition Tasks in Pre-LLM era.} presents decomposition benchmarks that investigate how complex code changes can be structured to better support human code review. 
Most of the studies~\citep{NP26,NP28,NP27,NP29} focus on decomposing a commits that contain multiple concerns into smaller logically related change slices to improve reviewer understanding and efficiency. One study~\citep{NP30} looked into tangled commit decomposition at the commit level, where the non-buggy changes are separated from buggy changes. We observe that the datasets are all open source projects, focus on Java and C\#, typically at the commit or PR level. Another study~\citep{NP55} aimed to automatically identify and separate behavior preserving refactoring code changes from other code changes so reviewers can focus on risky modifications.

\begin{table}[ht]
\centering
\scriptsize
\renewcommand{\arraystretch}{1} 
\resizebox{\columnwidth}{!}{%
\begin{tabular}{|>{\scriptsize\arraybackslash}p{1.4cm}|>{\scriptsize\arraybackslash}p{2cm}|>{\scriptsize\arraybackslash}p{1cm}|>{\scriptsize\arraybackslash}p{1.2cm}|>{\scriptsize\arraybackslash}p{2cm}|>{\scriptsize\arraybackslash}p{1.7cm}|>{\scriptsize\arraybackslash}p{1.3cm}|>{\scriptsize\arraybackslash}p{4.5cm}|}
\hline
\textbf{Literature} &
\textbf{Task} &
\textbf{Source} &
\textbf{Project No.} &
\textbf{Language} &
\textbf{Granularity} &
\textbf{Data Points} &
\textbf{Evaluation Metrics} \\
\hline

\citet{NP26} & Independent change decomposition &
Private Projects &
13 &
C\# &
Commit level &
1{,}000 code changes &
Classification Metrics: Accuracy. 

User Study: Qualitative Interview Feedback. \\
\hline

\citet{NP27} & Independent change decomposition &
Open Source &
4&
Java &
Commit level &
453 revisions &
Text-Matching/Generation Metrics: Exact-Match.\par Classification Metrics: Accuracy. \par
User Study:  Review Time. \\
\hline

\citet{NP28} & Independent change decomposition &
Open Source &
10  &
Java &
PR level &
1{,}000 pull requests &
Classification: False Positives and Error Types. \\
\hline

\citet{NP29} & Independent change decomposition &
Open Source &
4  &
Java &
Chunk level &
28 composite changes &
Classification Metrics: Precision; Recall; F-measure. \\
\hline

\citet{NP30} & Tangled commit decomposition &
Open Source &
7 &
Java &
Commit level &
50 tangled commits &
Classification Metrics: Rand Index; MAP; MRR. \\
\hline

\citet{NP55} & Refactoring change decomposition &
Open Source &
2 &
Java &
Commit level &
3,100 commits  &
Classification Metrics: Precision; Recall. \\ \hline

\end{tabular}
}
\caption{ Datasets and Evaluation Metrics for Change Decomposition in Pre-LLM era.}
\label{tab:Benchmarks for Change Decomposition Tasks in Pre-LLM era.}
\end{table}

\subsubsection {Impact Analysis:} 
Impact Analysis estimates how much a code change affects the system by predicting whether changes are risky or impactful and by identifying which parts of the code, design, or build may require closer review. Table~\ref{tab:impact-analysis-pre-llm} presents datasets for Impact Analysis proposed in the Pre-LLM era. 
\citet{NP18} introduced a revision-level design impact classification task in an open-source setting, where each PR revision is classified based on whether it negatively impacts software design. The task is formulated at the PR level and evaluated on 57{,}498 reviewed code changes from seven Java projects using classification metrics including Precision, Recall, F-measure, and AUC, demonstrating that review and change features can effectively flag design-impactful revisions for reviewer attention. 
\citet{NP19} aimed to identify the build deliverables that are affected by a given patch. \citet{NP20} constructed a dataset where impacted code regions for a given commit are annotated. The datasets for this task are typically at the PR or commit level. The datasets span various programming languages, although they are sourced from a limited number of projects. 

\begin{table}[ht]
\centering
\scriptsize
\renewcommand{\arraystretch}{1} 
\resizebox{\columnwidth}{!}{%
\begin{tabular}{|>{\scriptsize\arraybackslash}p{1.4cm}|>{\scriptsize\arraybackslash}p{2cm}|>{\scriptsize\arraybackslash}p{1cm}|>{\scriptsize\arraybackslash}p{1.2cm}|>{\scriptsize\arraybackslash}p{2cm}|>{\scriptsize\arraybackslash}p{1.7cm}|>{\scriptsize\arraybackslash}p{1.3cm}|>{\scriptsize\arraybackslash}p{4.5cm}|}
\hline
\textbf{Literature} &
\textbf{Task} &
\textbf{Source} &
\textbf{Project No.} &
\textbf{Language} &
\textbf{Granularity} &
\textbf{Data Points} &
\textbf{Evaluation Metrics} \\
\hline

\citet{NP18} & Design impact identification &
Open Source &
7 &
Java &
PR level &
57{,}498 reviewed code changes&
Classification Metrics: Precision; Recall; F-measure; AUC. \\
\hline

\citet{NP19} & Build deliverables impact analysis &
Private Projects &
1 &
C, C++, Java and C\# &
PR level &
Over 10 million lines of code &
User Study:  speed and accuracy of identifying the set of deliverables. \\
\hline

\citet{NP20} & Semantic change impact annotation&
open Source &
3  &
JavaScript (Node.js) &
Commit level &
Mining study over $\sim$2{,}000 commits &
Classification Metrics: Reductions in False Positives; Reductions in Change-Impact Set Size.  \par 
User Study: Search Time; Success Rate \\
\hline

\end{tabular}
}
\caption{Datasets and Evaluation Metrics for Impact Analysis in Pre-LLM era.}
\label{tab:impact-analysis-pre-llm}
\end{table}

\subsubsection{Visualization:} 
Visualization helps reviewers understand code changes using clear and easy-to-understand visuals. It shows risk, effort, structure, or behavior so reviewers can quickly spot important, risky, or complex changes. Table~\ref{tab:Benchmarks for Visualization Tasks in Pre-LLM era.} discusses representative visualization-based benchmarks proposed in the Pre-LLM era to support change understanding during code review. \citet{NP57} studied review risk, effort, and impact analysis at the commit level in open-source Java systems, where the task is to help reviewers assess how difficult and risky a commit is to review by analyzing evolutionary coupling information across 10 projects with 41,855 commit counts. Their work evaluates reviewer-oriented skills related to effort estimation and impact awareness using behavioral metrics: Effort, Risk, and Impact, complemented by visualization methods such as Clustering (KMeans), Spider Charts, Coupling Charts, and Density Maps to summarize change characteristics. \citet{NP58} focused on chunk-level risk visualization for code review, formulating the task as understanding the risk and nature of fine-grained change chunks in Java code, evaluated on one open-source project using 7 predefined change scenarios; the study emphasizes reviewer comprehension skills and reports results through a user study using Risk Percentages per Commit along with Readability Ratings, Understandability Ratings, and Practicality Ratings. \citet{NP59} presented an execution-trace–augmented code diff visualization task at the chunk level, where reviewers inspect both code differences and execution-trace differences to better understand behavioral changes; the approach is demonstrated on a single Java project using Math-57 from Defects4J, and no quantitative evaluation metrics are reported, relying instead of illustrative analysis through the visualization. Finally, \citet{NP60} proposed a graph-based change-set visualization task at the PR level to help reviewers navigate and understand large PRs in Java projects, evaluated on 138,452 PRs from 208 private projects; their task is framed as supporting reviewer decision-making and change comprehension, with results reported using classification metrics such as FP and success rate.

\begin{table}[ht]
\centering
\scriptsize
\renewcommand{\arraystretch}{1} 
\resizebox{\columnwidth}{!}{%
\begin{tabular}{|>{\scriptsize\arraybackslash}p{1.4cm}|>{\scriptsize\arraybackslash}p{2cm}|>{\scriptsize\arraybackslash}p{1cm}|>{\scriptsize\arraybackslash}p{1.2cm}|>{\scriptsize\arraybackslash}p{2cm}|>{\scriptsize\arraybackslash}p{1.7cm}|>{\scriptsize\arraybackslash}p{1.3cm}|>{\scriptsize\arraybackslash}p{4.5cm}|}
\hline
\textbf{Literature} &
\textbf{Task} &
\textbf{Source} &
\textbf{Project No.} &
\textbf{Language} &
\textbf{Granularity} &
\textbf{Data Points} &
\textbf{Evaluation Metrics} \\
\hline

\citet{NP57} & Review risk, effort, impact analysis and visualization &
Open Source &
10 &
Java &
Commit level &
41,855 commit counts &
Behavioral  Metrics: Effort (C), Risk, Impact 
\par 
Visualization Methods: Clustering (KMeans), Spider Charts, Coupling Charts, Density Maps \\
\hline

\citet{NP58} & Chunk-level risk visualization &
Open Source &
1  &
Java &
Chunk level &
7 predefined change scenarios &
User study: Risk Percentages per Commit; Readability Ratings; Understandability Ratings; Practicality Ratings \\
\hline

\citet{NP59} & Execution-trace–augmented code diff visualization &
Open Source &
1 &
Java &
Chunk level &
Math-57 from Defects4J &
N/A \\
\hline

\citet{NP60} & Graph-based change-set visualization &
Private Projects &
208 &
Java &
PR level &
138{,}452 pull requests  &
Classification: FP; Success rate.\\
\hline

\end{tabular}
}
\caption{Datasets and Evaluation Metrics for Visualization in Pre-LLM era.}
\label{tab:Benchmarks for Visualization Tasks in Pre-LLM era.}
\end{table}

\subsection{Peer Review}
\subsubsection{ Review Comment Generation:}
Review Comment Recommendation or Generation helps reviewers by suggesting helpful comments for new code changes, leveraging past reviews to identify recurring issues, and automatically producing clear and useful feedback. Table~\ref{tab:Benchmarks for Review Comment Recommendation.} discusses representative benchmarks from the Pre-LLM era that study automated review comment recommendation, generation, and analysis tasks in code review workflows by modeling reviewer feedback using historical data from both private and open-source projects. 
\citet{NP40} presented an automated code review comment generation task in which the system learns from repository history and pull request activity to automatically produce natural language review comments for new changes, with a particular focus on maintainability and architectural issues. 

The majority of prior studies in the Pre-LLM era focus on review comment recommendation~\citep{NP41,NP42,NP43, NP45}, which is typically formulated as an information retrieval problem, where similar changes are retrieved and their associated review comments are reused for a given code change.
\citet{NP34} automatically classifies code-review comments by their focus (e.g., design, style, logic, data, API, I/O, documentation, configuration, build/install, patch review), and constructed a dataset. 
The datasets for review recommendation/generation could span board granularity, from chunk level to PR level. The programming languages primarily focus on popular ones, e.g., Java, Python, C\#. The dataset size ranges from 8 to 11,289.

\begin{table}[ht]
\centering
\scriptsize
\renewcommand{\arraystretch}{1} 
\resizebox{\columnwidth}{!}{%
\begin{tabular}{|>{\scriptsize\arraybackslash}p{1.4cm}|>{\scriptsize\arraybackslash}p{2cm}|>{\scriptsize\arraybackslash}p{1cm}|>{\scriptsize\arraybackslash}p{1.2cm}|>{\scriptsize\arraybackslash}p{2cm}|>{\scriptsize\arraybackslash}p{1.7cm}|>{\scriptsize\arraybackslash}p{1.3cm}|>{\scriptsize\arraybackslash}p{4.5cm}|}
\hline
\textbf{Literature} &
\textbf{Task} &
\textbf{Source} &
\textbf{Project No.} &
\textbf{Language} &
\textbf{Granularity} &
\textbf{Data Points} &
\textbf{Evaluation Metrics} \\
\hline

\citet{NP40} & NL Review comment generation &
Private Projects &
21 &
Ruby &
Chunk level &
Almost 150k LOC Ruby repo &
User study:Comment Resolve Rate. \\
\hline

\citet{NP41} & Review comment recommendation &
Open Source &
19&
Java &
PR level &
57,260 $\langle$code change &
Classification Metrics: Recall; MRR.\\
\hline

\citet{NP42} & Review comment recommendation &
Private Projects &
208 &
C\# &
Chunk level &
22,435 completed PRs&
Classification Metrics: AUC; MRR; Recall\par
User Study: Acceptance Rate. \\
\hline

\citet{NP43} & Review comment recommendation &
Open Source &
11,289 &
Java &
Method level &
151,019 $\langle$changed method &
Text-Matching / Generation Metrics: Perfect Prediction; BLEU-4. \\
\hline

\citet{NP45} & Review comment generation &
Open Source &
8 &
Python and Java &
Chunk level &
56,068 review comments &
Text-Matching / Generation Metrics: BLEU; Exact Match; Sentence-BERT Cosine Similarity. \\
\hline

\citet{NP34} & Review comment classification &
Open Source &
3  &
C &
Chunk level &
2{,}672 comments &
Classification Metrics: Accuracy; MCC; AUC.\par
\\ \hline

\end{tabular}
}
\caption{Datasets and Evaluation Metrics for Review Comment Generation in Pre-LLM era.}
\label{tab:Benchmarks for Review Comment Recommendation.}
\end{table} 

\subsubsection{Defective Change Prediction:}
Defective Change Prediction estimates whether a code change is likely to contain bugs. It helps reviewers find risky changes early and decide which ones need careful review first. Table~\ref{tab:Benchmarks for Defective Change Prediction Tasks in Pre-LLM era.} presents Pre-LLM datasets that focus on predicting defective or risky changes to support code review decision making. 
Most of these studies~\citep{NP12, NP15, NP16, NP17} focus on defect prediction and construct corresponding datasets, where the goal is to predict whether a code change will later introduce a defect. For instance, \citet{NP15} proposed an approach to predict defective code changes by retrieving similar code from Stack Overflow and assigning defective labels to source files that exhibit high similarity. Therefore, they collected millions of Stack Overflow posts and a curated set of 370 GitHub files. 
\citet{NP14} introduced a rework risk prediction task in an industrial setting that classifies file changes as Correct or Rework in order to help reviewers and quality assurance teams focus on high-risk modifications. 
\citet{LP21} performed snippet-level defect detection, classifying Python functions from general defect data as clean or buggy with pre-trained code models (e.g., CodeBERT). Unlike the other studies in this category, it does not use code review artifacts, but we include it because the authors frame defect detection as part of automated code review. The dataset in this task spans various programming languages, such as Java, Python, C\#. It spans both open source and private projects. The granularity primarily focuses on file-level prediction; one dataset is at the commit level and one at the method level.

\begin{table}[ht]
\centering
\scriptsize
\renewcommand{\arraystretch}{1} 
\resizebox{\columnwidth}{!}{%
\begin{tabular}{|>{\scriptsize\arraybackslash}p{1.4cm}|>{\scriptsize\arraybackslash}p{2cm}|>{\scriptsize\arraybackslash}p{1cm}|>{\scriptsize\arraybackslash}p{1.2cm}|>{\scriptsize\arraybackslash}p{2cm}|>{\scriptsize\arraybackslash}p{1.7cm}|>{\scriptsize\arraybackslash}p{1.3cm}|>{\scriptsize\arraybackslash}p{4.5cm}|}
\hline
\textbf{Literature} &
\textbf{Task} &
\textbf{Source} &
\textbf{Project No.} &
\textbf{Language} &
\textbf{Granularity} &
\textbf{Data Points} &
\textbf{Evaluation Metrics} \\
\hline

\citet{NP12} & Defect prediction &
Open Source &
1 &
N/A &
Commit level &
4{,}750 review requests &
Classification Metrics: Recall; F-measure; AUC.\par
Statistical / Behavioral Analysis Metrics: \\ \hline

\citet{NP15} & Defect prediction &
Open Source &
3 &
Java, JavaScript and 5 others &
File level &
370 GitHub files &
Classification Metrics: Matching Ratio; F-Measure\\ \hline

\citet{NP16}& Defect prediction &
Open Source &
2 &
Java, Python and 3 others &
File level &
188{,}200 GitHub files &
Classification Metrics: Accuracy; F-Measure; Precision; Recall.\\ \hline

\citet{NP17} & Defect prediction &
Private Projects &
14 &
N/A &
File level &
2 industrial systems; 12 programs &
Classification Metrics: Accuracy; False positives; Defect-detection rate\\ \hline

\citet{NP14} & Risk prediction  &
Private Projects &
1 &
Java and .NET &
File level &
237{,}128 file-change records  &
Classification Metrics: AUC; Accuracy; Precision; Recall; PPV (Positive Predictive Value) ;NPV (Negative Predictive Value)
\\ \hline
{\citet{LP21}} &
{Defect detection} &
{Open Source} &
{N/A} &
{Python} &
{Method level} &
{N/A} &
{Classification Metrics: Precision; Recall; F-measure; Accuracy.} \\ \hline

\end{tabular}
}
\caption{Datasets and Evaluation Metrics for Defective Change Prediction in Pre-LLM era.}
\label{tab:Benchmarks for Defective Change Prediction Tasks in Pre-LLM era.}
\end{table}

\subsubsection{Refactoring Identification:}

\citet{NP56} focused on refactoring detection in real world pull request review scenarios at PR level using four private Go projects where the dataset contains 325 pull requests 374 commits approximately 754K LOC and 685 refactoring instances and the evaluation relies on Classification Metrics Precision and Recall to measure how accurately refactoring can be detected and surfaced to reviewers during code review.
\citet{NP61} studied method name consistency checking in large-scale open source projects. The datasets are sourced from both open source and private projects, and covered various programming languages. The granularity mainly focus on the chunk level, and size of the datasets from 85 to 17,000, depending on whether the evaluation is automated or manual. 

\begin{table}[ht]
\centering
\scriptsize
\renewcommand{\arraystretch}{1} 
\resizebox{\columnwidth}{!}{%
\begin{tabular}{|>{\scriptsize\arraybackslash}p{1.4cm}|>{\scriptsize\arraybackslash}p{2cm}|>{\scriptsize\arraybackslash}p{1cm}|>{\scriptsize\arraybackslash}p{1.2cm}|>{\scriptsize\arraybackslash}p{2cm}|>{\scriptsize\arraybackslash}p{1.7cm}|>{\scriptsize\arraybackslash}p{1.3cm}|>{\scriptsize\arraybackslash}p{4.5cm}|}
\hline
\textbf{Literature} &
\textbf{Task} &
\textbf{Source} &
\textbf{Project No.} &
\textbf{Language} &
\textbf{Granularity} &
\textbf{Data Points} &
\textbf{Evaluation Metrics} \\
\hline

\citet{NP61}& Refactoring detection in pull requests
& Open Source 
& 66
& Java 
& Method level 
& 400 reviewed method 
& Classification Metrics: Precision; Recall; F-measure; Exact Match Accuracy (EMAcc). 
\\ \hline

\citet{NP56} & Inconsistent method name detection &
Private Projects &
4 &
Go &
PR level &
325 pull requests, 374 commits, $\sim$754K LOC, 685 refactoring instances. &
Classification Metrics: Precision; Recall. \\ \hline


\citet{NP61} & Inconsistent method name detection &
Open Source &
66 &
Java &
Method level &
 13{,}537  methods &
Classification Metrics: Precision; Recall; F-measure; Accuracy. \\ \hline

\end{tabular}
}
\caption{Datasets and Evaluation Metrics for Refactoring Identification in Pre-LLM era.}
\label{tab:Refactoring Identification}
\end{table}

\subsection{Review Assessment and Analysis:}
\subsubsection{Review Quality Evaluation:}
Review Quality measures how good a code review is. It checks whether comments are useful, clear, and relevant to the code, and whether the review properly addresses important changes. Table~\ref{tab:Benchmarks for Review Quality Tasks in Pre-LLM era.} discusses the datasets and evaluation setup used in previous work in this line. 
Two studies~\citep{NP31,NP44} constructed datasets for clarity classification by predicting whether a review comment is clear or unclear and ranking similar examples to support explanation and reviewer understanding. 
Another group of studies~\citep{NP32,NP33} constructed a dataset for review usefulness prediction, where their goal is to predict whether a review comment is useful or not for developers. We also observe studies to investigate quality from various aspects.   
\citet{NP35} aim to evaluate a comment from four attributes, i.e., emotion, question, evaluation, and suggestion. For this purpose, they manually curated a dataset, that contains comments and the annotations in four attributes. 
Two studies~\citep{NP51,NP52} constructs datasets for low quality review classification task where the code is reviewed without sufficient attention from reviewers. The datasets coverage various languages such as Java, Python, C, Kotlin, etc. Most of the studies focus on the chunk-level, and the dataset ranges from small size of 84 code reviews from programming course to large size of 140,000 code review from open source repositories. 

\begin{table}[ht]
\centering
\scriptsize
\renewcommand{\arraystretch}{1} 
\resizebox{\columnwidth}{!}{%
\begin{tabular}{|>{\scriptsize\arraybackslash}p{1.4cm}|>{\scriptsize\arraybackslash}p{2cm}|>{\scriptsize\arraybackslash}p{1cm}|>{\scriptsize\arraybackslash}p{1.2cm}|>{\scriptsize\arraybackslash}p{2cm}|>{\scriptsize\arraybackslash}p{1.7cm}|>{\scriptsize\arraybackslash}p{1.3cm}|>{\scriptsize\arraybackslash}p{4.5cm}|}
\hline
\textbf{Literature} &
\textbf{Task} &
\textbf{Source} &
\textbf{Project No.} &
\textbf{Language} &
\textbf{Granularity} &
\textbf{Data Points} &
\textbf{Evaluation Metrics} \\
\hline

\citet{NP44} & Clarity classification &
Open Source &
3 &
N/A &
Commit level &
3,722 code reviews &
Classification Metrics: Accuracy, Precision, Recall, F-measure\par
Ranking Evaluation: Top-5 accuracy / Top-5 recall\\
\hline

\citet{NP31} & Clarity classification &
Open Source &
N/A &
N/A &
Chunk level &
140{,}006 code reviews &
Classification Metrics: Precision; Recall; F-measure.\par \\ \hline

\citet{NP32} & Usefulness prediction &
Private Projects &
4 &
Python &
Chunk level &
1{,}116 inline comments &
Classification Metrics: Precision; Recall; F-measure; Accuracy; ROC; Precision; Recall\par
\\ \hline

\citet{NP33} & Usefulness prediction &
Private Projects &
7&
Java, Kotlin and 4 others &
Chunk level &
9{,}477 code reviews &
Classification Metrics: Accuracy; Precision; Recall; F-measure
\\ \hline

\citet{NP35} & Review quality attribute classification &
Private Projects &
N/A &
N/A &
Chunk level &
17{,}000 review comments &
Classification Metrics: Hamming Loss; Subset 0/1 Loss; F-measure; AUC; Precision; Recall \\ \hline

\citet{NP51} & Low-quality review detection &
Progra- mming course / competition &
4 &
C &
Chunk level &
84 code reviews &
Classification Metrics: Accuracy; Precision; Recall; F-measure \\ \hline

\citet{NP52} & Low-quality review detection &
Private Projects &
N/A &
JavaScript &
Chunk level &
N/A &
User Study: Likert ratings \\ \hline

\end{tabular}
}
\caption{Datasets and Evaluation Metrics for Review Quality Evaluation in Pre-LLM era.}
\label{tab:Benchmarks for Review Quality Tasks in Pre-LLM era.}
\end{table}

\subsubsection{Code Compliance:}
Table~\ref{tab:Benchmarks for Code Compliance Tasks in Pre-LLM era.} presents the Pre-LLM dataset for code compliance, which examines whether code is consistent with a natural-language artifact or coding convention that describes or constrains it. \citet{NP50} presented policy based code compliance classification as a supervised learning task where the system receives a natural language policy description together with an isolated Java code snippet and predicts whether the snippet is compliant, non-compliant or irrelevant with respect to the policy.

\begin{table}[ht]
\centering
\scriptsize
\renewcommand{\arraystretch}{1} 
\resizebox{\columnwidth}{!}{%
\begin{tabular}{|>{\scriptsize\arraybackslash}p{1.4cm}|>{\scriptsize\arraybackslash}p{2cm}|>{\scriptsize\arraybackslash}p{1cm}|>{\scriptsize\arraybackslash}p{1.2cm}|>{\scriptsize\arraybackslash}p{2cm}|>{\scriptsize\arraybackslash}p{1.7cm}|>{\scriptsize\arraybackslash}p{1.3cm}|>{\scriptsize\arraybackslash}p{4.5cm}|}
\hline
\textbf{Literature} &
\textbf{Task} &
\textbf{Source} &
\textbf{Project No.} &
\textbf{Language} &
\textbf{Granularity} &
\textbf{Data Points} &
\textbf{Evaluation Metrics} \\
\hline

\citet{NP50}& Policy-based code compliance classification &
Open Source \& Private Projects &
N/A &
Java &
Chunk level &
231 policies; 28000 code examples. &
Classification Metrics: Accuracy; MRR. \\ \hline

\end{tabular}
}
\caption{Dataset for Code Compliance in Pre-LLM era.}
\label{tab:Benchmarks for Code Compliance Tasks in Pre-LLM era.}
\end{table}

\subsubsection{Sentiment/Toxicity Analysis}

Table~\ref{tab:Benchmarks for Comment Sentiment / Toxicity Analysis Tasks in Pre-LLM era.} presents benchmarks that focus on reviewers' emotions, attitudes, and interpersonal signals expressed through code review comments in the Pre-LLM era. 
\citet{NP36} presented the task of review comment sentiment classification and constructed the corresponding dataset. 
{\citet{LP45}} manually labeled 13,557 review comments from three open-source GitHub projects into four categories which the authors treat as the sentiment polarity of a comment, and compared traditional machine-learning classifiers using accuracy, precision, recall, and F-measure.
\citet{NP38} addressed pushback detection, which aims to detect the reviews that lead to developers' negative feelings, and eventually cause pushback. Another two studies~\cite{NP37,NP53} proposed approaches to detect toxic review comments and construct corresponding datasets. Lastly,~\cite{NP39} constructed a dataset for detecting interpersonal conflict across code review and issue discussion threads. The datasets in this task are typically at a high-level, i.e., commit/PR level, and are sourced from both open source and private projects.

\begin{table}[ht]
\centering
\scriptsize
\renewcommand{\arraystretch}{1} 
\resizebox{\columnwidth}{!}{%
\begin{tabular}{|>{\scriptsize\arraybackslash}p{1.4cm}|>{\scriptsize\arraybackslash}p{2cm}|>{\scriptsize\arraybackslash}p{1cm}|>{\scriptsize\arraybackslash}p{1.2cm}|>{\scriptsize\arraybackslash}p{2cm}|>{\scriptsize\arraybackslash}p{1.7cm}|>{\scriptsize\arraybackslash}p{1.3cm}|>{\scriptsize\arraybackslash}p{4.5cm}|}
\hline
\textbf{Literature} &
\textbf{Task} &
\textbf{Source} &
\textbf{Project No.} &
\textbf{Language} &
\textbf{Granularity} &
\textbf{Data Points} &
\textbf{Evaluation Metrics} \\
\hline

\citet{NP36} & Review comment sentiment classification &
Open Source &
20 &
N/A &
Commit level &
2{,}000 labeled comments &
Classification Metrics: Precision; Recall; F-measure; Accuracy. \\ \hline

\citet{NP38} & Pushback (negative feeling) detection &
Private Projects &
N/A &
N/A &
PR level &
1{,}250  change requests &
Classification Metrics: Precision; Recall. \\ \hline

\citet{NP53} & Review comment toxicity classification &
Open Source &
5 &
N/A &
Commit level &
19{,}651 labeled review comments &
Classification Metrics: Accuracy; Precision; Recall; F-measure; confusion matrix.\\ \hline

\citet{NP37} & Review comment toxicity classification &
Open Source &
4 &
N/A &
Commit level &
19{,}651 review comment &
Classification Metrics: Precision; Recall; F-measure; span-matching metric.\\ \hline

\citet{NP39} & Interpersonal conflict detection &
Private Projects &
4 &
N/A &
Commit level &
2{,}372 labeled discussions &
Classification Metrics: Precision–Recall; AUC; Precision; Recall; F-measure.
\\ \hline
{\citet{LP45}} &
{Review comment sentiment classification} &
{Open Source} &
{3} &
{HTML, CSS, C++, Java, Python} &
{Commit level} &
{13{,}557 labeled review comments} &
{Classification Metrics: Accuracy; Precision; Recall; F-measure.} \\ \hline

\end{tabular}
}
\caption{Datasets and Evaluation Metrics for Sentiment/Toxicity Analysis in the Pre-LLM era.}
\label{tab:Benchmarks for Comment Sentiment / Toxicity Analysis Tasks in Pre-LLM era.}
\end{table}

\subsection{Code Refinement}

\subsubsection{Code Revision:}

This task uses the original code and review feedback to generate or suggest a revised version or fixed patterns that match what reviewers expect. Table~\ref{tab:Benchmarks for Code Revision / Suggestion Tasks in Pre-LLM era.} presents Pre-LLM benchmarks that study automated code revision and suggestion skills in the context of code review. The papers related to this task can be grouped into two families: 1) revision pattern suggestion, and 2) revised code generation. In the first family, the study~\citep{NP21} extracted the code revision patterns from similar patches and suggested relevant patterns when a new review comment is received. In the second family, the studies~\citep{NP22,NP23,NP24,NP25,NP47,NP48} typically formulated the problem as a code generation task, where the model is trained to generate the revised code given a piece of code and its review comment. For instance, \citet{NP47} proposed an approach to generate revised Java methods from original code and optionally review comments. As we can observe, almost all datasets are at a low level, i.e., method- and chunk-level. Only two popular programming languages, Java and Python, are tested, and all datasets are sourced from open source projects. 

\begin{table}[ht]
\centering
\scriptsize
\renewcommand{\arraystretch}{1} 
\resizebox{\columnwidth}{!}{%
\begin{tabular}{|>{\scriptsize\arraybackslash}p{1.4cm}|>{\scriptsize\arraybackslash}p{2cm}|>{\scriptsize\arraybackslash}p{1cm}|>{\scriptsize\arraybackslash}p{1.2cm}|>{\scriptsize\arraybackslash}p{2cm}|>{\scriptsize\arraybackslash}p{1.7cm}|>{\scriptsize\arraybackslash}p{1.3cm}|>{\scriptsize\arraybackslash}p{4.5cm}|}
\hline
\textbf{Literature} &
\textbf{Task} &
\textbf{Source} &
\textbf{Project No.} &
\textbf{Language} &
\textbf{Granularity} &
\textbf{Data Points} &
\textbf{Evaluation Metrics} \\
\hline

\citet{NP21} & Revision pattern suggestion &
Open Source &
1 &
Python &
Chunk level &
616,723 code changes &
Classification Metrics: Accuracy; Confidence Ratio\\ \hline

\citet{NP22} & Revised code generation &
Open Source &
8,954 &
Java &
Method level &
17,194 methods &
Text-Matching / Generation Metrics: Exact-Match; BLEU-4; Levenshtein Distance. \\ \hline

\citet{NP23} & Revised code generation &
Open Source &
3  &
Java &
Method level &
630,858 methods &
Text-Matching / Generation Metrics: Exact Matches\\ \hline

\citet{NP24} & Revised code generation &
Open Source &
20,243 &
Java &
Method level &
332,429 methods &
Text-Matching / Generation Metrics: Exact Match; MRR ; Dataflow Match (DFM). \\ \hline

\citet{NP25} & Revised code generation &
Open Source &
3  &
Java &
Method level &
57,615 methods &
Text-Matching / Generation Metrics: Exact Match. \\ \hline

\citet{NP47} & Revised code generation &
Open Source &
N/A &
Java &
Method level &
15,909 methods &
Text-Matching / Generation Metrics: BLEU (BLEU-1\ldots BLEU-4 Average); ROUGE-L; Normalized Levenshtein Distance. \\ \hline

\citet{NP48} & Revised code generation &
Open Source &
8,954 &
Java &
Method level &
17,194  methods &
Text-Matching / Generation Metrics: Perfect Prediction Rate; BLEU-4; Levenshtein Distance. \\ \hline

\end{tabular}
}
\caption{Datasets and Evaluation Metrics for Code Revision in the Pre-LLM era.}
\label{tab:Benchmarks for Code Revision / Suggestion Tasks in Pre-LLM era.}
\end{table} 

\subsubsection{Code Refactoring (Rework):} 

Table~\ref{tab:Benchmarks for Code Refactoring Tasks in Pre-LLM era.} presents Pre-LLM benchmarks related to Code Refactoring. 
\citet{NP61} investigated refactoring related to method naming recommendation where models learn reviewer preferred method names from reviewed code and comments, and they curated 400 reviewed method instances from 66 open source Java projects. 

\begin{table}[ht]
\centering
\scriptsize
\renewcommand{\arraystretch}{1} 
\resizebox{\columnwidth}{!}{%
\begin{tabular}{|>{\scriptsize\arraybackslash}p{1.4cm}|>{\scriptsize\arraybackslash}p{2cm}|>{\scriptsize\arraybackslash}p{1cm}|>{\scriptsize\arraybackslash}p{1.2cm}|>{\scriptsize\arraybackslash}p{2cm}|>{\scriptsize\arraybackslash}p{1.7cm}|>{\scriptsize\arraybackslash}p{1.3cm}|>{\scriptsize\arraybackslash}p{4.5cm}|}
\hline
\textbf{Literature} &
\textbf{Task} &
\textbf{Source} &
\textbf{Project No.} &
\textbf{Language} &
\textbf{Granularity} &
\textbf{Data Points} &
\textbf{Evaluation Metrics} \\
\hline


\citet{NP61} & Method naming recommendation
& Open Source 
& 66
& Java 
& Method level 
& 400 reviewed method 
& Classification Metrics: Precision; Recall; F-measure; Accuracy \\ \hline


\end{tabular}
}
\caption{Dataset for Code Refactoring (Rework) in Pre-LLM era.}
\label{tab:Benchmarks for Code Refactoring Tasks in Pre-LLM era.}
\end{table}

\section{RQ2: Code Review Datasets and Evaluation in LLM era}\label{sec:rq2}

\begin{figure}[!ht]
    \centering
    \includegraphics[width=0.7\textwidth]{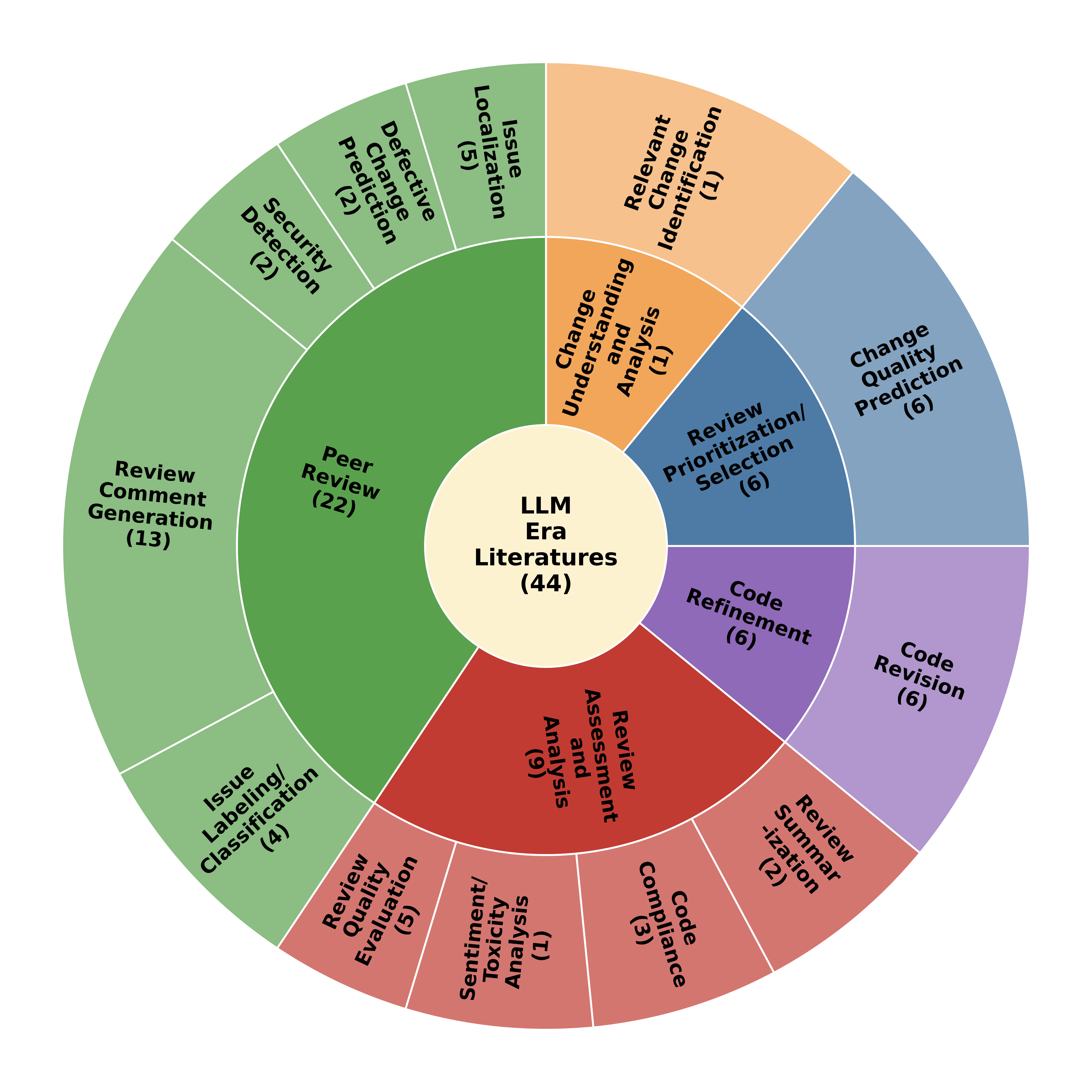}
    \caption{Distribution of literature across different code review-related tasks in the LLM era. }
    \label{fig:llm-sunburst}
\end{figure}

In this research question, we investigate the range of tasks and datasets introduced for evaluating LLM-driven code-review systems. Figure~\ref{fig:llm-sunburst} presents an overview of the literature taxonomy in the LLM era. 

\subsection{Review Prioritization/Selection}
\subsubsection{Change Quality Prediction:}

In the LLM era, this task primarily involves leveraging language models to predict whether a code change requires human review and to assess its quality. The table~\ref{tab:Datasets and Evaluation Metrics for Change Quality Prediction in LLM era.} summarizes representative LLM era datasets for the task.
Most of the studies~\citep{LP1,LP3,LP15} focus on review necessity prediction, where the goal is to determine whether a code change should receive human review. These approaches typically operate at the chunk or file level, analyzing code diffs to filter out changes that do not need attention, allowing reviewers to concentrate on necessary reviews rather than uniformly inspecting all submissions.
Two studies~\citep{LP5,LP25} address quality estimation from a reviewer’s perspective, aiming to judge whether a code change is of acceptable quality or likely to require comments. These tasks often leverage code diffs enriched with contextual information such as issue descriptions, pull request metadata and surrounding code to improve decision-making. 
Most of the studies use of open-source repositories, multi-language coverage (e.g. Python, Java, C++, JavaScript), and granularity ranging from chunk-level diffs to method-level changes. Evaluation is generally based on classification metrics. One study~\citep{LP15} uniquely incorporates usability-oriented measures (e.g., Blocking Rate and Review-Omission Rate) and relies on private data from student-written code samples to decide whether a submitted solution meaningfully attempts a given task.

\begin{table*}[ht]
\centering
\resizebox{1\columnwidth}{!}{
\renewcommand{\arraystretch}{1}
\begin{tabular}{|>{\scriptsize\arraybackslash}p{1.4cm}|>{\scriptsize\arraybackslash}p{1cm}|>{\scriptsize\arraybackslash}p{1cm}|>{\scriptsize\arraybackslash}p{1.2cm}|>{\scriptsize\arraybackslash}p{1.3cm}|>{\scriptsize\arraybackslash}p{1.66cm}|>{\scriptsize\arraybackslash}p{1cm}|>{\scriptsize\arraybackslash}p{2.2cm}|>{\scriptsize\arraybackslash}p{2cm}|>{\scriptsize\arraybackslash}p{2cm}|>{\scriptsize\arraybackslash}p{2cm}|}
\hline
\textbf{Literature} &
\textbf{Task} &
\textbf{Source} &
\textbf{Project No.} &
\textbf{Language} &
\textbf{Granularity} &
\textbf{Data Points} &
\textbf{Evaluation Metrics} &
{\textbf{Model(s) Used}} &
{\textbf{Adaptation Strategy}} &
{\textbf{Evaluation Approach}} \\
\hline

\citet{LP1} & Review necessity prediction &
Open Source &
1,161 &
C, C++ and 7 others &
Chunk level &
328{,}000 diff hunks &
Classification Metrics: Accuracy; Precision; Recall; F-measure. &
{4 models, including open-weight pre-trained models (e.g., CodeReviewer)} &
{Fine-tuning type: Full} &
{LLM-as-judge: No \newline Human validation: No} \\ \hline

\citet{LP3} & Review necessity prediction &
Open Source &
N/A &
Go, Java and 7 others &
Chunk level &
$288,000$ samples &
Classification Metrics: Precision; Recall; F-measure. &
{8 models, including open-weight models (e.g., LLaMA)} & 
{Fine-tuning type: Instruction tuning; Parameter-efficient; LoRA} & 
{LLM-as-judge: No \newline Human validation: No} \\ \hline

\citet{LP15} & Review necessity prediction &
Private projects &
27 &
Python &
File level &
93 submission of 27 coding test questions  &
Classification Metrics: Blocking Rate; Review-Omission Rate; Usability Ratings. &
{3 models, including proprietary models(e.g., GPT-4o)} & 
{Prompting technique: Few-shot} & 
{LLM-as-judge: No \newline Human validation: Yes (3 programming education experts; 6 software education experts; 11 K-12 learners)} \\ \hline

{\citet{LP44}} &
{Request necessity prediction} &
{Open Source} &
{1} &
{Java, Python and C++} &
{PR level} &
{76{,}161 requests} &
{Classification Metrics: Accuracy; Precision; Recall; F-measure} &
{15 models (e.g., DeepSeek-7B)} &
{
Prompting technique: Hard prompt; Soft prompt (prefix tuning)
\newline
Fine-tuning type: N/A
} &
{
LLM-as-judge: No
\newline
Human validation: No
} \\ \hline

\citet{LP25} & Change quality prediction&
Open Source &
90  &
Java, JavaScript and 7 others &
Method level &
67,910 changed methods. &
Classification Metrics: F-measure; Accuracy; Precision. &
{9 models, including both proprietary (GPT-4) and open-weight models (Qwen2.5-Coder-Instruct 32B)} &
{Prompting technique: Few-shot; Chain-of-thought} &
{LLM-as-judge: Yes \newline Human validation: Yes (3 experts)} \\ \hline

\citet{LP5} & Change quality prediction &
Open Source &
N/A &
Python, Java and 7 others &
Chunk level &
1,800 code changes. &
Classification Metrics: Accuracy; Precision; Recall; F-measure. &
{8 models, including both proprietary (GPT-4) and open-weight models (LLaMA 2)} &
{N/A} &
{LLM-as-judge: No \newline Human validation: Yes} \\ \hline
\end{tabular}
}
\caption{Datasets and Evaluation Metrics for Change Quality Prediction in LLM era.}
\label{tab:Datasets and Evaluation Metrics for Change Quality Prediction in LLM era.}
\end{table*}

\subsection{Change Understanding and Analysis: }

\subsubsection{Relevant
Change Identification:}

Table~\ref{tab:Change Identification LLM} presents the dataset used to evaluate this capability in the LLM era. Unlike earlier Pre-LLM approaches that concentrated on detecting inconsistencies or missed edits, the dataset of the LLM era treats the task as change type recognition (i.e., delete, modify, and add), where models classify the intended action using both code fragments and review context. The dataset introduced by \citet{LP9} spans a diverse set of open-source projects across multiple languages that operate at the chunk level. Evaluation relies on classification-based measures (eg. MCQA accuracy and PPA) to assess correctness and agreement in multi-choice settings.



\begin{table*}[ht]
\centering
\resizebox{1\columnwidth}{!}{
\renewcommand{\arraystretch}{1}
\begin{tabular}{|>{\scriptsize\arraybackslash}p{1.4cm}|>{\scriptsize\arraybackslash}p{1cm}|>{\scriptsize\arraybackslash}p{1cm}|>{\scriptsize\arraybackslash}p{1.2cm}|>{\scriptsize\arraybackslash}p{1.3cm}|>{\scriptsize\arraybackslash}p{1.66cm}|>{\scriptsize\arraybackslash}p{1cm}|>{\scriptsize\arraybackslash}p{2.2cm}|>{\scriptsize\arraybackslash}p{2cm}|>{\scriptsize\arraybackslash}p{2cm}|>{\scriptsize\arraybackslash}p{2cm}|}
\hline
\textbf{Literature} &
\textbf{Task} &
\textbf{Source} &
\textbf{Project No.} &
\textbf{Language} &
\textbf{Granularity} &
\textbf{Data Points} &
\textbf{Evaluation Metrics} &
{\textbf{Model(s) Used}} &
{\textbf{Adaptation Strategy}} &
{\textbf{Evaluation Approach}} \\
\hline

\citet{LP9} &
Change Type
Recognition &
Open Source &  
199 &
PHP, Ruby, and 7 others 
&
Chunk level &
900 code review examples. &
Classification Metrics: MCQA accuracy; PPA. &
{72 models, including open-weight models (e.g., Llama-3.1-70B-Instruct)} &
{Prompting technique: Zero-shot; Few-shot; Chain-of-thought} &
{LLM-as-judge: No; \newline Human validation: Yes (2 authors)} \\ \hline
\end{tabular}
}
\caption{Datasets and Evaluation Metrics for Relevant Change Identification in LLM era.}
\label{tab:Change Identification LLM}
\end{table*}

\subsection{Peer Review }
\subsubsection{Defective Change Prediction:}
Defective Change Prediction in the LLM era leverages language models to identify buggy or vulnerable code by reasoning about semantic correctness and security weaknesses which enable automated systems to detect potentially risky code artifacts for further review. 
Table~\ref{tab:Defective Change Prediction LLM} summarizes representative datasets used to evaluate LLM-driven defective change prediction and vulnerability identification. \citet{LP15} address defect prediction in an educational context, where LLMs analyze programming exercise submissions to detect semantic issues (e.g., unmet requirements, hard-coded solutions, or logical mistakes) that standard test cases may overlook. Beyond defect prediction, one dataset extends this research toward security-oriented analysis through CWE prediction~\citep{LP24}, identifying standardized vulnerability types in C and C++ source code. Together, these datasets cover both private and open-source settings. 






\begin{table*}[ht]
\centering
\resizebox{1\columnwidth}{!}{
\renewcommand{\arraystretch}{1}
\begin{tabular}{|>{\scriptsize\arraybackslash}p{1.4cm}|>{\scriptsize\arraybackslash}p{1cm}|>{\scriptsize\arraybackslash}p{1cm}|>{\scriptsize\arraybackslash}p{1.2cm}|>{\scriptsize\arraybackslash}p{1.3cm}|>{\scriptsize\arraybackslash}p{1.66cm}|>{\scriptsize\arraybackslash}p{1cm}|>{\scriptsize\arraybackslash}p{2.2cm}|>{\scriptsize\arraybackslash}p{1.7cm}|>{\scriptsize\arraybackslash}p{2cm}|>{\scriptsize\arraybackslash}p{2.3cm}|}
\hline
\textbf{Literature} &
\textbf{Task} &
\textbf{Source} &
\textbf{Project No.} &
\textbf{Language} &
\textbf{Granularity} &
\textbf{Data Points} &
\textbf{Evaluation Metrics} &
{\textbf{Model(s) Used}} &
{\textbf{Adaptation Strategy}} &
{\textbf{Evaluation \newline Approach}} \\
\hline

\citet{LP15} &
Defect prediction&
Private Projects &
27 &
Python &
File level &
93 answers for coding test &
Classification Metrics: Relative Error Detection Rate (REDR); Fisher’s Exact Test &
{3 models, including proprietary models(e.g., GPT-4o)} &
{Prompting technique: Few-shot} &
{LLM-as-judge: No; Human validation: Yes (3 programming education experts; 6 software education experts; 11 K-12 learners)}\\ \hline

\citet{LP24} &
CWE prediction&
Open Source &
26 &
C, C++ &
Method level & 251{,}000 lines of code &
Classification Metrics: True Positives (TP); False Positives (FP); Precision; Recall &
{4 models (e.g., Claude 3.5 Sonnet)} &
{Prompting technique: Few-shot; Chain-of-thought; Instruction prompting} &
{LLM-as-judge: Yes; Human validation: Yes} \\ \hline

\end{tabular}
}
\caption{Datasets and Evaluation Metrics for Defective Change Prediction in LLM era.}
\label{tab:Defective Change Prediction LLM}
\end{table*}

\subsubsection{Issue Labeling / Classification:}
Issue Labeling and Classification assigns clear problem types to review comments or code changes so that reviewers can quickly understand what kind of issue is present (e.g., bugs, readability or maintainability issues) and focus on the most important parts of the code. Table~\ref{tab:Issue Labeling Classification} summarizes datasets constructed by studies in the LLM era. Most studies focus on issue categorization, where the goal is to classify review comments or code changes into predefined issue categories~\citep{LP17,LP13,LP27,LP2}. For example, \citet{LP17} constructed a dataset from open-source projects by injecting issues such as code duplication, missing documentation, and logic bugs into small Java and Python programs at the file level, in order to study how automated reviews affect human reviewer performance. 
At an industrial scale, \citet{LP13} built a dataset for a production-ready system that categorizes issues such as security vulnerabilities, functional bugs, and performance problems in private projects using a two-stage LLM pipeline. 
These studies are mainly evaluated using standard classification metrics. Beyond simple categorization, \citet{LP2} extended the task to multi-perspective labeling, which includes predicting issue types from review comments, classifying code changes, and locating problematic lines in source code. 

\begin{table*}[ht]
\centering
\resizebox{1\columnwidth}{!}{
\renewcommand{\arraystretch}{1}
\begin{tabular}{|>{\scriptsize\arraybackslash}p{1.25cm}|>{\scriptsize\arraybackslash}p{1.75cm}|>{\scriptsize\arraybackslash}p{0.85cm}|>{\scriptsize\arraybackslash}p{0.75cm}|>{\scriptsize\arraybackslash}p{1.15cm}|>{\scriptsize\arraybackslash}p{1.15cm}|>{\scriptsize\arraybackslash}p{1.25cm}|>{\scriptsize\arraybackslash}p{1.85cm}|>{\scriptsize\arraybackslash}p{1.55cm}|>{\scriptsize\arraybackslash}p{1.65cm}|>{\scriptsize\arraybackslash}p{2.05cm}|}
\hline
\textbf{Literature} &
\textbf{Task} &
\textbf{Source} &
\textbf{Project No.} &
\textbf{Language} &
\textbf{Granularity} &
\textbf{Data Points} &
\textbf{Evaluation Metrics} &
{\textbf{Model(s) Used}} &
{\textbf{Adaptation Strategy}} &
{\textbf{Evaluation \newline Approach}} \\
\hline

\citet{LP17} &
Issue categorization (e.g., code duplication, documentation issues, and logic bugs)&
Open Source &
6 &
Java, Python &
File level &
12 programs with 48 injected issues &
Classification Metrics: Accuracy. \par
User Study: Time spent; Reviewer confidence &
{1 model (e.g., GPT-4)} &
{N/A} &
{LLM-as-judge: No \newline Human validation: Yes (29 professional developers)} \\ \hline

\citet{LP13} & 
Issue categorization
using comment &
Private Projects &
N/A &
Go, JavaScript, and 3 others &
Chunk level &
120{,}000 review comments &
Classification Metrics: Precision; Recall.&
{4 models (e.g., Doubao-Pro-32K-0828)} &
{Fine-tuning type: LoRA} &
{LLM-as-judge: Yes \newline Human validation: Yes (137 users; 12 expert developers)} \\ \hline

\citet{LP2} &
Issue categorization
using comment &
Open Source &
19 &
N/A &
Chunk level &
262 comments &
Classification Metrics: Precision; Recall; F-measure &
{6 models (e.g., RoBERTa)} &
{N/A} &
{LLM-as-judge: No \newline Human validation: Yes (31 researchers)}\\ \cmidrule{2-11}

 &
 Issue categorization using code change &
Open Source &
19 &
N/A
&
Chunk level &
94{,}121 code review pairs  &
Classification Metrics: Precision; Recall; F-measure; Accuracy &
{6 models (e.g., RoBERTa)} &
{N/A} &
{LLM-as-judge: No \newline Human validation: Yes (31 researchers)}\\ \hline

\citet{LP27} &
Issue categorization (e.g., Readability, Bugs, Maintainability, Design, etc) &
Open-source \& Private Project &
N/A &
N/A &
Chunk level &
94{,}121 review comments&
Classification Metrics: Accuracy &
{2 models (e.g., gpt-4.1-2025-04-14)} &
{Prompting technique: Chain-of-thought; Instruction prompting} &
{LLM-as-judge: Yes \newline Human validation: Yes (3 annotators)} \\ \hline

\end{tabular}
}
\caption{Datasets and Evaluation Metrics for Issue Labeling/Classification in the LLM era.}
\label{tab:Issue Labeling Classification}
\end{table*}

\subsubsection{Issue Localization:}

Table~\ref{tab:Issue Localization Peer Review} summarizes datasets that link reviewer feedback or intent to exact code locations, enabling precise issue localization in real-world code review workflows. The majority of these studies constructed a dataset to address line-level issues or defect localization, where models are required to predict or rank specific faulty lines within a pull request or code diff by jointly reasoning over code context and reviewer feedback~\citep{LP2,LP19,LP25,LP26}.  While some approaches concentrate on ranking suspicious lines within modified methods using both textual context and before-and-after code structure~\citep{LP25}, others aim to detect exact faulty regions across entire diffs by analyzing code logic, recent changes, edge cases, resource handling, and API usage based on real-world reviewer comments~\citep{LP26}. Change localization focuses on determining where a requested modification should be applied in response to a natural-language review comment by reasoning over reviewer intent and pre-change code context~\citep{LP9}. 
Across these datasets, issue localization is studied using large-scale open-source code review scenarios, based on pull requests, diffs, and review comments. The tasks focus on precisely identifying faulty code regions at fine granularity, such as line, chunk, or method level, rather than whole files. Evaluation in these datasets mainly uses classification-based accuracy and ranking metrics.

\begin{table*}[ht]
\centering
\resizebox{1\columnwidth}{!}{
\renewcommand{\arraystretch}{1}
\begin{tabular}{|>{\scriptsize\arraybackslash}p{1.4cm}|>{\scriptsize\arraybackslash}p{1cm}|>{\scriptsize\arraybackslash}p{1cm}|>{\scriptsize\arraybackslash}p{1.2cm}|>{\scriptsize\arraybackslash}p{1.3cm}|>{\scriptsize\arraybackslash}p{1.66cm}|>{\scriptsize\arraybackslash}p{1cm}|>{\scriptsize\arraybackslash}p{2.2cm}|>{\scriptsize\arraybackslash}p{2cm}|>{\scriptsize\arraybackslash}p{2cm}|>{\scriptsize\arraybackslash}p{2cm}|}
\hline
\textbf{Literature} &
\textbf{Task} &
\textbf{Source} &
\textbf{Project No.} &
\textbf{Language} &
\textbf{Granularity} &
\textbf{Data Points} &
\textbf{Evaluation Metrics} &
{\textbf{Model(s) Used}} &
{\textbf{Adaptation Strategy}} &
{\textbf{Evaluation Approach}} \\
\hline

\citet{LP19} &
Line-level issue localization  
&
Open Source &
12 &
Python &
PR level &
1{,}000 pull requests &
Classification Metrics: True Positive; False Positive; False Negative; Precision; Recall; F-measure. &
{16 models, including both proprietary (o3-0416) and open-weight models (DeepSeek-R1-0120)} &
{Prompting technique: Instruction prompting} &
{LLM-as-judge: Yes \newline Human validation: Yes (3 experts; 5 graduate students)}\\ \hline

\citet{LP2} &
Line-level issue localization &
Open Source &
19 &
N/A &
Chunk level &
2{,}349{,}102 lines of code &
Classification Metrics: Precision; Recall; F-measure &
{6 models (e.g., RoBERTa)} &
{N/A} &
{LLM-as-judge: No \newline Human validation: Yes (31 researchers)}\\ \hline

\citet{LP25} &
Line-level defect localization &
Open Source &
90 &
Python, Java, and 7 others &
Method level &
13,512 entries. &
Classification Metrics: F-measure; Accuracy; Precision. &
{9 models, including both proprietary (GPT-4) and open-weight models (Qwen2.5-Coder-Instruct 32B)} &
{Prompting technique: Few-shot; Chain-of-thought} &
{LLM-as-judge: Yes \newline Human validation: Yes (3 experts)}\\ \hline

\citet{LP26} &
Line-level issue localization &
Open Source &
11,324 &
Java, Python, and 3 others &
Chunk level &
12,881 code review examples &
Text-Matching / Generation Metrics: IoU (Intersection over Union). &
{5 models, including open-weight models (e.g., Qwen2.5-14B)} &
{Prompting technique: Chain-of-thought \newline Fine-tuning type: Full; Instruction tuning} &
{LLM-as-judge: Yes \newline Human validation: Yes (2 students)}\\ \hline

\citet{LP9} &
Change Localization (where to apply changes) &
Open Source &
199 &
C, C++ and 7 others &
Chunk level &
900 code review examples. &
Classification Metrics:MCQA accuracy; PPA; Perplexity and 5-gram accuracy. &
{72 models, including open-weight models (e.g., Llama-3.1-70B-Instruct)} &
{Prompting technique: Zero-shot; Few-shot; Chain-of-thought} &
{LLM-as-judge: No; Human validation: Yes (2 authors)} \\ \hline

\end{tabular}
}
\caption{Datasets and Evaluation Metrics for Issue Localization in the LLM era.}
\label{tab:Issue Localization Peer Review}
\end{table*} 

\subsubsection{Review Comment Generation:}

Review Comment Generation in the LLM era focuses on automatically producing helpful, human-like review comments for code changes by leveraging large language models’ code understanding and natural-language generation capabilities. Table~\ref{tab:Benchmarks for Review Comment Recommendation / Generation Tasks in LLM era.} summarizes datasets for this task. Most of the datasets are constructed for natural language (NL) comment generation~\citep{LP1,LP3,LP5,LP12,LP14,LP15,LP23,LP25}. Beyond free-form comments, several studies introduce structured review comment generation, requiring models to produce organized feedback across multiple dimensions such as functionality, complexity, style, documentation, and defects~\citep{LP10,LP26}, and some further reframe the task as defect identification and explanation with emphasis on fault localization and impact analysis in high-risk contexts~\citep{LP14,LP22}. Tasks are primarily defined at chunk and method levels, with specialized line-level settings demanding precise, localized feedback~\citep{LP25}. The evaluation typically combines text-generation metrics with human or LLM-as-a-Judge assessments to measure clarity, usefulness, and alignment with reviewer intent~\citep{LP1,LP3,LP6,LP10,LP22,LP25,LP26}.
Most datasets are derived from open-source projects, enabling models to learn realistic reviewer behavior from historical comments and diffs~\citep{LP1,LP3,LP6,LP23}, while a smaller but important subset extends to private or industrial settings, including multilingual review generation~\citep{LP12}, educational code review on student submissions~\citep{LP15}, and defect-focused industrial reviews~\citep{LP22}. ~\citep{LP6} relies on a dataset which is reused by several subsequent studies ~\citep{LP33, LP34, LP35, LP36, LP37, LP38, LP39, LP40, LP41}.
Similarly, ~\citep{LP3} belongs to another group of works ~\citep{LP29, LP30, LP31, LP32} that employ closely related or overlapping datasets originating from the same CodeReviewer-style corpus.

\begin{table*}[ht]
\centering
\begin{adjustbox}{width=\columnwidth, max totalheight=0.94\textheight}
\renewcommand{\arraystretch}{1}
\begin{tabular}{|>{\scriptsize\arraybackslash}p{1.25cm}|>{\scriptsize\arraybackslash}p{1.35cm}|>{\scriptsize\arraybackslash}p{1cm}|>{\scriptsize\arraybackslash}p{1.05cm}|>{\scriptsize\arraybackslash}p{1.0cm}|>{\scriptsize\arraybackslash}p{1.66cm}|>{\scriptsize\arraybackslash}p{1cm}|>{\scriptsize\arraybackslash}p{2.25cm}|>{\scriptsize\arraybackslash}p{2.3cm}|>{\scriptsize\arraybackslash}p{1.7cm}|>{\scriptsize\arraybackslash}p{2.2cm}|}
\hline
\textbf{Literature} &
\textbf{Task} &
\textbf{Source} &
\textbf{Project No.} &
\textbf{Language} &
\textbf{Granularity} &
\textbf{Data Points} &
\textbf{Evaluation Metrics} &
{\textbf{Model(s) Used}} &
{\textbf{Adaptation Strategy}} &
{\textbf{Evaluation \newline Approach}} \\
\hline

\citet{LP1} &
NL Review comment generation &
Open Source &
1{,}161 &
C, C++ and 7 others &
Chunk level &
138{,}000 hunk - comment pairs &
Text-Matching / Generation Metrics: BLEU &
{5 models, including both proprietary (e.g. Codex/Copilot) and open-weight models (CodeReviewer)} &
{Prompting technique: Few-shot \newline Fine-tuning type: Full} &
{LLM-as-judge: No \newline Human validation: Yes (6 professional programmers)} \\ \hline

\citet{LP3}
&
NL Review comment generation &
Open Source &
CodeReviewer + Tufano datasets &
C, C++ and 7 others &
Method level &
Tufano: $168,000$ functions\par
CRer: 138000 diffs&
Text-Matching / Text-Generation Metrics: BLEU, top-1 results. &
{8 models, including open-weight models (e.g., LLaMA)} &
{Fine-tuning type: Instruction tuning; Parameter-efficient; LoRA} &
{LLM-as-judge: No \newline Human validation: No} \\ \hline

\citet{LP5} &
NL Review comment generation &
Open Source &
N/A &
Python, Java and 7 others &
Chunk level &
1{,}800 code samples. &
Text-Matching / Generation Metrics: BLEU; ROUGE; BERTScore.  &
{8 models, including both proprietary (GPT-4) and open-weight models (LLaMA 2)} &
{N/A} &
{LLM-as-judge: No \newline Human validation: Yes} \\ \hline

\citet{LP6};
&
NL Review comment generation  &
Open Source &
CodeReviewer dataset &
C, C++ and 7 others &
Chunk level &
$143,500$ samples &
Text-Matching / Generation Metrics: BLEU; BERTScore. &
{10 models, including both proprietary (GPT-3.5 Turbo) and open-weight models (Llama 2)} &
{Prompting technique: Few-shot; Instruction prompting \newline Fine-tuning type: Parameter-efficient; LoRA} &
{LLM-as-judge: No \newline Human validation: Yes (8 software practitioners)} \\ \hline

\citet{LP12} &
NL review comment generation (Russian) &
Private projects &
N/A &
Java, Python, Go, Scala &
Chunk level &
689 pairs of merge-request&
Text-Matching / Generation Metrics: Judge@k (LLM-as-a-Judge); BLEU; chrF. &
{27 models (e.g., GPT-4.1)} &
{Prompting technique: Instruction prompting} &
{LLM-as-judge: Yes \newline Human validation: Yes (2 experienced software developers)} \\ \hline

\citet{LP14} & 
NL Review comment generation &
Open Source &
N/A &
Java &
Chunk level &
27{,}267 Java examples &
Classification Metrics: Accuracy Labels (accurate / partially accurate / not accurate) &
{2 models (e.g., CodeLlama-7b)} &
{Fine-tuning type: Parameter-efficient} &
{LLM-as-judge: Yes \newline Human validation: Yes (2 evaluators)} \\ \hline

\citet{LP15} &
NL review comment generation &
Private projects &
27  &
Python &
File level &
93 test cases for 27 questions written by 72 students &
Text-Matching / Generation Metrics: BERTScore. &
{3 models, including proprietary models(e.g., GPT-4o)} &
{Prompting technique: Few-shot} &
{LLM-as-judge: No \newline Human validation: Yes (3 programming education experts; 6 software education experts; 11 K-12 learners)} \\ \hline

\citet{LP23} &
NL review comment generation &
Open Source &
N/A &
PHP, Ruby and 7 others &
Chunk level &
176{,}613 samples &
Text-Matching / Generation Metrics: BLEU. &
{2 models (e.g., Llama-3.1-70B-Instruct)} &
{Prompting technique: Few-shot; Chain-of-thought \newline Fine-tuning type: LoRA} &
{LLM-as-judge: Yes \newline Human validation: Yes (2 authors)} \\ \hline

\citet{LP25} &
NL review comment generation  &
Open Source &
90 &
Python, Java and 7 others &
Method level &
13{,}512 entries. &
Text-Matching / Generation Metrics: ROUGE-1; ROUGE-L; Edit Similarity. \par
Classification Metrics: F-measure; Accuracy; Precision. &
{9 models, including both proprietary (GPT-4) and open-weight models (Qwen2.5-Coder-Instruct 32B)} &
{Prompting technique: Few-shot; Chain-of-thought} &
{LLM-as-judge: Yes \newline Human validation: Yes (3 experts)} \\ \hline

\citet{LP22} &
Defect identification and explanation &
Private projects &
N/A &
C++ &
PR level &
45 merge requests &
Classification Metrics: False Alarm Rate;
LSR (Line Success Rate);
CPI1/CPI2 (Comprehensive Performance Index 1/2) &
{9 models, including open-weight models (e.g., LLaMA-3.1 (70B))} &
{Prompting technique: Chain-of-thought} &
{LLM-as-judge: Yes \newline Human validation: Yes} \\ \hline

\citet{LP26} &
Structured review comment generation &
Open Source &
11{,}324 &
Java, Python and 3 others &
Chunk level &
2{,}000 test examples &
Retrieval Metric: Hit Rate &
{5 models, including open-weight models (e.g., Qwen2.5-14B)} &
{Prompting technique: Chain-of-thought \newline Fine-tuning type: Full; Instruction tuning} &
{LLM-as-judge: Yes \newline Human validation: Yes (2 students)}\\ \hline

\citet{LP10} &
Structured review generation (e.g., function, complexity, style, documentation, and defects) &
Open Source &
70 &
Python &
Chunk level &
601 code review examples &
Text-Matching / Generation Metrics: BLEU, LLM-as-a-Judge \par
Classification Metrics: Precision, Recall, and F-Measure. &
{10 models, including both proprietary (GPT-5) and open-weight models (Qwen3-235B-A22B)} &
{Fine-tuning type: LoRA} &
{LLM-as-judge: Yes \newline Human validation: Yes (2 authors)} \\ \hline

{\citet{LP46}} &
{NL review comment generation} &
{Open Source} &
{5{,}973 repositories} &
{Java} &
{Commit level} &
{35{,}950 commits} &
{Text-Matching / Generation Metrics: BLEU-4} &
{7 models (e.g., GPT-4o)} &
{Prompting technique: Zero-shot; Chain-of-thought; Self-reflection}
& {LLM-as-judge: No} \par {Human validation: Yes (2 authors)} \\ \hline

\end{tabular}
\end{adjustbox}
\caption{Datasets and Evaluation Metrics for Review Comment Generation in LLM era.}
\label{tab:Benchmarks for Review Comment Recommendation / Generation Tasks in LLM era.}
\end{table*}

\subsubsection{Security Detection:}
Table~\ref{tab:Security Peer Review} summarizes datasets that formulate this task as vulnerability analysis during peer review. The dataset by \citet{LP4} models security-focused peer review as automated analysis of code changes, where LLMs inspect commits to detect a broad range of security issues introduced during development. 





\begin{table*}[ht]
\centering
\resizebox{1\columnwidth}{!}{
\renewcommand{\arraystretch}{1}
\begin{tabular}{|>{\scriptsize\arraybackslash}p{1.4cm}|>{\scriptsize\arraybackslash}p{1cm}|>{\scriptsize\arraybackslash}p{1cm}|>{\scriptsize\arraybackslash}p{1.2cm}|>{\scriptsize\arraybackslash}p{1.3cm}|>{\scriptsize\arraybackslash}p{1.66cm}|>{\scriptsize\arraybackslash}p{1cm}|>{\scriptsize\arraybackslash}p{2.2cm}|>{\scriptsize\arraybackslash}p{2cm}|>{\scriptsize\arraybackslash}p{2cm}|>{\scriptsize\arraybackslash}p{2cm}|}
\hline
\textbf{Literature} &
\textbf{Task} &
\textbf{Source} &
\textbf{Project No.} &
\textbf{Language} &
\textbf{Granularity} &
\textbf{Data Points} &
\textbf{Evaluation Metrics} &
{\textbf{Model(s) Used}} &
{\textbf{Adaptation Strategy}} &
{\textbf{Evaluation Approach}} \\
\hline

\citet{LP4} &
Vulnerability detection &
Open Source &
180 &
Python, Java and 7 others &
Commit level &
3{,}545 commits &
Classification Metrics: Hit Rate &
{14 models (e.g., CodeAgent)} &
{Prompting technique: Chain-of-thought; Instruction prompting} &
{LLM-as-judge: Yes \newline Human validation: Yes (2 people per sample: a researcher and an engineer)}\\ \hline

{\citet{LP42}} &
{Security defect detection} &
{Open Source} &
{4} &
{Python, C and C++} &
{File level} &
{534 code files} &
{Rating Metrics: I-Score; M-Score} &
{7 models, including both proprietary (GPT-4) and open-weight models (Llama 2 7B)} &
{Prompting technique: Chain-of-thought; Instruction prompting} &
{LLM-as-judge: No \newline Human validation: Yes (3 authors)} \\ \hline

\end{tabular}
}
\caption{Dataset for Security Detection in LLM era.}
\label{tab:Security Peer Review}
\end{table*}

\subsection{Review Assessment and Analysis}

\subsubsection{Sentiment / Toxicity Analysis}

Table~\ref{tab:CommentToxicityAnalysis} presents an LLM era dataset for analyzing toxicity in code review comments, focusing on identifying harmful or unprofessional language. The task is framed at the pull-request level within an educational and competition-oriented framework, where LLMs automatically evaluate review comments and assign toxicity scores to discourage negative behavior~\citep{LP7}. Unlike pre-LLM approaches that relied on static label prediction, this dataset uses LLMs for context-aware interpretation of reviewer language, prioritizing collaboration quality over code correctness. 





\begin{table*}[ht]
\centering
\resizebox{1\columnwidth}{!}{
\renewcommand{\arraystretch}{1}
\begin{tabular}{|>{\scriptsize\arraybackslash}p{1.4cm}|>{\scriptsize\arraybackslash}p{1cm}|>{\scriptsize\arraybackslash}p{1cm}|>{\scriptsize\arraybackslash}p{1.2cm}|>{\scriptsize\arraybackslash}p{1.3cm}|>{\scriptsize\arraybackslash}p{1.66cm}|>{\scriptsize\arraybackslash}p{1cm}|>{\scriptsize\arraybackslash}p{2.2cm}|>{\scriptsize\arraybackslash}p{2cm}|>{\scriptsize\arraybackslash}p{2cm}|>{\scriptsize\arraybackslash}p{2cm}|}
\hline
\textbf{Literature} &
\textbf{Task} &
\textbf{Source} &
\textbf{Project No.} &
\textbf{Language} &
\textbf{Granularity} &
\textbf{Data Points} &
\textbf{Evaluation Metrics} &
{\textbf{Model(s) Used}} &
{\textbf{Adaptation Strategy}} &
{\textbf{Evaluation Approach}} \\
\hline

\citet{LP7} &
Review comment toxicity analysis&
Educational / Competition &
2 &
Java, Python &
PR level &
86 students in 30 teams reviewed 40 pull requests &
Rating Metrics 
: LLM toxicity scores (rated on a negative scale from $-1$ (least toxic) to $-10$ (very toxic)) &
{1 model (e.g., ChatGPT 3.5 Turbo)} &
{N/A} &
{LLM-as-judge: Yes \newline Human validation: Yes (86 students)} \\ \hline

\end{tabular}
}
\caption{Dataset for Sentiment / Toxicity Analysis in the LLM era.}
\label{tab:CommentToxicityAnalysis}
\end{table*} 

\subsubsection{Code Compliance}
Table~\ref{tab:Review Comment Code Compliance} presents datasets for code compliance tasks in the LLM era. Most studies examine whether natural-language artifacts that accompany a change—such as review comments, commit messages, or linked issues—accurately describe the related code changes (e.g., semantic relevance or missing/tangled edits)~\citep{LP7,LP18,LP4}. This task appears in both educational and open-source contexts and requires reasoning over code diffs and text to detect mismatches or mixed changes. An important extension is coding format consistency analysis, which verifies whether code modifications follow existing style conventions, showing that consistency checks now go beyond semantic intent~\citep{LP4}. These datasets cover pull-request and commit-level analysis, include programming course projects and large-scale open-source repositories, and span multiple languages, reflecting the diversity of evaluation settings. Common evaluation approaches use classification metrics (e.g., precision, recall, F-measure) or rating-based metrics, combining automated correctness checks with human-like judgment of semantic alignment.

\begin{table*}[ht]
\centering
\resizebox{1\columnwidth}{!}{
\renewcommand{\arraystretch}{1}
\begin{tabular}{|>{\scriptsize\arraybackslash}p{1.4cm}|>{\scriptsize\arraybackslash}p{1cm}|>{\scriptsize\arraybackslash}p{1cm}|>{\scriptsize\arraybackslash}p{1.2cm}|>{\scriptsize\arraybackslash}p{1.3cm}|>{\scriptsize\arraybackslash}p{1.66cm}|>{\scriptsize\arraybackslash}p{1cm}|>{\scriptsize\arraybackslash}p{2.2cm}|>{\scriptsize\arraybackslash}p{2cm}|>{\scriptsize\arraybackslash}p{2cm}|>{\scriptsize\arraybackslash}p{2cm}|}
\hline
\textbf{Literature} &
\textbf{Task} &
\textbf{Source} &
\textbf{Project No.} &
\textbf{Language} &
\textbf{Granularity} &
\textbf{Data Points} &
\textbf{Evaluation Metrics} &
{\textbf{Model(s) Used}} &
{\textbf{Adaptation Strategy}} &
{\textbf{Evaluation Approach}} \\
\hline

\citet{LP7} &
Review comment-code consistency analysis & 
Programming course / competition &
2 &
Java, Python. &
PR-level &
40 pull requests 
&
Rating Metrics: LLM-as-a-Judge relevance scores  &
{1 model (e.g., ChatGPT 3.5 Turbo)} &
{N/A} &
{LLM-as-judge: Yes \newline Human validation: Yes (86 students)}\\
\hline

\citet{LP18} &
Issue– pull request consistency analysis &
Open Source &
1 &
Python &
PR level &
194 pull-request&
Classification Metrics: Accuracy; Precision; Recall;  F-measure; &
{4 models, including proprietary models (GPT-4o)} &
{Prompting technique: Instruction prompting} &
{LLM-as-judge: No \newline Human validation: Yes (3 authors)}\\
\hline

\multirow{2}{=}{\citet{LP4}}
&
Commit message -code consistency analysis &
Open Source &
180 & Python, Java and 7 others &
Commit level &
3{,}545 labeled commits &
Classification Metrics: F-measure; Recall. &
{14 models (e.g., CodeAgent)} &
{Prompting technique: Chain-of-thought; Instruction prompting} &
{LLM-as-judge: Yes \newline Human validation: Yes (2 people per sample: a researcher and an engineer)}\\
\cmidrule{2-11}

& 
Coding format consistency analysis &
Open Source &
180 & Python, Java and 7 others &
Commit level &
3{,}545 labeled commits &
Classification Metrics: F-measure; Recall. &
{14 models (e.g., CodeAgent)} &
{Prompting technique: Chain-of-thought; Instruction prompting} &
{LLM-as-judge: Yes \newline Human validation: Yes (2 people per sample: a researcher and an engineer)}\\
\hline

\end{tabular}
}
\caption{Datasets and Evaluation Metrics for Code Compliance Tasks in the LLM era.}
\label{tab:Review Comment Code Compliance}
\end{table*} 

\subsubsection{Review Quality Evaluation:} 

Table~\ref{tab:Review Quality} summarizes the datasets in the LLM era that primarily address review comment helpfulness analysis, which can be grouped into two closely related task settings: educational review quality assessment and real-world development review effectiveness analysis.
In educational contexts, \citet{LP8} and \citet{LP28} study review helpfulness as a learning support task in programming courses, where LLM-generated feedback is provided on student submissions and evaluated at the file level. These studies focus on understanding how automated reviews support learning, clarity, and defect comprehension, using relatively small-scale datasets (e.g., AI suggestions, review text pages, and student assignment responses) and relying on Rating Metrics such as Likert-scale user studies, which require skills in human-centered evaluation and qualitative analysis.
In contrast, studies targeting professional or mixed development settings emphasize whether review comments lead to concrete code changes. \citet{LP16} investigates review helpfulness in private industrial projects across multiple programming languages at the PR level, using large-scale pull request data and modeling reviewer impact with Classification Metrics based on Poisson Regression, which demands statistical modeling and empirical software engineering expertise. Similarly, \citet{LP27} analyzes review helpfulness across open-source and private projects at a finer chunk-level granularity, leveraging millions of labeled code lines to measure Resolution Rate with review comments, thereby emphasizing large-scale data mining, change tracking, and causal reasoning about review effectiveness.

\begin{table*}[ht]
\centering
\resizebox{1\columnwidth}{!}{
\renewcommand{\arraystretch}{1}
\begin{tabular}{|>{\scriptsize\arraybackslash}p{1.4cm}|>{\scriptsize\arraybackslash}p{1cm}|>{\scriptsize\arraybackslash}p{1cm}|>{\scriptsize\arraybackslash}p{1.2cm}|>{\scriptsize\arraybackslash}p{1.3cm}|>{\scriptsize\arraybackslash}p{1.66cm}|>{\scriptsize\arraybackslash}p{1cm}|>{\scriptsize\arraybackslash}p{2.2cm}|>{\scriptsize\arraybackslash}p{2cm}|>{\scriptsize\arraybackslash}p{2cm}|>{\scriptsize\arraybackslash}p{2cm}|}
\hline
\textbf{Literature} &
\textbf{Task} &
\textbf{Source} &
\textbf{Project No.} &
\textbf{Language} &
\textbf{Granularity} &
\textbf{Data Points} &
\textbf{Evaluation Metrics} &
{\textbf{Model(s) Used}} &
{\textbf{Adaptation Strategy}} &
{\textbf{Evaluation Approach}} \\
\hline

\citet{LP8} &
Review comment helpfulness analysis & 
Programming course / competition &
N/A &
Java &
File level &
173 AI suggestions and 58 pages review text. &
Rating Metrics:  Likert-scale ratings &
{1 model (e.g., GPT)} &
{N/A} &
{LLM-as-judge: No \newline Human validation: Yes (students)}\\ \hline

\citet{LP16} &
Review comment helpfulness analysis &
Private projects &
3 &
Java, JavaScript and 5 others &
PR level &
4,335 pull requests  &
Classification Metrics: Poisson Regression; &
{2 models (e.g., GPT-4 Turbo)} &
{N/A} &
{LLM-as-judge: No \newline Human validation: Yes (22 developers)}\\ \hline

\citet{LP27} &
 Review comment helpfulness analysis &
Open-source \& private project &
N/A &
N/A &
Chunk level &
2,349,102 labeled code lines &
Classification Metrics: Resolution Rate with review comments.&
{2 models (e.g., gpt-4.1-2025-04-14)} &
{Prompting technique: Chain-of-thought; Instruction prompting} &
{LLM-as-judge: Yes \newline Human validation: Yes (3 annotators)} \\ \hline

\citet{LP28} &
Review helpfulness analysis &
Programming course / competition &
N/A &
C++ &
File level & 40 responses of student assignments &
Rating Metrics: Likert-scale (user study) &
{1 model (e.g., GPT 3.5-turbo)} &
{Prompting technique: Instruction prompting} &
{LLM-as-judge: No \newline Human validation: Yes (40 students)}\\ \hline

{\citet{LP43}} &
{Review helpfulness analysis} &
{Open Source \& private project} &
{6} &
{N/A} &
{Chunk level} &
{7{,}929 comments} &
{Classification Metrics: Accuracy; Precision; Recall; F-measure} &
{23 models, including both proprietary (GPT-4o) and open-weight models (CodeReviewer)} &
{Prompting technique: Instruction prompting \newline Fine-tuning type: Full} &
{LLM-as-judge: No; Human validation: No}\\ \hline

\end{tabular}
}
\caption{Datasets and Evaluation Metrics for Review Quality Evaluation in the LLM era. }
\label{tab:Review Quality}
\end{table*}

\subsubsection{Review Summarization: }
Table~\ref{tab:Benchmarks for Other Techniques on Code Review in LLM era.} summarizes LLM era datasets that explore pull-request link summarization as a supporting technique for code review. Rather than revising code directly, this task aims to assist reviewers by generating concise summaries of external resources (e.g., documentation, issue pages, or web links) referenced within pull requests, thereby reducing context switching and improving review efficiency. The existing studies address both educational and open-source datasets where models are expected to interpret the surrounding pull-request context and produce informative link summaries that help reviewers quickly grasp the relevance of referenced materials. While educational settings emphasize reviewer engagement and perceived usefulness through human-centered evaluations~\citep{LP7}, open-source datasets focus more on the quality and semantic faithfulness of generated summaries using automated text similarity measures~\citep{LP20}. Despite differences in evaluation emphasis, both settings treat link summarization as a PR-level auxiliary task designed to support, rather than replace human review activities ~\citep{LP7,LP20}.





\begin{table*}[ht]
\centering
\resizebox{1\columnwidth}{!}{
\renewcommand{\arraystretch}{1}
\begin{tabular}{|>{\scriptsize\arraybackslash}p{1.4cm}|>{\scriptsize\arraybackslash}p{1cm}|>{\scriptsize\arraybackslash}p{1cm}|>{\scriptsize\arraybackslash}p{1.2cm}|>{\scriptsize\arraybackslash}p{1.3cm}|>{\scriptsize\arraybackslash}p{1.66cm}|>{\scriptsize\arraybackslash}p{1cm}|>{\scriptsize\arraybackslash}p{2.2cm}|>{\scriptsize\arraybackslash}p{2cm}|>{\scriptsize\arraybackslash}p{2cm}|>{\scriptsize\arraybackslash}p{2cm}|}
\hline
\textbf{Literature} &
\textbf{Task} &
\textbf{Source} &
\textbf{Project No.} &
\textbf{Language} &
\textbf{Granularity} &
\textbf{Data Points} &
\textbf{Evaluation Metrics} &
{\textbf{Model(s) Used}} &
{\textbf{Adaptation Strategy}} &
{\textbf{Evaluation Approach}} \\
\hline

\citet{LP7} & Pull-request link summarization &
Programming course/competition &
2 &
Java , Python &
PR-level &
86 students, working in 30 teams, reviewed 40 pull requests &
User studies  &
{1 model (e.g., ChatGPT 3.5 Turbo)} &
{N/A} &
{LLM-as-judge: Yes \newline Human validation: Yes (86 students)}\\ \hline

\citet{LP20} & Pull-request link summarization &
Open Source &
50 &
N/A &
PR level &
365 sample links &
Text-Matching / Generation Metrics: BLEU; METEOR; ROUGE-1; ROUGE-2; Sentence Similarity; BERTScore (Precision, Recall, F-measure &
{2 models (e.g., DeepSeek)} &
{N/A} &
{LLM-as-judge: No \newline Human validation: Yes (7 participants)}\\ \hline

\end{tabular}
}
\caption{Datasets and Evaluation Metrics for Review Summarization in the LLM era.}
\label{tab:Benchmarks for Other Techniques on Code Review in LLM era.}
\end{table*}

\subsection{Code Refinement }

\subsubsection{Code Revision:} 

Table~\ref{tab:Code Revision Suggestion} summarizes datasets for code revision. The datasets of existing studies can be categorized into two primary tasks, e.g., revised code generation ( where models update code to satisfy reviewer feedback) and best candidate selection (where models identify the correct revision among alternatives). The majority of LLM era datasets focus on revised code generation, where models must understand natural-language review comments and apply the requested changes to produce a corrected, review-compliant version of the code~\citep{LP1,LP3,LP4,LP9,LP11,LP23}. Those datasets emphasize end-to-end code refinement, requiring joint reasoning over reviewer intent and code context. Most studies rely on open-source repositories, while a smaller number extend to private industrial settings and evaluate generated patches in real-world deployment ~\citep{LP11}, where build, lint, and test checks are applied before a patch is shown to the author. In contrast, best candidate selection is explored in a smaller set of datasets, where the goal is to identify the correct revision from multiple candidate changes instead of generating code directly, emphasizing discriminative reasoning over change intent and location~\citep{LP9}. Overall, LLM era datasets reflect a clear shift toward reviewer-aware code revision, with broader language coverage, more diverse data sources, and an increased focus on producing or selecting revisions that align with human reviewer expectations in realistic review scenarios.

\begin{table*}[ht]
\centering
\resizebox{1\columnwidth}{!}{
\renewcommand{\arraystretch}{1}
\begin{tabular}{|>{\scriptsize\arraybackslash}p{1.4cm}|>{\scriptsize\arraybackslash}p{1cm}|>{\scriptsize\arraybackslash}p{1cm}|>{\scriptsize\arraybackslash}p{1.2cm}|>{\scriptsize\arraybackslash}p{1.3cm}|>{\scriptsize\arraybackslash}p{1.66cm}|>{\scriptsize\arraybackslash}p{1cm}|>{\scriptsize\arraybackslash}p{2.2cm}|>{\scriptsize\arraybackslash}p{2cm}|>{\scriptsize\arraybackslash}p{2cm}|>{\scriptsize\arraybackslash}p{2cm}|}
\hline
\textbf{Literature} &
\textbf{Task} &
\textbf{Source} &
\textbf{Project No.} &
\textbf{Language} &
\textbf{Granularity} &
\textbf{Data Points} &
\textbf{Evaluation Metrics} &
{\textbf{Model(s) Used}} &
{\textbf{Adaptation Strategy}} &
{\textbf{Evaluation Approach}} \\
\hline

\citet{LP1} &
Revised code generation &
Open Source &
1,161 &
C, C++ and 7 others &
Chunk level &
176{,}000 code revisions &
Text-Matching / Generation Metrics: BLEU; Exact Match. &
{3 open-weight models (e.g., CodeReviewer)} &
{Fine-tuning type: Full} &
{LLM-as-judge: No \newline Human validation: No} \\ \hline

\citet{LP3} &
Revised code generation &
Open Source &
N/A &
Go, Java and 7 others&
Method level &
Tufano: $168,000$ functions\par
CRer: 138000 diffs &
Text-Matching / Text-Generation Metrics: BLEU &
{8 models, including open-weight models (e.g., LLaMA)} &
{Fine-tuning type: Instruction tuning; Parameter-efficient; LoRA} &
{LLM-as-judge: No \newline Human validation: No} \\ \hline

\citet{LP4} &
Revised code generation &
Open Source &
N/A 
&
N/A &
Commit level &
3,545 &
Classification Metrics: Edit Progress. &
{14 models (e.g., CodeAgent)} &
{Prompting technique: Chain-of-thought; Instruction prompting} &
{LLM-as-judge: Yes \newline Human validation: Yes (2 people per sample: a researcher and an engineer)} \\ \hline

\citet{LP11} &
Revised code generation &
Private projects &
N/A &
Hack; PHP &
Chunk level &
206 benchmark pairs &
Text-Matching / Generation Metrics: Exact Match; Patch applicability: Successful Patch Generation (SPG) &
{3 models, including proprietary (GPT-4o) and fine-tuned open-weight models (Llama-3-70B-based)} &
{Prompting technique: Zero-shot (GPT-4o) \newline Fine-tuning type: Continual pre-training and supervised fine-tuning} &
{LLM-as-judge: No \newline Human validation: Yes (engineers)} \\ \hline

\citet{LP9} &
Revised code generation &
Open Source &
199 &
C, C++ and 7 others &
Chunk level &
900 code review examples &
Text-Matching / Generation Metrics: Exact Match; Perplexity; 5-gram accuracy. &
{72 models, including open-weight models (e.g., Llama-3.1-70B-Instruct)} &
{Prompting technique: Zero-shot; Few-shot; Chain-of-thought} &
{LLM-as-judge: No; Human validation: Yes (2 authors)}\\
\cmidrule{2-11}

&
Best candidate selection &
Open Source &
199 &
C, C++ and 7 others &
Chunk level &
900 code review examples &
Classification Metrics: Invariant MCQA accuracy; PPA; Perplexity; 5-gram accuracy. &
{72 models, including open-weight models (e.g., Llama-3.1-70B-Instruct)} &
{Prompting technique: Zero-shot; Few-shot; Chain-of-thought} &
{LLM-as-judge: No; Human validation: Yes (2 authors)} \\
\hline

\citet{LP23} &
Revised code generation &
Open Source &
N/A &
PHP, Ruby and 7 others&
Chunk level &
20{,}000 pairs review comments &
Text-Matching / Generation Metrics: CodeBLEU; Exact Match. &
{2 models (e.g., Llama-3.1-70B-Instruct)} &
{Prompting technique: Few-shot; Chain-of-thought \newline Fine-tuning type: LoRA} &
{LLM-as-judge: Yes \newline Human validation: Yes (2 authors)}  \\ \hline

\end{tabular}
}
\caption{Datasets and Evaluation Metrics for Code Revision in LLM era.}
\label{tab:Code Revision Suggestion}
\end{table*}

\section{RQ3: Comparison between Pre-LLM and LLM eras}

\begin{figure}[t]
    \centering
    \includegraphics[width=0.9\textwidth]{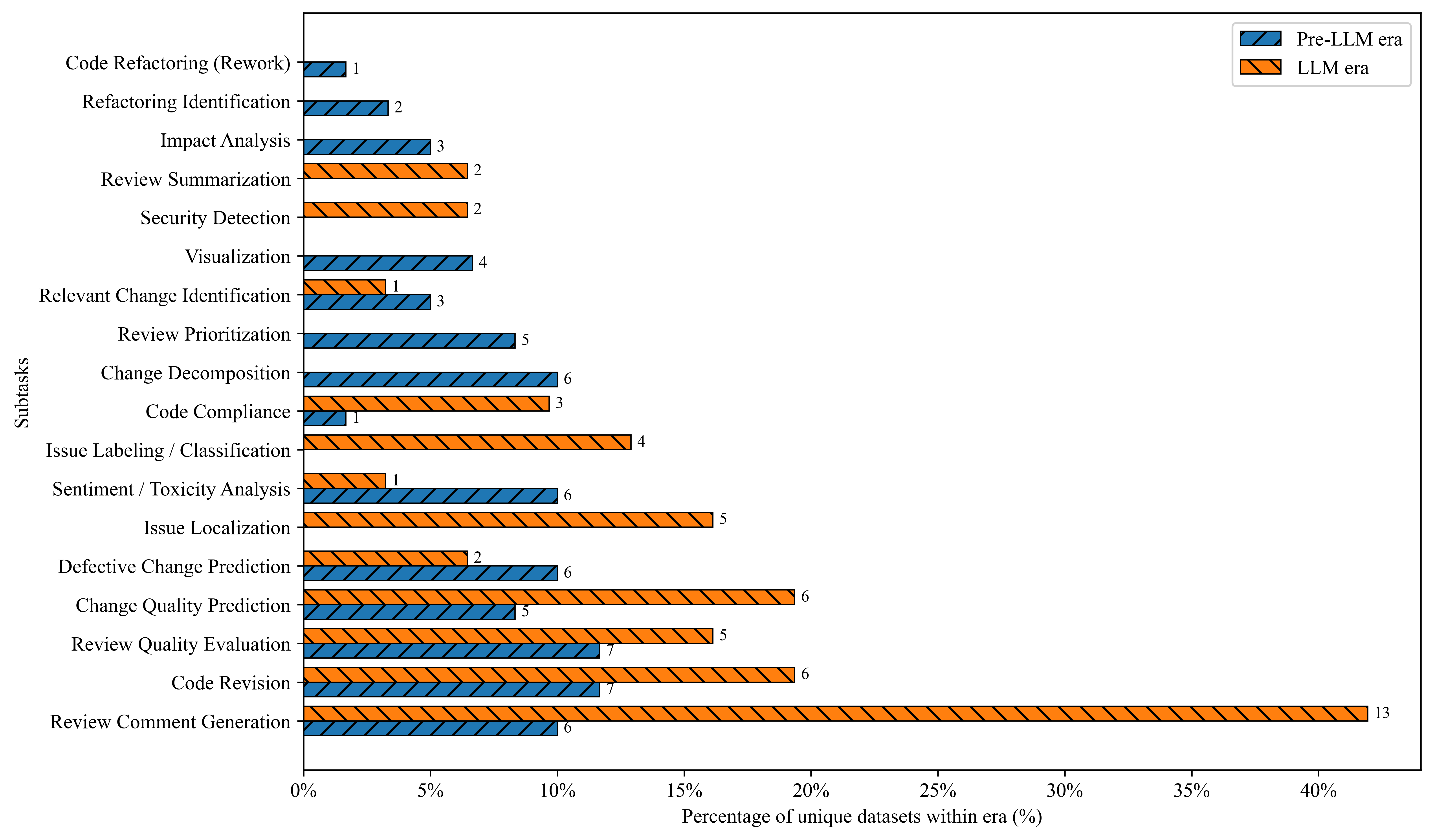}
    \caption{Comparative distribution of unique datasets across code review sub-tasks in the Pre-LLM and LLM eras. Values next to bars indicate the absolute number of datasets.}
    \label{fig:subtask_bar_chart}
\end{figure}

\subsection{Distribution of tasks}

Figure~\ref{fig:subtask_bar_chart} provides a complementary view of how datasets are distributed across code review-related sub-tasks in the Pre-LLM and LLM eras. The unit of analysis in Figure 7 is the unique dataset where each bar reports the number of unique datasets that support a sub-task, and percentages are computed over the 60 Pre-LLM and 31 LLM-era unique datasets. Because our coding is multi-label, a dataset that supports more than one sub-task appears in the bar of each of those sub-tasks. The bar values therefore should not be added together to obtain category-level totals; instead, we compute category-level percentage by counting each dataset once if it supports at least one sub-task in the category. In the Pre-LLM era, the datasets are more evenly distributed across sub-tasks, whereas the LLM era exhibits a more uneven distribution.

\textbf{The Change Understanding and Analysis category shows the clearest decline between the two eras.} In the Pre-LLM era, 16 of the 60 datasets (27\%) support at least one change-understanding sub-task, whereas only 1 of the 31 LLM-era datasets (3\%) evaluates change understanding as a standalone task. For instance, Change Decomposition and Impact Analysis do not appear as standalone benchmark tasks in the LLM-era corpus, although related capabilities are sometimes evaluated implicitly within broader tasks such as issue localization and security detection (e.g., CodeReviewQA decomposes review reasoning into discrete sub-capabilities). This suggests a shift in how understanding-related skills are packaged into benchmarks, rather than their disappearance as a reviewer competency.

\textbf{Peer Review is the largest category in the LLM era} where 22 of the 31 LLM-era datasets (71\%) support at least one Peer Review sub-task, compared with 14 of the 60 Pre-LLM datasets (23\%). Within this category, LLM-era benchmarks place more emphasis on end-to-end review generation and several studies evaluate reasoning quality, contextual understanding and adaptability across review scenarios. In addition to text-matching and generation metrics (e.g., BLEU and ROUGE), a growing number of studies use LLM-as-Judge as an evaluation method. 
One possible interpretation of the growth of Peer Review alongside the decline in standalone Change Understanding datasets is consolidation since in some LLM-era benchmarks, change understanding is no longer evaluated as a separate task but is treated as part of producing a review. This interpretation is drawn from descriptive dataset counts and is not tested in our analysis.

\subsection{Programming Language}

\begin{figure}[t] 
\centering \includegraphics[width=0.9\textwidth]{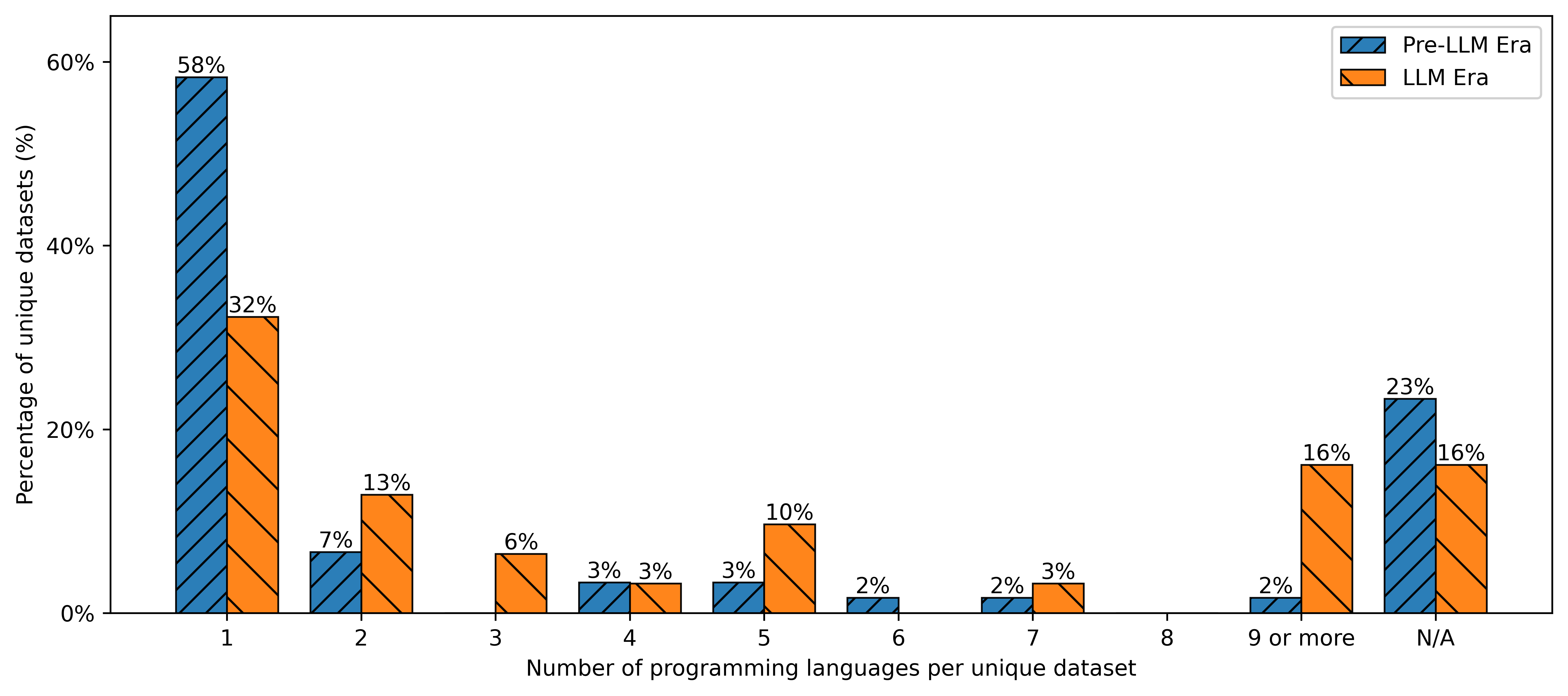} 
\caption{Distribution of programming language coverage in datasets across Pre-LLM and LLM eras. Values indicate the percentage of datasets within each era.} \label{fig:langugaeCompare.png} \end{figure}

\textbf{The transition to the LLM era is associated with a shift from predominantly language-specific datasets toward broader multilingual coverage.} Figure~\ref{fig:langugaeCompare.png} illustrates the percentage distribution of programming-language coverage across unique datasets, while Table~\ref{tab:language-coverage-counts} reports the corresponding underlying counts for direct verification. Consistent with the counting policy described in Section~3.1, this analysis is performed at the dataset level rather than the paper level: reused or overlapping dataset lineages are counted once and therefore the number of unique datasets differs from the total number of studies in each era.

In the Pre-LLM era, single-language datasets dominate the distribution, with 35 of 60 unique dataset entries (58\%) restricted to one programming language. Only 11 of 60 datasets (18\%) cover more than one programming language, while the programming language is not specified for 14 datasets (23\%). Broad multilingual coverage is uncommon: only 5 datasets (8\%) cover five or more languages, and just 1 dataset (2\%) covers nine or more languages.

In contrast, the de-duplicated LLM era dataset corpus exhibits a more dispersed and multilingual distribution. Single-language datasets account for 10 of 31 unique datasets (32\%), substantially lower than the 58\% observed in the Pre-LLM era. Datasets covering five or more programming languages account for 9 of 31 unique datasets (29\%), compared with 5 of 60 (8\%) in the Pre-LLM era. Similarly, 5 of 31 LLM era datasets (16\%) cover nine or more languages, whereas only 1 of 60 Pre-LLM datasets (2\%) falls into this category. These de-duplicated counts provide a more conservative view than paper-level counting because descendant studies reusing the same underlying dataset lineage do not create additional unique dataset entries. Overall, the results indicate greater representation of multilingual benchmark settings in the LLM era, although single-language datasets remain a substantial part of the landscape.

\begin{table*}[ht]
\centering
\resizebox{1\columnwidth}{!}{
\renewcommand{\arraystretch}{1}
\begin{tabular}{|>{\scriptsize\arraybackslash}p{5.10cm}|>{\scriptsize\arraybackslash}p{5cm}|>{\scriptsize\arraybackslash}p{5cm}|}

\hline
{\textbf{Languages Covered}} &
{\textbf{Pre-LLM Era}} &
{\textbf{LLM Era}} \\
\hline

{1} &
{35} &
{10} \\ \hline

{2} &
{4} &
{4} \\ \hline

{3} &
{0} &
{2} \\ \hline

{4} &
{2} &
{1} \\ \hline

{5} &
{2} &
{3} \\ \hline

{6} &
{1} &
{0} \\ \hline

{7} &
{1} &
{1} \\ \hline

{8} &
{0} &
{0} \\ \hline

{9 or more} &
{1} &
{5} \\ \hline

{Not specified} &
{14} &
{5} \\ \hline

{\textbf{Total unique datasets}} &
{\textbf{60}} &
{\textbf{31}} \\ \hline

\end{tabular}
}
\caption{{Number of unique datasets across programming languages}}
\label{tab:language-coverage-counts}
\end{table*}

\begin{figure}[t] \centering \includegraphics[width=0.9\textwidth]{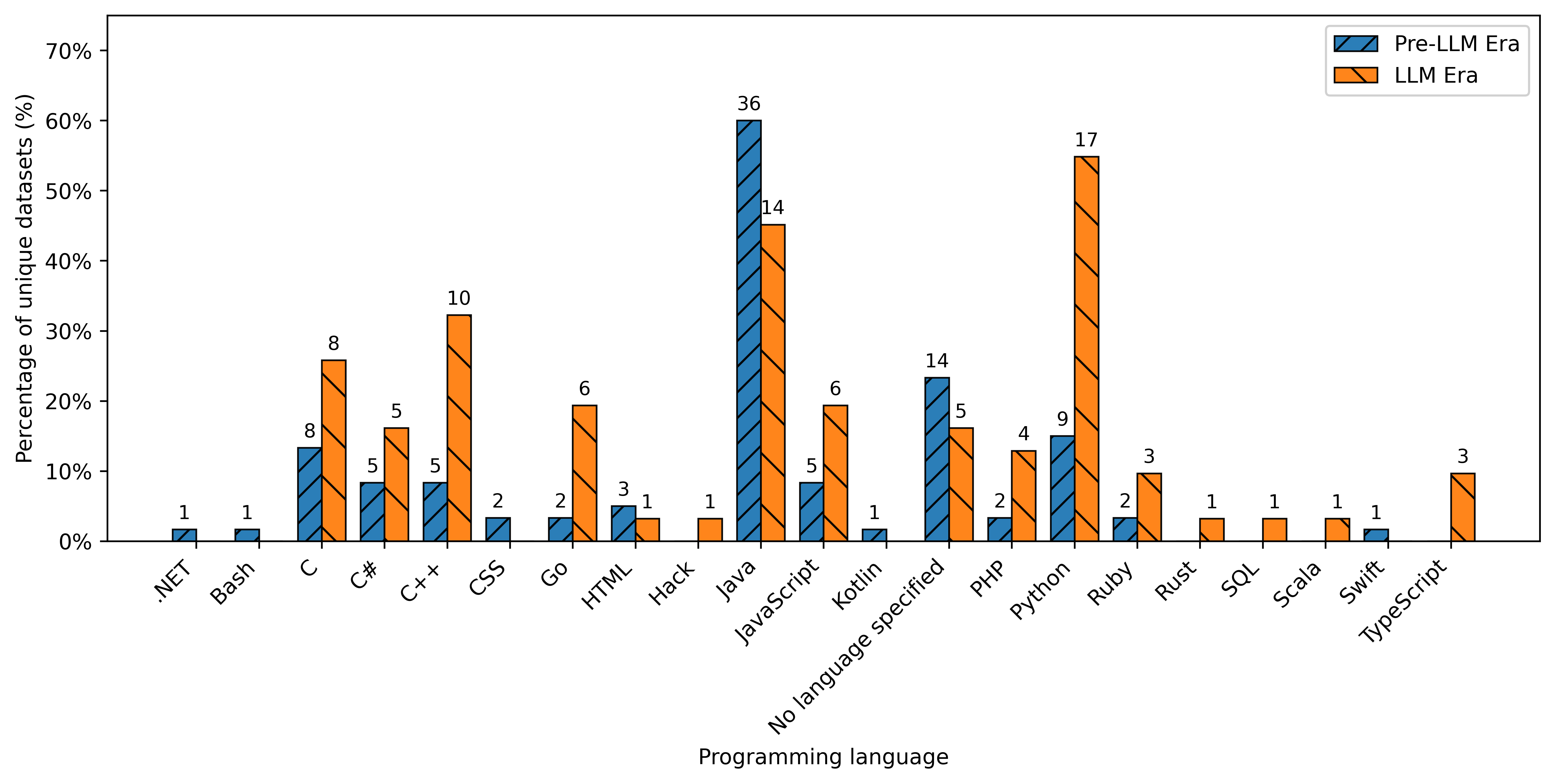} \caption{A specific programming language is covered by the datasets in the Pre-LLM and LLM eras. The values above each bar indicate the absolute number of datasets for each language.} \label{fig:SpecificlangugaeCompare.png} \end{figure}

Figure~\ref{fig:SpecificlangugaeCompare.png} presents the specific programming languages used across unique datasets in both eras. In the Pre-LLM era, Java dominates the datasets, accounting for approximately 60\% of all unique datasets, making it by far the most frequently used language. Python represents the second most common language at roughly 15\%, followed by C (about 13\%). Most other languages, including C++, PHP, Ruby, Go, and HTML, individually account for only a small fraction of datasets. In contrast, the LLM era exhibits a substantially more diverse language landscape. Although Java remains prominent, appearing in approximately 45\% of unique datasets, Python becomes the most widely represented language in this era, appearing in about 55\% of datasets. Additionally, other languages such as C++ (around 32\%), C (about 26\%), JavaScript (19\%), Go (19\%), C\# (16\%), PHP (13\%), and Ruby (10\%) contribute to the broader multilingual coverage of the LLM era dataset landscape. The LLM era also includes languages that were absent or negligible in the Pre-LLM corpus, including Rust, TypeScript, Scala, SQL, and Hack, reflecting an expansion in experimental scope. The proportion of unique datasets that do not specify a programming language decreases from approximately 23\% in the Pre-LLM era to about 16\% in the LLM era, indicating fewer unspecified-language entries in the LLM era dataset corpus.

\section{Discussion}\label{sec:dis}

\subsection{Limitations of Existing Benchmarks in the LLM Era and Future Directions}

Our survey reveals that while the volume and diversity of code review benchmarks have grown substantially in the LLM era, significant gaps remain in both task coverage and evaluation practice. In this section, we move beyond descriptive mapping to articulate what current benchmarks fail to measure, why current evaluation practices are inadequate, and how future datasets and protocols should be redesigned. We organize this discussion around three dimensions and derive four concrete benchmark design principles from our taxonomy and gap analysis.

\subsubsection{What Current Benchmarks Fail to Measure}
\label{sec:What Current Benchmarks Fail to Measure}

While LLM era benchmarks have made notable progress in evaluating comment generation and code revision, they leave large portions of the real-world code review process unmeasured. Our taxonomy-driven analysis reveals three categories of missing coverage.

\textbf{\textit{Macro-level review responsibilities.}} As shown in Figure~\ref{fig:subtask_bar_chart}, Change Decomposition and Impact Analysis do not appear as standalone benchmark tasks in the LLM-era corpus; no LLM-era dataset we surveyed covers either sub-task. Yet these tasks represent fundamental reviewer cognition skills that operate at a higher level of abstraction than comment generation. Change Decomposition requires a reviewer to recognize that a pull request conflates unrelated concerns and should be split into smaller, logically independent units before merging. Impact Analysis requires a reviewer to reason about which downstream components, dependencies, or build artifacts a change may affect. We did not identify an LLM-era benchmark in our corpus that measures whether a model can perform either of these responsibilities, so the macro-level judgment that human reviewers exercise before reading individual lines of code remains largely unevaluated.

\textbf{\textit{Refactoring and structural review skills.}} Code Refactoring has one Pre-LLM dataset in our corpus and none in the LLM era, and Refactoring Identification likewise appears only in the Pre-LLM era. The ability to distinguish behavior-preserving restructuring from behavior-changing edits is a critical reviewer skill, particularly in industrial settings where large refactoring changes are common and carry significant maintenance risk. The absence of these tasks from the LLM-era corpus means that we did not identify an LLM-era benchmark that measures whether a model can correctly classify, flag, or respond to refactoring-type changes during peer review.

\textbf{\textit{Dynamic functional correctness.}} The code revision benchmarks in our corpus evaluate model outputs mainly against a reference revision at the textual level (e.g., BLEU, Exact Match and CodeBLEU). Such metrics do not show whether a revision is functionally correct: a model can achieve a high BLEU or Exact Match score while proposing a revision that fails to compile, breaks existing tests or introduces a new runtime error. We did not identify a code revision benchmark in our corpus that reports build success or test pass rates as an offline evaluation metric, so functional correctness remains largely unmeasured in current benchmark evaluation.

\subsubsection{Why Current Evaluation Practices Are Inadequate}
\label{sec:Why Current Evaluation Practices Are Inadequate}

Beyond missing task coverage, our analysis reveals four structural failure modes in how existing benchmarks measure model performance, even within the tasks they do cover. We examine each evaluation paradigm in turn, assessing not just which metrics appear in the literature but whether and when they actually capture meaningful code review quality.

\textbf{\textit{Lexical similarity metrics: widespread but misaligned.}} Metrics such as BLEU, ROUGE, CodeBLEU, and Exact Match are by far the most commonly used evaluation tools in both Pre-LLM and LLM era benchmarks, appearing in the majority of datasets for Review Comment Generation and Code Revision tasks. However, there is a fundamental mismatch between what these metrics measure and what meaningful code review quality requires. Lexical similarity metrics reward outputs that are surface-level similar to a reference string, regardless of whether the output is semantically correct, actionable, or useful to the developer receiving it. For instance, in the context of review comment generation, a comment that uses similar vocabulary to the reference but points to the wrong line of code, misidentifies the root cause of a defect, or omits the most important issue in the change may score well on BLEU while being entirely unhelpful or even misleading. Conversely, a comment that correctly identifies a critical security vulnerability using different terminology than the reference may score poorly despite being the more valuable output.

The problem is compounded by the single-reference design of most existing benchmarks. Code review is inherently multi-perspective: two experienced reviewers may leave equally valid but lexically dissimilar comments on the same change. Evaluating against a single reference conflates lexical divergence from the reference with actual incorrectness, systematically underestimating model capability when the model produces valid but differently worded feedback. For Code Revision tasks specifically, the inadequacy of lexical metrics is even more severe. A model can achieve a high Exact Match or BLEU score while proposing a revision that fails to compile, introduces a new bug, or does not actually address the issue raised in the review comment. The functional correctness of a code revision — arguably its most important property — is entirely invisible to text-matching metrics. We therefore argue that lexical similarity metrics are necessary as a lightweight baseline but are insufficient as primary evaluation criteria for any code review task, and that the field's continued reliance on them as the dominant evaluation signal represents a significant methodological limitation.

\textbf{\textit{Human evaluation: necessary but underused and underspecified.}} Human evaluation appears in a subset of benchmarks across both eras, primarily through user studies, Likert-scale ratings, and acceptance rate measures. When properly designed, human evaluation is the most direct way to measure whether a review comment is useful, clear, and actionable from the perspective of the developer receiving it, which no automated metric currently captures reliably. However, our analysis reveals that human evaluation in existing benchmarks is frequently underspecified and difficult to replicate. Many studies report human evaluation results without disclosing the number of evaluators, their level of expertise, the rating rubric applied, or inter-rater agreement statistics. This makes it impossible to assess the reliability of reported results or to compare human evaluation findings across studies. Furthermore, most human evaluations in the surveyed literature assess perceived usefulness or clarity at a coarse granularity, without measuring whether reviewers actually acted on the feedback, whether the feedback led to code changes, or whether it improved the quality of the final merged code. Concretely, human evaluation should report evaluator expertise and count, inter-rater agreement, and a clearly defined rubric.

\textbf{\textit{LLM-as-Judge: promising but uncalibrated. }} The LLM-as-Judge evaluation paradigm has emerged rapidly in the LLM era as a scalable alternative to human evaluation, and several recent benchmarks in our survey use it as a primary evaluation signal for review comment generation, including structured review comment generation. While LLM-as-Judge offers practical advantages in terms of cost and scalability, our analysis reveals that its adoption in code review benchmarks has largely proceeded without calibration against human judgments. The validity of LLM-as-Judge as an evaluation method depends on whether the judging model's assessments correlate with human expert opinions on review quality. In domains where this correlation has been empirically verified, LLM-as-Judge can serve as a reliable proxy for human evaluation at scale. However, in the current code review literature, most benchmarks that use LLM-as-Judge do not report this correlation, making it unclear whether the scores reflect genuine review quality or artifacts of the judging model's preferences and biases. There is also a risk of self-evaluation bias when the model being evaluated and the model acting as judge share the same architecture or pre-training data. We therefore recommend that benchmarks adopting LLM-as-Judge evaluation establish its reliability by reporting the correlation between judge scores and human expert ratings on a held-out calibration set, and that this calibration be treated as a required quality criterion for benchmark acceptance in the field.

\textbf{\textit{Execution-based evaluation: rare but essential.}} The code revision benchmarks in our survey mainly evaluate a model's revision by comparing it with a reference revision at the textual level, using metrics such as BLEU, Exact Match and CodeBLEU, which do not show whether the revision is functionally correct. \citet{LP11} illustrate why different evaluation settings need to be distinguished. Their offline evaluation reports Exact Match together with Successful Patch Generation (SPG). The paper's introduction describes SPG as the rate of patches that compile, pass lint checks and pass all tests, whereas its formal definition in the evaluation methodology measures whether a patch can be generated and applied to the file. Build, lint and test checks are applied in the paper's production deployment, where a patch is shown to the author only after passing these validity checks or being approved by the reviewer. Execution-related checks therefore appear in deployment settings, but we did not identify a code revision benchmark in our corpus that reports build success or test pass rates as an offline evaluation metric. These measures are among the most direct indicators of whether a revision works, because they require the revised code to compile and pass existing tests. A model can achieve a high Exact Match or BLEU score while proposing a revision that fails to compile, introduces a new bug, or does not address the issue raised in the review comment. We therefore recommend that future code revision benchmarks adopt execution-based evaluation as a required component, implemented through sandboxed build and test environments such as Docker-based build and test pipelines, and report text-matching metrics only as secondary measures. This approach has been successfully adopted in code generation benchmarks ~\citep{related1} and should be extended to the code review setting. 

Table~\ref{tab:Critical assessment of evaluation paradigms across code review benchmark tasks.} summarizes the commonly used tasks and limitations for each evaluation paradigm. 







\begin{table*}[ht]
\centering
\resizebox{1\columnwidth}{!}{
\renewcommand{\arraystretch}{1}
\begin{tabular}{|>{\scriptsize\arraybackslash}p{3.10cm}|>{\scriptsize\arraybackslash}p{5cm}|>{\scriptsize\arraybackslash}p{7cm}|}

\hline
{\textbf{Evaluation Paradigm}} &
{\textbf{Commonly Used For}} &
{\textbf{Key Limitations}}\\
\hline

{Lexical similarity metrics} &
{Review comment generation, review summarization, code revision} &
{Does not measure semantic correctness, functional validity, or usefulness. Single-reference evaluation may penalize valid but differently worded outputs.} \\
\hline

{Human evaluation} &
{Comment quality, review helpfulness, applicability, semantic equivalence} &
{Underspecified and hard to replicate when evaluator count, expertise, rubric, or inter-rater agreement are not reported; rarely measures downstream impact.} \\
\hline

{Execution-based evaluation} &
{Code revision, code refinement} &
{Rarely used in current code review benchmarks; requires executable projects, tests, and sandboxed evaluation environments.} \\
\hline

{LLM-as-Judge} &
{Comment quality, structured review comment generation, generated review assessment} &
{Often uncalibrated against human judgment; may introduce judge bias or self-evaluation bias when the judging model is similar to the evaluated model.} \\
\hline

\end{tabular}
}
\caption{{Critical assessment of evaluation paradigms across code review benchmark tasks.}}
\label{tab:Critical assessment of evaluation paradigms across code review benchmark tasks.}
\end{table*}

\subsubsection{How Future Datasets and Protocols Should Be Redesigned.}

Drawing on the gap analysis in section~\ref{sec:What Current Benchmarks Fail to Measure} and Section~\ref{sec:Why Current Evaluation Practices Are Inadequate}, we propose four concrete benchmark design principles for the next generation of LLM-based code review evaluation. These principles are grounded in our taxonomy and are intended to guide dataset construction, metric selection, and evaluation protocol design.

\textbf{\textit{Principle 1 — Taxonomy-Aligned Task Coverage.}} New benchmarks should be designed with explicit reference to the 18 sub-tasks in our taxonomy, with clear justification when a sub-task is excluded. In particular, we recommend prioritizing the development of LLM era benchmarks for Change Decomposition, Impact Analysis, and Refactoring Identification. These sub-tasks are well-motivated by the real-world code review process, were actively studied in the Pre-LLM era, and have substantial existing datasets that could be adapted or extended for LLM evaluation. For Change Decomposition, future benchmarks could measure whether an LLM can correctly identify that a pull request should be split and propose a valid decomposition. For Impact Analysis, benchmarks could measure whether a model can correctly predict which files, modules, or build targets are affected by a given change, evaluated against ground-truth impact sets derived from build dependency graphs or test coverage data. Furthermore, while some datasets for these tasks exist in the Pre-LLM era, they are often limited to specific languages such as Java, C, and C\#. We encourage the development of new benchmarks that expand coverage to languages such as Python and Go to align with the current LLM landscape.

\textbf{\textit{Principle 2 — Execution-Grounded Evaluation. }}Future code revision benchmarks should integrate sandboxed runtime verification so that functional correctness can be measured alongside textual similarity. Concretely, we recommend a Docker-based evaluation protocol in which, after a model proposes a code revision, the benchmark automatically attempts to build the project with the revision applied and runs the existing regression test suite. The evaluation then reports not only BLEU or Exact Match scores but also build success rate, test pass rate, and regression failure rate. 

This recommendation is consistent with broader code generation evaluation practices, where execution-based metrics and test-based correctness measures have been used to assess generated code \citep{related1}. 

This approach has been successfully adopted in code generation benchmarks \citep{LP5, related1} and should be extended to the code review setting. Such dynamic evaluation would allow researchers to distinguish models that produce plausible-sounding but non-functional revisions from those that produce revisions that are genuinely correct. More broadly, future benchmarks across all code review tasks should incorporate runtime information where available, including CI/CD signals, static analysis outputs, and test coverage data, to better reflect the execution context that human reviewers routinely use when assessing code changes.

\textbf{\textit{Principle 3 — Multi-Reference and Calibrated LLM-as-Judge Evaluation.}} Future comment generation benchmarks should collect multiple reference reviews per example from different experienced reviewers, enabling evaluation metrics that reward outputs matching any valid reference rather than a single one. In parallel, benchmarks that use LLM-as-Judge as an evaluation method should establish its reliability through calibration against human judgments before adopting it as a primary metric. Concretely, we recommend that benchmark papers report the correlation between LLM-as-Judge scores and human expert ratings on a held-out calibration set, and include this correlation as part of the benchmark's quality characterization. Without such calibration, LLM-as-Judge scores are difficult to interpret and compare across studies. Human evaluation protocols should additionally report evaluator count, expertise level, inter-rater agreement, and a clearly defined rubric assessing at minimum correctness, actionability, and necessity of generated review comments.

\textbf{Principle 4 — Structured Cross-Language Generalization Testing.} Given the pronounced shift toward multilingual benchmarks in the LLM era, future benchmarks should include structured cross-language splits that allow explicit measurement of zero-shot transfer performance, rather than simply pooling all programming languages into a single undifferentiated evaluation set. A structured split would designate certain languages as held-out test languages during training and fine-tuning, enabling direct measurement of how well a model generalizes to programming languages it has not been explicitly trained on for the code review task. This is particularly important given that many real-world development environments involve less common or domain-specific languages that are unlikely to be well represented in pre-training corpora.

Table~\ref{tab:benchmark-gaps-design-principles} summarizes the relationship between the benchmark gaps identified in our analysis, the corresponding design principles, and the taxonomy sub-tasks most directly affected.

\begin{table*}[ht]
\centering
\resizebox{1\columnwidth}{!}{
\renewcommand{\arraystretch}{1}
\begin{tabular}{|>{\scriptsize\arraybackslash}p{3.10cm}|>{\scriptsize\arraybackslash}p{5cm}|>{\scriptsize\arraybackslash}p{7cm}|}

\hline
{\textbf{Benchmark Gap}} &
{\textbf{Design Principle}} &
{\textbf{Affected Sub-tasks}}\\
\hline

{Missing macro-level tasks} &
{Principle 1: Taxonomy-aligned coverage} &
{Change Decomposition, Impact Analysis} \\
\hline

{Missing structural review tasks} &
{Principle 1: Taxonomy-aligned coverage} &
{Refactoring Identification, Code Refactoring} \\
\hline

{Surface-level metrics} &
{Principle 2: Execution-grounded evaluation} &
{Code Revision, Review Comment Generation} \\
\hline

{Single-reference evaluation} &
{Principle 3: Multi-reference evaluation} &
{Review Comment Generation, Peer Review} \\
\hline

{No execution context} &
{Principle 2: Execution-grounded evaluation} &
{Code Revision, Defective Change Prediction} \\
\hline

{Uncalibrated LLM-as-Judge} &
{Principle 3: Calibrated evaluation} &
{Review Comment Generation (including structured comment generation)} \\
\hline

{Underspecified human evaluation} &
{Principle 3: Multi-reference evaluation} &
{All comment generation and quality tasks} \\
\hline

{Unstructured multilingual evaluation} &
{Principle 4: Cross-language generalization} &
{All Peer Review and Code Refinement sub-tasks} \\
\hline

\end{tabular}
}
\caption{{Relationship between benchmark gaps, design principles, and affected taxonomy sub-tasks.}}
\label{tab:benchmark-gaps-design-principles}
\end{table*}

\subsection{Threats to Validity}
    \textbf{Internal Validity:} One threat to our data collection is the potential omission of relevant studies. While we employed a snowballing approach starting from major SE (Software Engineering) and AI venues~\citep{parker2019snowball}, there is a risk that certain papers might have been missed. To minimize this, we utilized multiple iterations of both forward and backward snowballing. We also cross-referenced our initial seed set with well-known repositories and previous surveys to ensure the most influential datasets were captured. To extend our search space, we also collected papers from ArXiv. However, we acknowledge that the rapid growth of AI-driven code review means new datasets are published weekly, making absolute exhaustiveness a moving target.

    \textbf{Construct Validity:} The categorization of tasks is subject to human interpretation. Human bias or fatigue could lead to inconsistent labeling across the survey. We addressed this by implementing a dual-reviewer protocol. Each paper was labeled independently by at least two authors. In cases of disagreement, a third senior researcher acted as an arbitrator to reach a consensus. We report Cohen's Kappa as a corpus-level reliability measure but did not track arbitration frequency by high-level category; future replications of this taxonomy should log disagreement rates per category to support finer-grained reliability reporting.
    
    \textbf{External Validity:} Our findings are conditioned on the databases, venues, search terms, screening criteria, and time window described in Section~\ref{subsec:Paper Collection}. Although we searched multiple scholarly databases, including IEEE Xplore, ACM Digital Library, Elsevier ScienceDirect, ACL Anthology, arXiv, Google Scholar, and SpringerLink, and further applied snowballing, our corpus may still not include every relevant code review benchmark or evaluation study. Relevant work may have appeared in channels outside our search scope, such as non-English venues, non-indexed workshops, industry engineering blogs, or internal technical reports. This limitation is particularly relevant for industrial code review. Some industrial datasets, tools, and evaluation protocols are proprietary and may not be publicly released. Therefore, our findings may better reflect the published and publicly accessible benchmark literature than the full range of code review practices used in industry. In addition, our search period ends on December 31, 2025. Given the rapid pace of LLM-related research, benchmarks released after this cutoff may change the observed distribution of tasks, datasets, and evaluation practices. Therefore, our conclusions should be interpreted within the scope of the collected corpus rather than as universal claims about all code review research or practice.
    
    \textbf{Conclusion Validity:} Several of our comparative findings are based on counts within fine-grained sub-tasks. Because some sub-tasks contain only a small number of papers or datasets, sub-task-level comparisons can be sensitive to the inclusion, exclusion, or classification of individual studies. For example, the observed reduction in standalone Change Understanding and Analysis datasets from the Pre-LLM era to the LLM era should be interpreted as a descriptive trend within our corpus, rather than as definitive evidence that this task has disappeared from code review research. Likewise, the consolidation of review activities into broader review settings is one possible interpretation of these counts; it is not tested in our analysis and should not be read as a causal explanation. Similarly, percentage-based comparisons may appear large when they are calculated over modest category-level counts. Because our goal is descriptive synthesis rather than hypothesis testing, we do not present the observed cross-era differences as statistically significant effects. Moreover, because our coding scheme permits multi-label assignment, sub-task-level counts are not strictly independent observations. For these reasons, we present cross-era comparisons as descriptive corpus-level trends, not as statistically validated effects or causal claims. To reduce over-interpretation, we report counts and percentages where appropriate and use the taxonomy as an analytical framework for identifying coverage gaps and evaluation patterns.
    
\section{Conclusion}\label{sec:conclusion}
Our survey indicates a shift in the focus of code review benchmarks. Pre-LLM benchmarks more often targeted tasks that support human reviewers, such as review prioritization, change decomposition and visualization, whereas LLM-era benchmarks place greater emphasis on generating review outputs. Change understanding tasks, supported by 16 of the 60 Pre-LLM datasets (27\%), appear as a standalone task in only one of the 31 LLM-era datasets (3\%), while 22 of the 31 LLM-era datasets (71\%) support at least one Peer Review sub-task. These are descriptive counts from our corpus. 

Furthermore, benchmarks have evolved from language-specific studies toward broader multilingual coverage. The prevalence of single-language datasets decreased from 58\% in the Pre-LLM era to 32\% in the LLM era, while highly multilingual benchmarks covering nine or more languages increased to 16\%. Despite these shifts, future research should address current benchmark limitations by incorporating macro-level responsibilities, such as impact analysis, and transitioning toward dynamic evaluation methods. By providing a multi-level taxonomy and systematic classification of 18 sub-tasks, our work establishes a foundation for more rigorous and context-aware code review automation.
\section{Declarations}

\subsection{Funding}

The research is supported by the NSERC, and the project number is RGPIN-2021-03354. 

\subsection{Ethical Approval}
This study consists of a systematic survey of publicly available research papers. As the research involves the secondary analysis of published literature and does not involve human participants, animal subjects, or the collection of sensitive personal data, formal ethical approval from an Institutional Review Board (IRB) was not required. All data used in this study were obtained from open scholarly databases and are cited accordingly.

\subsection{Informed consent}
This study did not involve human participants, individual clinical data, or primary data collection from living individuals. As the analysis was limited to publicly available literature and previously published datasets, informed consent was not required.

\subsection{Author Contributions}

The authors confirm their contribution to the paper as follows: 
\begin{itemize}
    \item \textbf{Conceptualization:} SHAOWEI WANG, HAOXIANG ZHANG, TSE-HSUN CHEN;

    \item \textbf{Methodology and Search Strategy:} TAUFIQUL ISLAM KHAN, SHAOWEI WANG; 

    \item \textbf{Data Curation and Formal Analysis (Paper Review):} TAUFIQUL ISLAM KHAN, SHAOWEI WANG, HAOXIANG ZHANG; 

    \item \textbf{Writing - Original Draft Preparation:} TAUFIQUL ISLAM KHAN, SHAOWEI WANG,  HAOXIANG ZHANG, TSE-HSUN CHEN; 

    \item \textbf{Writing - Review \& Editing:} TAUFIQUL ISLAM KHAN, SHAOWEI WANG,  HAOXIANG ZHANG, TSE-HSUN CHEN;
\end{itemize}

\subsection{Data Availability}
All data analyzed during this study are included in this published article. The list of papers surveyed can be found in the References section, and the primary evaluation metrics extracted from the original paper.

\subsection{Conflict of Interest}

The authors declare that they have no competing interests, financial or otherwise, related to the research conducted or the papers surveyed in this study.

\subsection{Clinical Trial Number}
Clinical trial number: not applicable.

\newpage
\bibliographystyle{ACM-Reference-Format}
\IfFileExists{latex/references.bib}{\bibliography{latex/references}}{\bibliography{references}}

\end{document}